\documentclass{emulateapj}

\def\kms{\,km\,s$^{-1}$}
\def\mo{M$_\odot$\,}

\def\Dwa{$\,$\uppercase\expandafter{\romannumeral5}$\,$}

\def\sles{\lower2pt\hbox{$\buildrel {\scriptstyle <}
   \over {\scriptstyle\sim}$}}

\def\sgreat{\lower2pt\hbox{$\buildrel {\scriptstyle >}
   \over {\scriptstyle\sim}$}}
\def\sharpnull#1{}
\def\aa{Astron. Astrophys.}

\begin{document}

\title{Multi-Dimensional Simulations of the Accretion-Induced 
Collapse of White Dwarfs to Neutron Stars}

\author{L. Dessart\altaffilmark{1},
A. Burrows\altaffilmark{1},
C.D. Ott\altaffilmark{2},
E. Livne\altaffilmark{3},
S.-Y. Yoon\altaffilmark{4},
N. Langer\altaffilmark{5}
}
\altaffiltext{1}{Department of Astronomy and Steward Observatory,
                 The University of Arizona, Tucson, AZ \ 85721;
                 luc@as.arizona.edu,burrows@as.arizona.edu}
\altaffiltext{2}{Max-Planck-Institut f\"{u}r Gravitationsphysik,
Albert-Einstein-Institut, Golm/Potsdam, Germany; cott@aei.mpg.de}
\altaffiltext{3}{Racah Institute of Physics, The Hebrew University,
Jerusalem, Israel; eli@frodo.fiz.huji.ac.il}
\altaffiltext{4}{Astronomical Institute ``Anton Pannekoek'', University of Amsterdam, 
Kruislaan 403, 1098 SJ, Amsterdam, The Netherlands; scyoon@science.uva.nl}
\altaffiltext{5}{Astronomical Institute, Utrecht University, 
Princetonplein 5,3584 CC, Utrecht, The Netherlands; n.langer@astro.uu.nl}

\begin{abstract}

We present 2.5D radiation-hydrodynamics simulations of the 
accretion-induced collapse (AIC) of white dwarfs, starting from 2D 
rotational equilibrium configurations of a 1.46-M$_{\odot}$ 
and a 1.92-M$_{\odot}$ model. Electron capture leads to 
the collapse to nuclear densities of these cores within a few 
tens of milliseconds. The shock generated at bounce moves slowly, 
but steadily, outwards. Within 50-100\,ms, the stalled shock breaks 
out of the white dwarf along the poles. The blast is followed by 
a neutrino-driven wind that develops within the white dwarf, in 
a cone of $\sim$\,40$^{\circ}$ opening angle about the poles, with a 
mass loss rate of 5-8$\times$10$^{-3}$M$_{\odot}$yr$^{-1}$. 
The ejecta have an entropy on the order of 
20-50 k$_{\rm B}$/baryon, and an electron fraction distribution that is 
bimodal. By the end of the simulations, at $\ge$600\,ms after bounce, 
the explosion energy has reached 3-4$\times$10$^{49}$\,erg and the total ejecta 
mass has reached a few times 0.001\,M$_{\odot}$. We estimate the asymptotic 
explosion energies to be lower than 10$^{50}$\,erg, significantly lower 
than those inferred for standard core collapse. 
The AIC of white dwarfs thus represents one instance where a neutrino 
mechanism 
leads undoubtedly to a successful, albeit weak, explosion. 

We document in detail the numerous effects of the fast rotation of the 
progenitors: 
The neutron stars are aspherical; the ``$\nu_{\mu}$'' and $\bar{\nu}_e$ 
neutrino luminosities are reduced compared to the $\nu_e$ neutrino 
luminosity; the 
deleptonized region has a butterfly shape; the neutrino flux and 
electron fraction depend strongly upon latitude ({\it \`a la} von Zeipel); 
and a quasi-Keplerian 0.1-0.5-M$_{\odot}$ accretion disk is formed. 

\end{abstract}

\keywords{hydrodynamics -- neutrinos -- rotation -- stars: neutron -- 
stars: supernovae: general -- stars: white dwarfs}

\section{Introduction}

  Stars can follow a few special evolutionary routes to form an unstable 
Chandrasekhar mass core. A main-sequence star of more than 
$\sim$8\,\mo evolves to form either a degenerate O/Ne/Mg core (Barkat et al. 1974; 
Nomoto 1984,1987; Miyaji \& Nomoto 1987) or a degenerate Fe core (Woosley \& Weaver 1995), 
which, due to photodisintegration of heavy nuclei and/or electron capture, 
collapses to form a protoneutron star (PNS).
If an explosion ensues, the event is associated with a Type II supernovae (SN).
Less massive stars end their lives as white dwarfs. 
White dwarfs located in a binary system may accrete from a companion and
achieve the Chandrasekhar mass, triggering the thermonuclear runaway of the object and leading
to Type Ia SN, leaving no remnant behind.

However, a third class of objects is expected. Theoretically, massive white dwarfs with O/Ne/Mg cores, 
due to their high central density ($\sgreat$10$^{10}$\,g\,cm$^{-3}$), experience rapid electron 
capture that leads to the collapse of the core. This is accretion-induced collapse (AIC),
an alternative path to stellar disruption through explosive burning, currently associated with 
Type Ia SN (Nomoto \& Kondo 1991).
% Various paths may lead to a white dwarf with such properties. 
% LUC
It is presently unclear what fraction of all white dwarfs will lead to AICs,
but of those white dwarfs that evolve to form a Chandrasekhar-mass O/Ne/Mg core, 
all will necessarily undergo core collapse.
One formation channel is the coalescence of two white dwarfs (Mochkovitch \& Livio 1989), 
with either C/O or O/Ne/Mg cores, although few such binary systems have yet been 
observed with a cumulative mass above the Chandrasekhar mass.
There is still uncertainty as to whether such binary systems would not undergo 
thermonuclear runaway rather than collapse.
Since the coalescence of two white dwarfs requires a shrinking of the 
orbit through gravitational radiation, these systems will take many gigayears to coalesce.
An alternative formation mechanism is via single-degenerate systems, through a combination 
of high original white dwarf mass and mass and angular-momentum accretion by mass transfer from a 
(non-degenerate) H/He star (Nomoto \& Kondo 1991).
Binary star population synthesis codes predict the occurence of the AIC of white dwarfs 
with a galactic rate of 8$\times$10$^{-7}$ yr$^{-1}$ to 8$\times$10$^{-5}$ yr$^{-1}$, depending, 
amongst other things, on the treatment of the common-envelope phase and mass transfer 
(Yungelson \& Livio 1998).
The set of parameters leading to a Type Ia rate of 10$^{-3}$\,yr$^{-1}$ corresponds to 
an AIC rate of 5$\times$10$^{-5}$ yr$^{-1}$. The observed Type Ia rate of 
$\sim$3$\times$10$^{-3}$\,yr$^{-1}$ (Madau et al. 1998; Blanc et al. 2004; Manucci et al. 2005) 
would imply a galactic AIC rate of 1.5$\times$10$^{-4}$\,yr$^{-1}$.
These rates are likely functions of galaxy and metallicity (Yungelson \& Livio 2000; 
Belczynski et al. 2005; Greggio 2005; Scannapieco \& Bildsten 2005).
Based on r-process nucleosynthetic yields obtained from previous simulations of the AIC of white dwarfs, 
Fryer et al. (1999) inferred rates ranging from $\sim$10$^{-5}$ to $\sim$10$^{-8}$\,yr$^{-1}$.
Overall, AICs are not expected to occur more than once per 20--50 standard Type Ia events; because 
they are intrinsically rarer, they remain to be identified and observed in Nature.

Whatever their origin, their fundamental nature is to accrete both mass
and angular momentum from a companion object. Rotation is, therefore, a key physical component. 
Prior to core collapse, differential rotation acts as a stabilizing agent
for shell burning by widening its spatial extent through enhanced mixing and reducing the envelope 
density through centrifugal support (Yoon \& Langer 2004). 
Mass accretion also leads to an increase of the central density, which may rise up 
to a few $10^{10}$ g\,cm$^{-3}$, establishing suitable conditions for efficient electron capture 
on Mg/Ne nuclei.
By the time of core collapse and depending on the evolutionary path followed, such white dwarfs may cover a 
range of masses from $\sgreat$1.35\,\mo up to $\sim$2\,\mo (2.7\,\mo) in the case of 
a non-degenerate (degenerate) companion, potentially well in excess of the standard 
Chandrasekhar limit, and possessing an initial rotational energy up to $\sim$10\% of their 
gravitational binding energies.  
At such values, the centrifugal potential leads to a deformation of the white dwarf from 
spherical symmetry, equipotentials and isopressure surfaces adopting a peanut-like shape 
in cross section for the largest rotation rates.  Such structures are obtained in the 
2D differentially rotating equilibrium white dwarf models constructed by Yoon \& Langer 
(2005, YL05; see also Liu \& Lindblom 2001).

% Guided by 1D studies of rotating accreting white dwarfs (Yoon \& Langer 2004), 
% Yoon \& Langer (2005, YL05) constructed 2D differentially rotating equilibrium white dwarf models for a wide
% range of central densities, total angular momenta and masses. In the present work, we study 
% two progenitors with masses at the extremes of the spectrum computed by YL05, i.e., the models with 
% masses 1.46\,\mo and 1.92\,\mo. Both start with a central density of 5$\times$10$^{10}$\,g\,cm$^{-3}$.

In the past, the collapse of O/Ne/Mg cores originating from stars in the 8-10\,\mo 
range has been studied in 1D by Baron et al. (1987ab), Mayle \& Wilson (1988), and 
Woosley \& Baron (1992), who showed that the shock generated at core bounce stalls 
rather than leading to a prompt explosion.
However, Hillebrandt et al. (1984) and Mayle \& Wilson (1988) obtained delayed explosions,
the former after 20-30\,ms, and the latter after $\sim$200\,ms, supposedly driven by 
neutrino energy deposition behind the stalled shock.
Woosley \& Baron (1992) found the emergence of a sustained neutrino-driven wind, with a mass loss 
rate of 0.005\,\mo\,s$^{-1}$ and with an ejecta electron fraction of the order of 0.45, showing promise,
modulo uncertainties, for a contribution to the enrichment of the ISM in r-process elements.
Using 1D/2D SPH simulations, Fryer et al. (1999) reproduced the simulations of Woosley \& Baron (1992), 
confirmed that the shock stalls due to the copious neutrino losses associated with core bounce, 
did not find prompt explosion, and, depending on the equation of state (EOS) employed, observed a 
delayed explosion.
They focused mostly on the early phase, prior to the neutrino-driven wind, and found an ejected mass 
of low-$Y_{\rm e}$ material of $\sim$0.05\,\mo, depending on adopted model assumptions.
Their 2D simulations with solid-body rotation showed similar properties to their 1D equivalents,
the authors attributing the small differences to the different grid resolution.
This may partly stem from the essentially spherical explosion triggered just $\la$100\,ms
after core bounce, ejecting the fast-rotating material in the outer mantle.
Consequently, at the end of their 2D simulations, they obtain very slow rotation rates for the 
PNS, i.e., of $\sim$1\,s.
Recently, Kitaura et al. (2005) re-inspected the collapse of the progenitor model used by 
Hillebrandt et al. (1984). They confirmed again the by-now well-accepted idea that no prompt
explosion occurs, but instead obtain a successful, though 
sub-energetic, delayed explosion in spherical symmetry, powered by neutrino heating and a 
neutrino-driven wind that sets in $\sim$200\,ms after bounce.

A primary motivation for this work is to improve upon these former investigations that 
assumed one-dimensionality, sphericity, and/or zero-rotation, and start instead from the more
physically-consistent 2D models of YL05, thereby fully accounting for the effects of rotation 
on the collapse, bounce, and post-bounce evolution of the white dwarf core and envelope, as 
well as for the strong asphericity of the progenitor.
Our study uses VULCAN/2D (Livne et al. 2004; Walder et al. 2005) to perform 2D 
Multi-Group Flux Limited Diffusion (MGFLD) radiation hydrodynamics simulations.
As we will demonstrate, rotation plays a major role 
in the post-bounce evolution, making 1D investigations of such objects of limited utility. 
By carrying out the simulations from $\sim$30\,ms before bounce to $\sgreat$600\,ms after 
bounce, we capture a wide range of physical processes, including the establishment 
of a strong, fast, and aspherical neutrino-driven wind.
We model the centrifugally-supported equatorial regions and the large angular momentum budget 
leading to the formation of a sizable accretion disk.
% LUC
Moreover, an Eulerian investigation is better-suited than a Lagrangean approach to explore
the neutrino-driven wind that develops after $\sgreat$200\,ms.
Finally, despite the relative scarcity of AIC in Nature, these simulations represent interesting examples 
for the formation of disks around neutron stars.
% , and for the collapsar model of gamma-ray bursts.

The main findings of this work are the following:
We find that the AIC of white dwarfs forms $\sim$1.4-\mo neutron stars, expelling a modest 
mass of a few 10$^{-3}$\,\mo mostly through a neutrino-driven wind that develops $\sgreat$200\,ms
after bounce, and that they lead to very modest explosion energies of 5-10$\times$10$^{49}$erg.
Accounting for the rotation and the asphericity of the progenitor white dwarfs reveals a wealth of
phenomena. The shock wave generated at core bounce 
emerges through the poles rather than the equator, and it is in this excavated polar region
that the neutrino-driven wind develops. The strong asphericity of the newly-formed protoneutron stars
leads to a latitudinally-dependent neutrino flux, while the effects of rotation modify the relative 
flux magnitude of different neutrino flavors. 
Besides mass accretion by both the neutron star and mass ejection by the initial blast and the subsequent
wind, we find a sizable component that survives and resides in a quasi-Keplerian disk, which obstructs the wind flow
at low latitudes. This disk will be accreted by the neutron star only on longer, viscous timescales.

In the next two sections, we present the two selected progenitor models in more detail; we also discuss the 
radiation-hydrodynamics code VULCAN/2D and the various assumptions made. In \S\ref{sect_results}, 
we present
the simulation results, focusing on the general temporal evolution from the start until $\sgreat$600\,ms, 
a time by which the neutrino-driven wind has reached a steady state.
We then analyse in more detail the various components of these simulations.
In \S\ref{sect_pns}, we discuss the properties of the nascent neutron stars, 
with special attention paid to the geometry of the neutrinospheres.
In \S\ref{sect_nu}, we discuss the neutrino signatures, both in terms of luminosity and
energy distribution.
In \S\ref{sect_disk}, we focus on the residual material lying at low latitudes, forming a
quasi-Keplerian disk.
In \S\ref{sect_ener}, we turn to the energetics of the explosion and describe in detail
the main component of the simulations at late times, i.e., the neutrino-driven wind.
In \S\ref{sect_ye}, we analyse the electron fraction of the ejected material and address
the relevance of the AIC of white dwarfs for neutron-rich element pollution
of the interstellar medium.
In \S\ref{sect_gw}, we present the gravitational-wave signal predicted for the AIC of our
white dwarf models.
In \S\ref{sect_conc}, we wrap up with a discussion of the main results of this investigation
and present our conclusions.

% ~\ref{sect_progenitor}
% ~\ref{sect_VULCAN}
% ~\ref{sect_results}
% ~\ref{sect_pns}
% ~\ref{sect_nu}
% ~\ref{sect_disk}
% ~\ref{sect_ener}
% ~\ref{sect_ye}
% ~\ref{sect_conc}
 
\section{Initial models}
\label{sect_progenitor}

  In this section, we present the properties of the AIC progenitors selected in our study and 
summarize the presentation in \S2 of YL05.

  The general assumption for the construction of 2D progenitor models for the AIC of white 
dwarf (YL05) is that the resulting structure of the object is essentially independent 
of its evolutionary history, the 
only factors that matter being the given (final) mass, angular momentum, and central density, 
$\rho_{\rm c}$.      
Additionally, the angular velocity distribution $\Omega(r,z)$, where $r$ is the cylindrical radius and 
$z$ is the distance to the equator, is determined self-consistently, given a number of properties
identified in 1D models: 1) the role played by the dynamical shear instability, 2) the compression 
(and spin up) of the surface layers due to mass accretion which puts its peak angular velocity  
interior to the surface radius, 3) the adopted surface rotational velocity value (its fraction of 
the local Keplerian value), and 4) the geometry of the angular velocity profile, which we assume  
to be constant on cylinders, i.e., $\Omega = \Omega(r)$.
Together with the pressure/density dependence $P = P(\rho)$, such rotating stars are called barotropic.
Note that the criterion used in YL05 for the shear rate for the onset of the dynamical shear 
instability is determined using the EOS of Blinnikov et al. (1996).

   The 2D rotating models then correspond to equilibrium configurations iteratively found from trial
density and angular velocity distributions, using the Self-Consistent-Field method (Ostriker \& Mark 1968; 
Hachisu 1986), under the constraint that the density $\rho$ is solely a function of 
the effective potential $\Psi(r,z)$, given as the sum of the gravitational potential, i.e.,  
$$\Phi(r,z) = -G \int \frac{\rho(r',z')}{|{\bf R}-{\bf R'}|} d^3 R' \,, $$
and the centrifugal potential, i.e., $$\Theta(r) = -\int \Omega^2(r') r' dr' \,,$$ 
where $R' = \sqrt{r'^2 + z'^2}$ (Tassoul 2000, YL05).

In this paper, we select two progenitors with masses of 1.46 \mo and 1.92 \mo and present their global 
characteristics in Table~\ref{tab_aic}. 
Both models have an initial central density $\rho_{\rm c}$ equal to 5$\times$10$^{10}$\,g\,cm$^{-3}$.
The 1.46-\mo model serves as a reference for an object with a moderate initial rotational energy $T$
relative to gravitational energy $|W|$, i.e., $T/|W| = 0.0076$, and, indeed, shows only a modest 
initial departure from spherical symmetry, with a polar 
($R_{\rm p}$) to equatorial ($R_{\rm eq}$) radius ratio $R_{\rm p}/R_{\rm eq} = 0.7$. 
At the other end of the white dwarf mass spectrum, the 1.92-\mo model
has considerable rotational energy, with initial $T/|W| = 0.0833$, and the morphology of the 
star departs strongly from spherical symmetry, with $R_{\rm p}/R_{\rm eq} = 0.28$.
For later reference, we also provide in Table~\ref{tab_aic} the ratio $T/|W|$ at the end of the simulations.
In practice, the infall of the ambient material causes numerical difficulties soon after the 
start of the simulation. These difficulties were resolved by trimming the outer and low-density layers
of both white dwarf progenitors. While the polar radius is hardly affected, the equatorial radius 
is reduced to 980\,km (down from 1130\,km) for the 1.46-\mo model and 
1860\,km (down from 2350\,km) for the 1.92-\mo model.
The progenitor mass is, however, reduced by less than one part in 10$^6$. Hence, we do not expect 
any perceptible effect on the results.
%\clearpage
\begin{deluxetable}{lccccccc}
%\rotate
\tablewidth{8cm}
\tabletypesize{\scriptsize}
\tablecaption{Properties of selected AIC progenitors
\label{tab_aic}}
\tablehead{
\colhead{M}&
\colhead{$R_{\rm p}$}&
\colhead{$R_{\rm eq}$}&
\colhead{$J$}&
\colhead{$T$}&
\colhead{$|W|$}&
\colhead{$T/|W|$}&
\colhead{$T/|W|$} \\
\colhead{\mo}&
\colhead{km}&
\colhead{km}&
\colhead{erg$\cdot$s}&
\colhead{erg}&
\colhead{erg}&
\colhead{initial}&
\colhead{final} \\
\colhead{}&
\colhead{}&
\colhead{}&
\colhead{(10$^{50}$)}&
\colhead{(10$^{50}$)}&
\colhead{(10$^{50}$)}&
\colhead{}&
\colhead{}
}
\startdata
1.46 &  800      &  1130 &0.160 & 0.7   &  91.97 & 0.0076 &  0.059            \\
1.92 &  660      &  2350 &1.092 & 10.57 &  126.9 & 0.0833 &  0.262            \\
\enddata               
%\tablecomments{}
\end{deluxetable}
%\clearpage
\section{VULCAN/2D Simulation Code}
\label{sect_VULCAN}

  The simulations discussed in this paper were performed with the Newtonian hydrodynamic code 
VULCAN/2D (Livne 1993), supplemented with an algorithm for neutrino transport as described 
in Livne et al. (2004) and Walder et al. (2005).
The version of the code used here is the same as that discussed in Dessart et al. (2005) 
and Burrows et al. (2005), and uses the 2D Multi-Group Flux-Limited Diffusion (MGFLD) method to 
handle neutrino transport (see Appendix~A of Dessart et al. 2005).  The MGFLD variant of VULCAN/2D
is much faster than  
the more accurate, but considerably more costly, multi-angle $S_n$ variant.
Doppler velocity-dependent terms are not included in the transport, although advection terms are. 
Frequency redistribution due to the subdominant process of neutrino-electron scattering 
is neglected. Our calculations include 16 energy groups logarithmically distributed in energy 
from 2.5 to 220\,MeV, take into account the electron and anti-electron neutrinos, and 
bundle the four additional neutrino and anti-neutrino flavors into a ``$\nu_{\mu}$'' component.
%\clearpage
\begin{figure*}
\vspace{-1.0cm}
% \plottwo{aic_m1pt46_rho_equipot_t=-0.037s.ps}{aic_m1pt92_rho_equipot_t=-0.037s.ps}
\plottwo{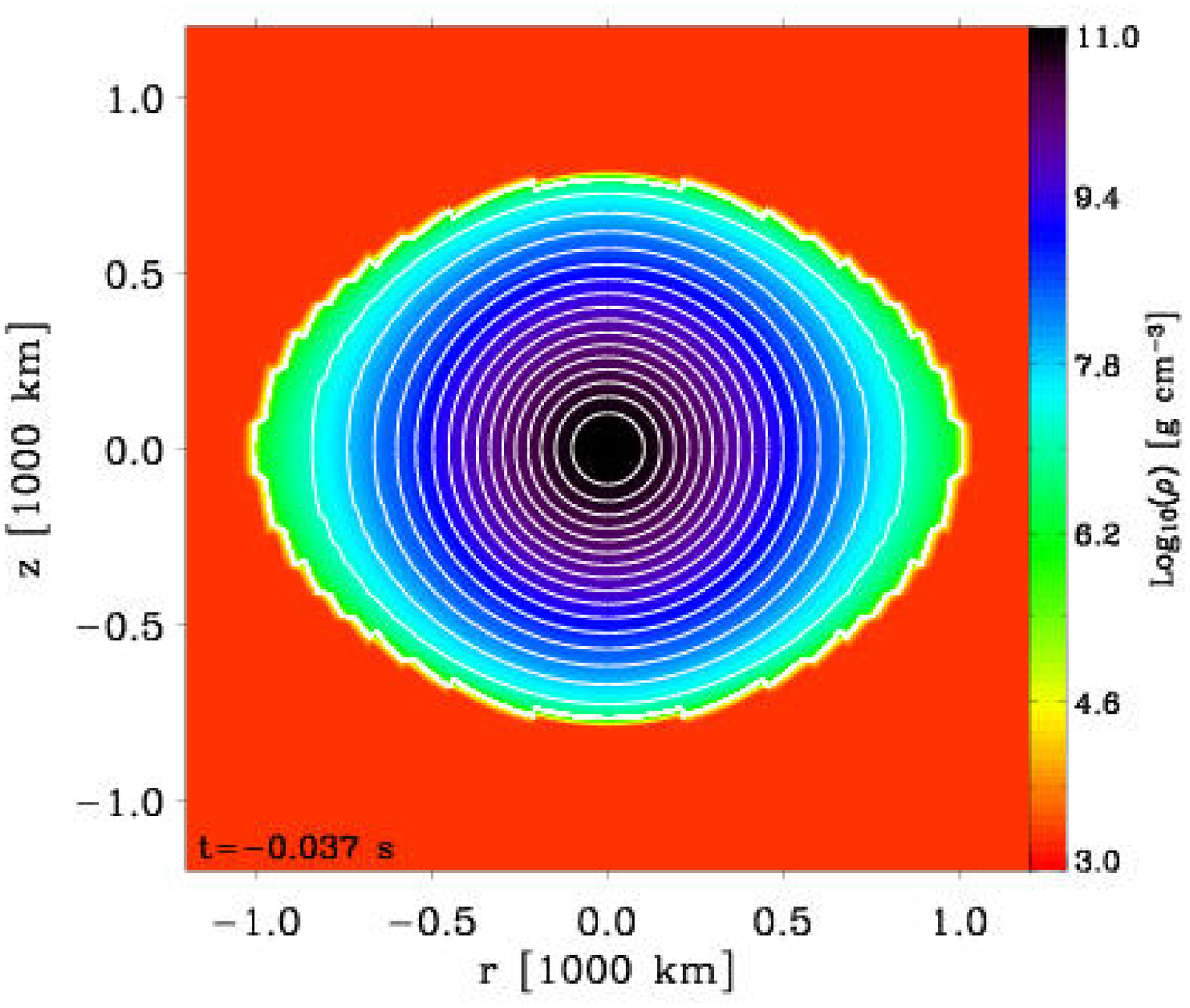}{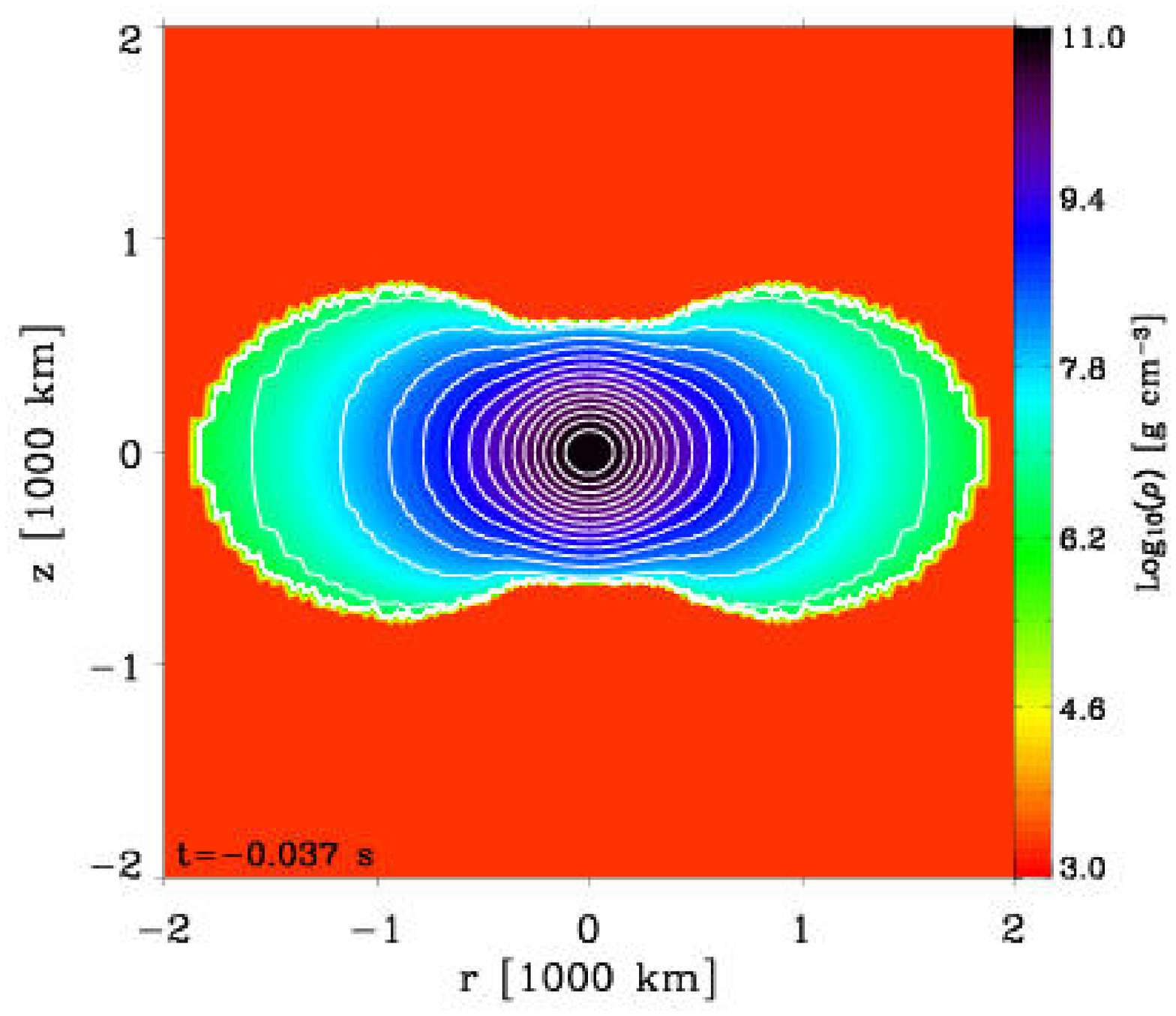}
\vspace{-0.3cm}
% \plottwo{aic_m1pt46_omega_t=-0.037s.ps}{aic_m1pt92_omega_t=-0.037s.ps}
\plottwo{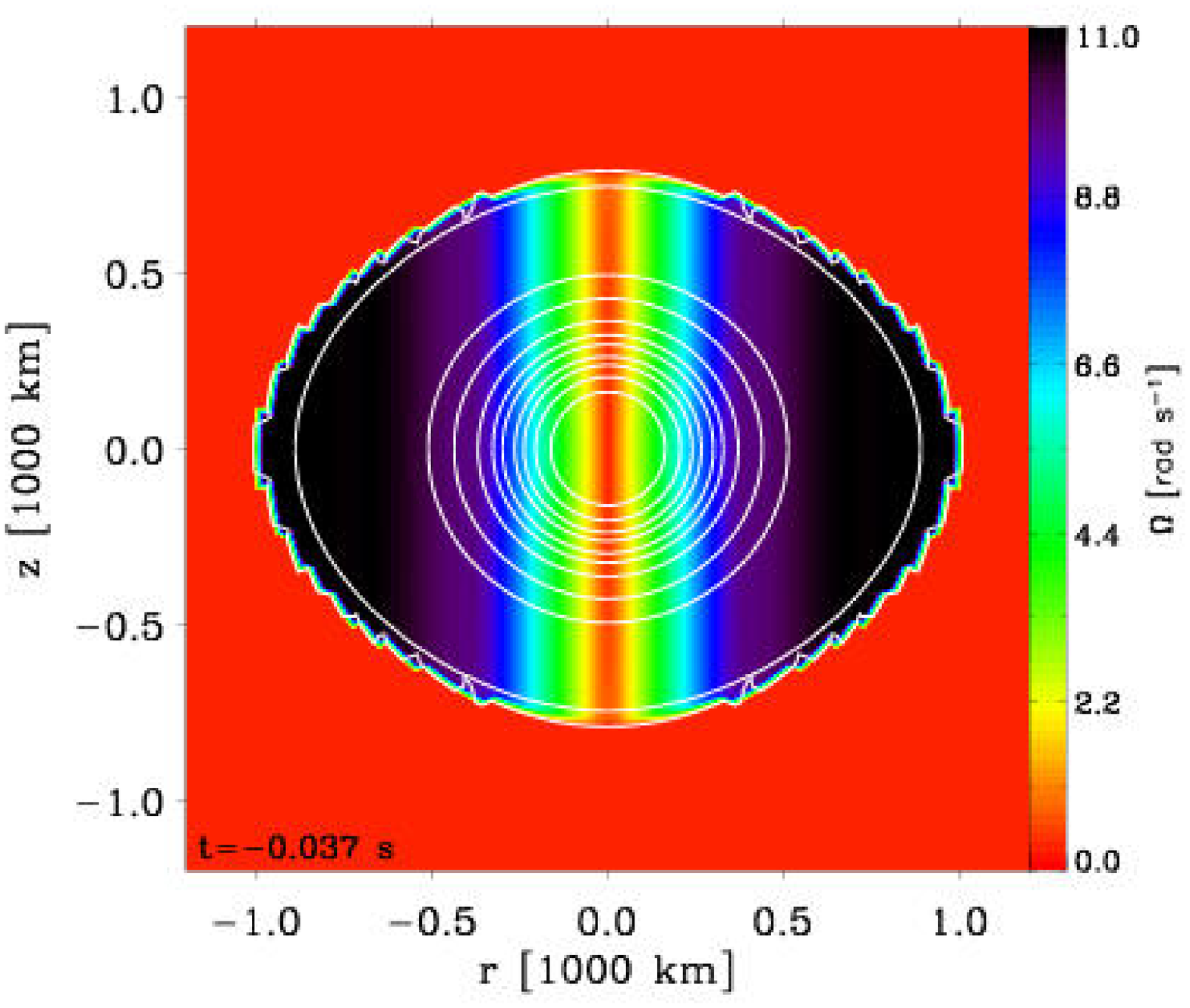}{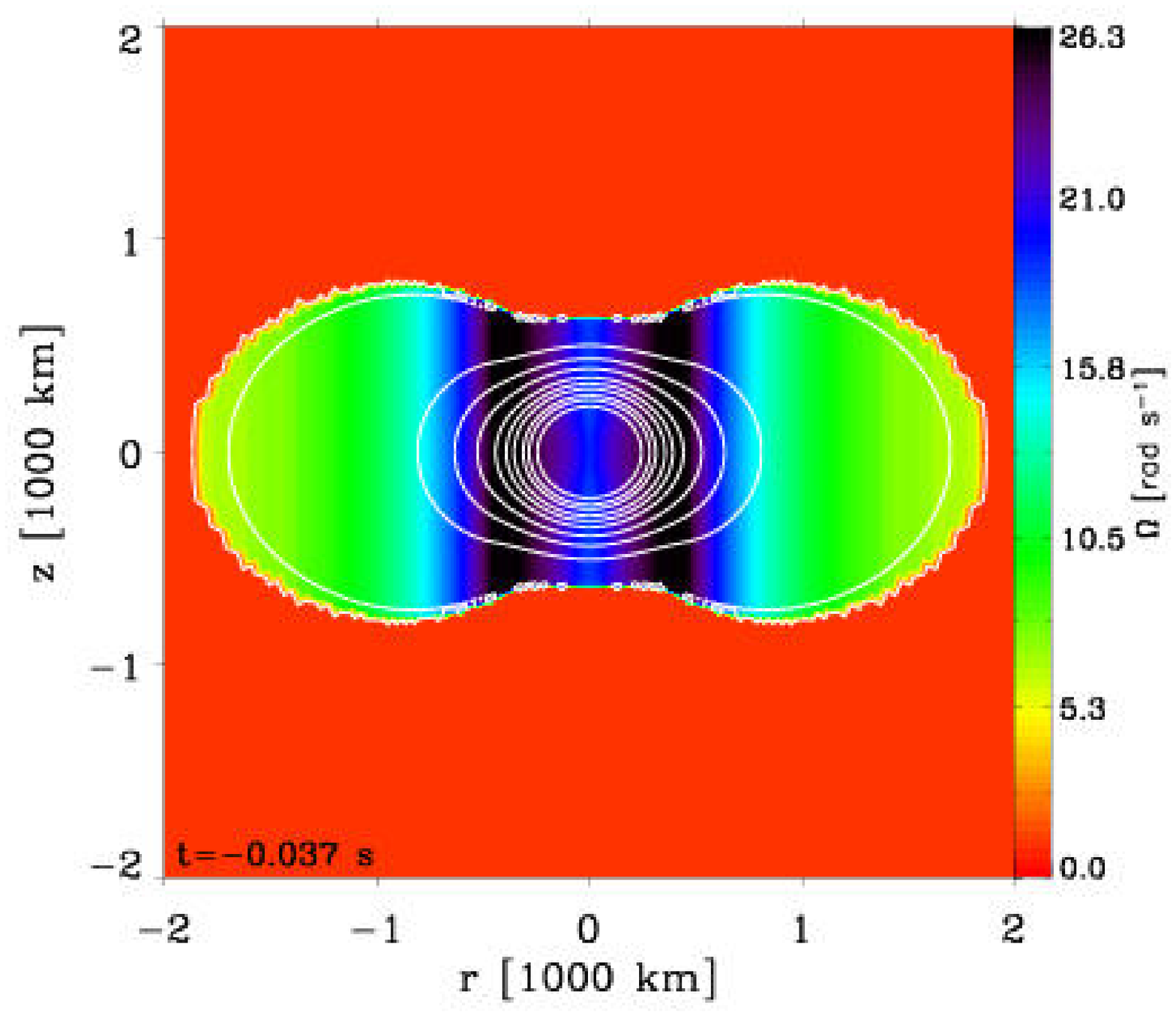}
\caption{
{\it Top}: Color map of the density, $\rho(r,z)$ (log-scale), at the 
start of the simulations for the 1.46-\mo model (left) and the
1.92-\mo model (right). We overplot
the equipotentials corresponding to this equilibrium configuration, whereby the
density $\rho(r,z)$ is solely a function of the effective potential $\Psi(r,z)$ given 
as the sum of the gravitational potential, i.e.,  
$\Phi(r,z) = -G \int \frac{\rho(r,z)}{|{\bf R}-{\bf R'}|} d^3 R'  $\,,
and the centrifugal potential, i.e., 
$\Theta(r) = -\int \Omega^2(r') r' dr'$ (see text and YL05 for details).
{\it Bottom}: Same as above, but this time for the angular velocity, $\Omega(r,z)$.
We overplot contours of the temperature $T(r,z)$, chosen to follow the relation
$T(r,z) = T_{\rm c} (\rho_{\rm c}/\rho(r,z))^{0.35}$, where $T_{\rm c}=10^{10}$\,K 
($T_{\rm c}=1.3 \times 10^{10}$\,K) for the 1.46-\mo (1.92-\mo) model. 
Both models have an initial central density $\rho_{\rm c}$ equal to 5$\times$10$^{10}$\,g\,cm$^{-3}$.
Contours shown are linearly distributed between 4$\times$10$^8$\,K and 10$^{10}$\,K.
}
\label{fig_init}
\end{figure*}
%\clearpage
VULCAN/2D uses a hybrid grid, switching from Cartesian in the inner 20\,km to spherical-polar 
further out.  
%In preliminary investigations, we adopted a grid covering 180$^{\circ}$ about the rotation axis 
%coincident with the $z$ coordinate. However, we realised that in our AIC simulations
%nearly perfect top-bottom symmetry was maintained.
%Hence, only the 1.46-\mo model was mapped onto a 180$^{\circ}$ wedge,
%covering both hemispheres about the equatorial plane, while for the 1.92-\mo, we limited the 
%computational domain to just one hemisphere.
%Information on potential neutron star kicks is thus not available for this latter model, 
%but the upshot is that we could run the model with a higher spatial resolution.
% LUC
In the simulation of the 1.46-\mo model, mapped onto a 180$^{\circ}$ wedge,
nearly perfect top-bottom symmetry about the equatorial plane was maintained 
during the pre- and post-bounce evolution.
Thus, for the 1.92-\mo, we limited the computational domain to just one hemisphere.
The thus reduced number of zones was used to increase the resolution.
To summarize, the 1.46-\mo model uses a grid with a maximum resolution in the Cartesian inner 
region of 0.56\,km, and a minimum resolution of 150\,km at a maximum radius of 5000\,km, with 121 
regularly spaced angular zones to cover 180$^{\circ}$.
The 1.92-\mo model uses a grid with a maximum resolution in the Cartesian inner region of 
0.48\,km, and a minimum resolution of 100\,km at a maximum radius of 4000\,km, with 71 
regularly-spaced angular zones covering 90$^{\circ}$.
% 1.46 0.56km, 150km (37+181) point in radial direction. rmax = 5000km. 121 angles for 180deg.
% 1.92 0.48km  100km (65, 198) point in radial direction rmax=4000km. 71angles pver 90deg.

A tricky part of the set up was to choose the properties of the ``ambient'' medium surrounding the
AIC model. This need arises because our inputs are Lagrangean in spirit, while VULCAN/2D
employs a Eulerian grid.
Ideally, one would like to have material that merely occupies the space that will soon, after bounce,
be replaced by the ejected material following the explosion. 
A key requirement is, thus, that this material have 
a very low pressure, to influence as little as possible the properties of the blast, but 
to allow for a smooth transition from circumstellar to ejected material in a given
region of the Eulerian grid.
To achieve this, we extended our SHEN EOS (Shen et al. 1998) down to very low 
densities (10\,g\,cm$^{-3}$) and 
low temperatures (10$^8$ K), conditions in which the medium is actually 
radiation-dominated and, thus, has a pressure that depends mostly on temperature.
A successful choice was to adopt a density of 1000\,g\,cm$^{-3}$ and a temperature of 
4$\times$10$^{8}$\,K for the ambient medium surrounding both white dwarf models.

A major deficiency of the white dwarf progenitor models used here is their unknown initial 
thermal structure, which YL05 did not provide.
Given the additional difficulty in handling low temperatures and high densities, 
we resorted to using a parameterized function of the local density, i.e.,
$$ T(r,z) = T_{\rm c} (\rho_{\rm c}/\rho(r,z))^{0.35}\,,$$ with 
$T_{\rm c} = 10^{10}$\,K for 
the 1.46-\mo model, and  $T_{\rm c} = 1.3 \times 10^{10}$\,K for the 1.92-\mo model.
Note that, similarly, Woosley \& Baron (1992) were forced to set a central temperature of 
1.2$\times$10$^{10}$\,K at the start of their simulation.

Figure~\ref{fig_init} recapitulates the basic properties of the white dwarf 
progenitors (left column: 1.46-\mo model; right column: 1.92-\mo model) as 
mapped onto our Eulerian grid. In the top panel, we show a color map of the 
density on which we superpose
line contours of the effective potential $\Psi$ (defined above), as computed by VULCAN/2D.
Our computation of the gravitational potential is based on a multipole expansion in spherical
harmonics up to $l=33$.
We reproduce the fundamental property of these fast-rotating white dwarf progenitors, in that 
the isopressure surfaces and the equipotentials coincide. In the bottom panel, we plot the 
initial angular velocity field. To avoid the distortion of streamlines of the infalling ambient
material, the angular velocity is set to zero outside the WD progenitor.  
Note how, from zero, the angular velocity rises to a maximum value near the outer
equatorial radius in the 1.46-\mo model (left column), while it is much higher in the center
of the white dwarf in the 1.92-\mo model (right column). As we will show below, 
rotation has a much bigger 
impact on the collapse phase for the latter configuration. We also overplot line 
contours of the temperature for each model, following the prescription outlined in 
the above paragraph.

Finally, we adopt an initial electron fraction of 0.5; the high initial 
central density permits fast electron capture which soon decreases the $Y_{\rm e}$
in the core, leading to bounce on a timescale ten times shorter than typically 
experienced in the collapse of the core of massive star progenitors 
(Woosley \& Weaver 1995; Heger et al. 2000; Woosley et al. 2002). 

\section{Simulation results}
\label{sect_results}

    In this section, we discuss the general properties of the pre- and post-collapse 
phases for both models at the same time. We follow 
the post-bounce evolution of the 1.46-\mo (1.92-\mo) model for 550\,ms (780\,ms), for a total of 
520\,000 (820\,000) timesteps.

\begin{figure*}
%\clearpage
\epsscale{.9}
\plotone{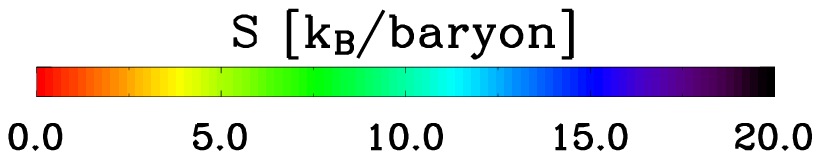}
%\vspace{-0.5cm}
% \plottwo{aic_m1pt46_S_t=0.0160s.ps}{aic_m1pt46_S_t=0.1000s.ps}
\plottwo{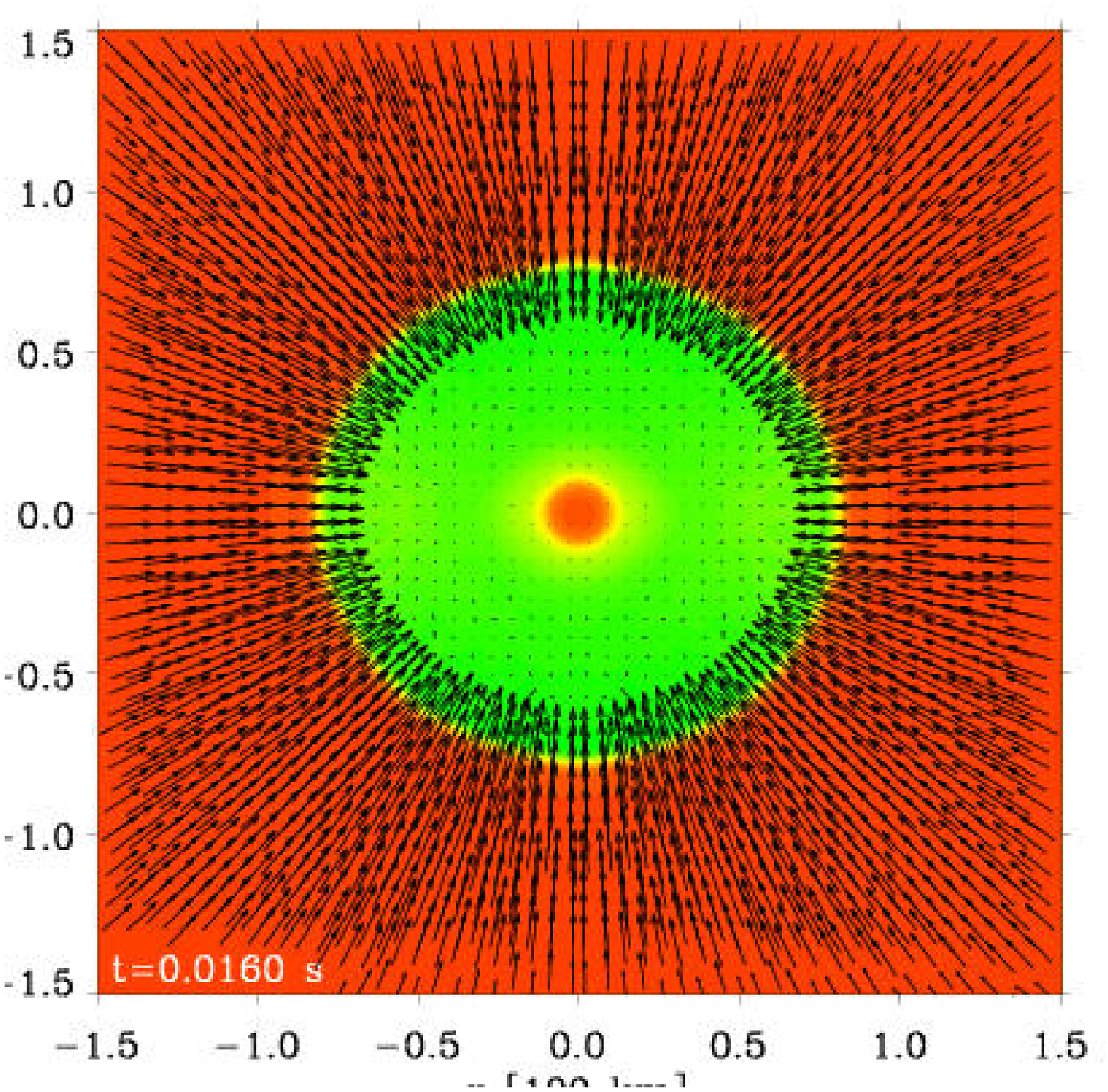}{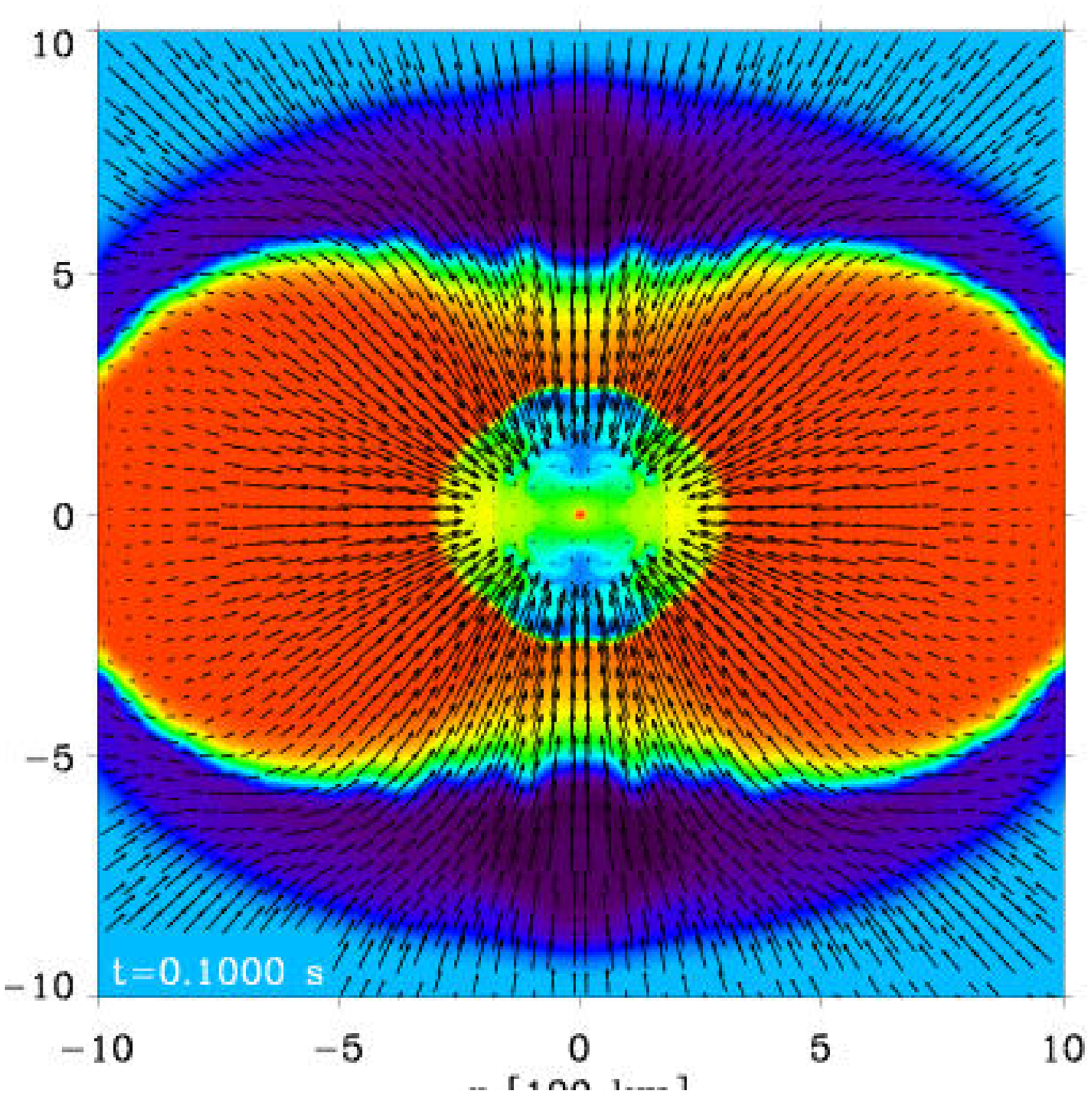}
\vspace{-0.3cm}
% \plottwo{aic_m1pt46_S_t=0.1060s.ps}{aic_m1pt46_S_t=0.1465s.ps}
\plottwo{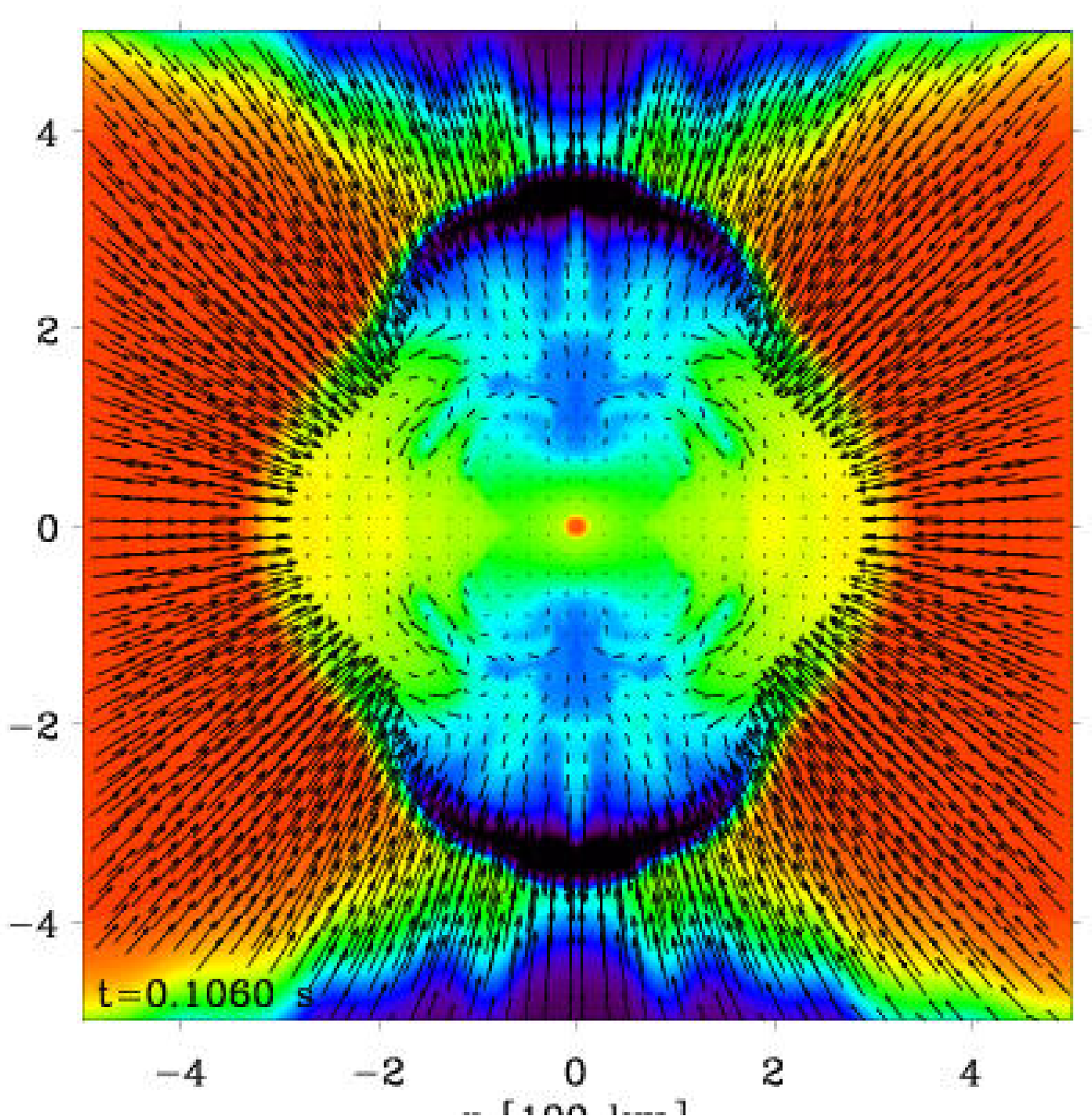}{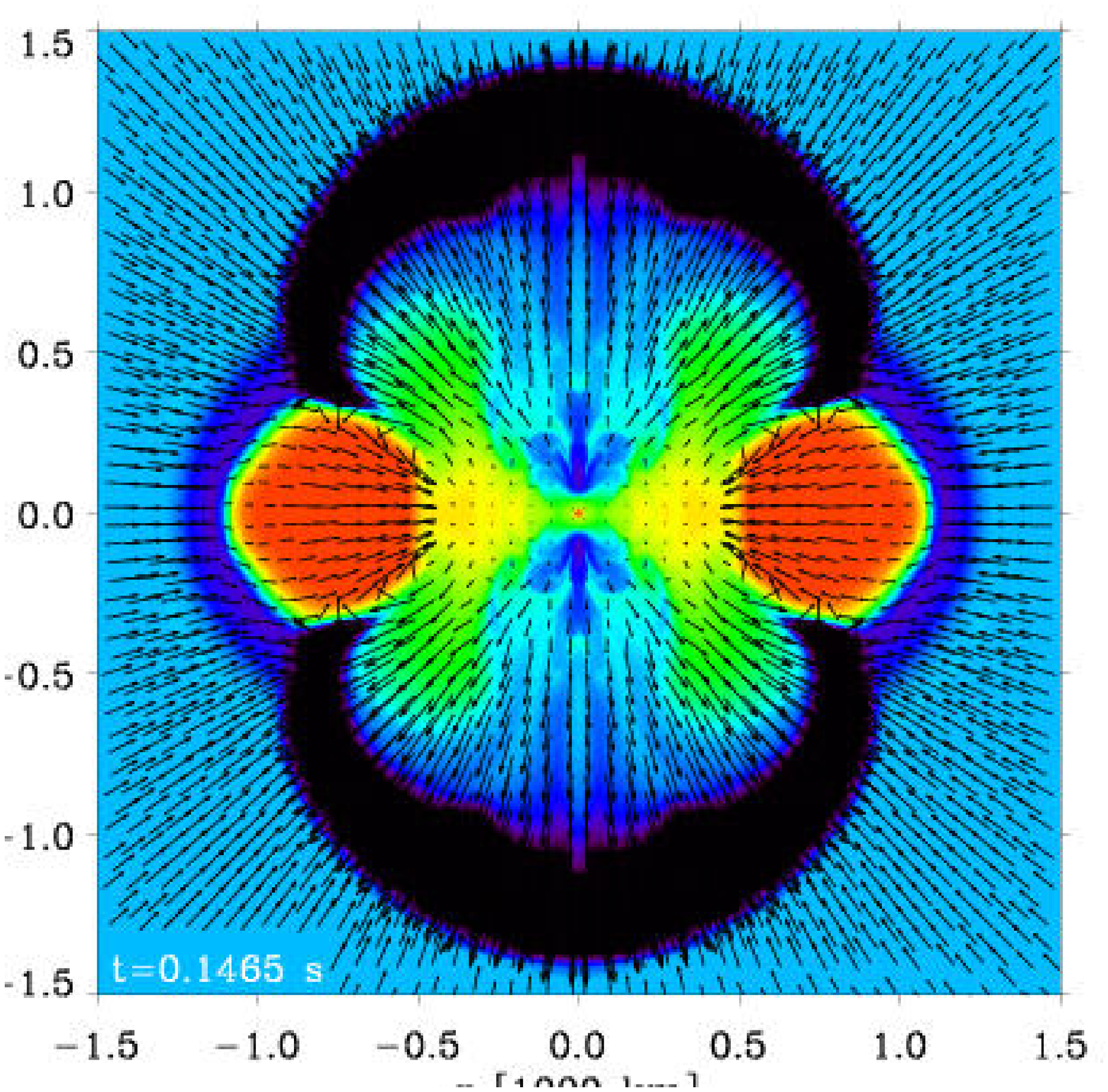}
\vspace{-0.3cm}
% \plottwo{aic_m1pt46_S_t=0.2420s.ps}{aic_m1pt46_S_t=0.5700s_white_vec.ps}
\plottwo{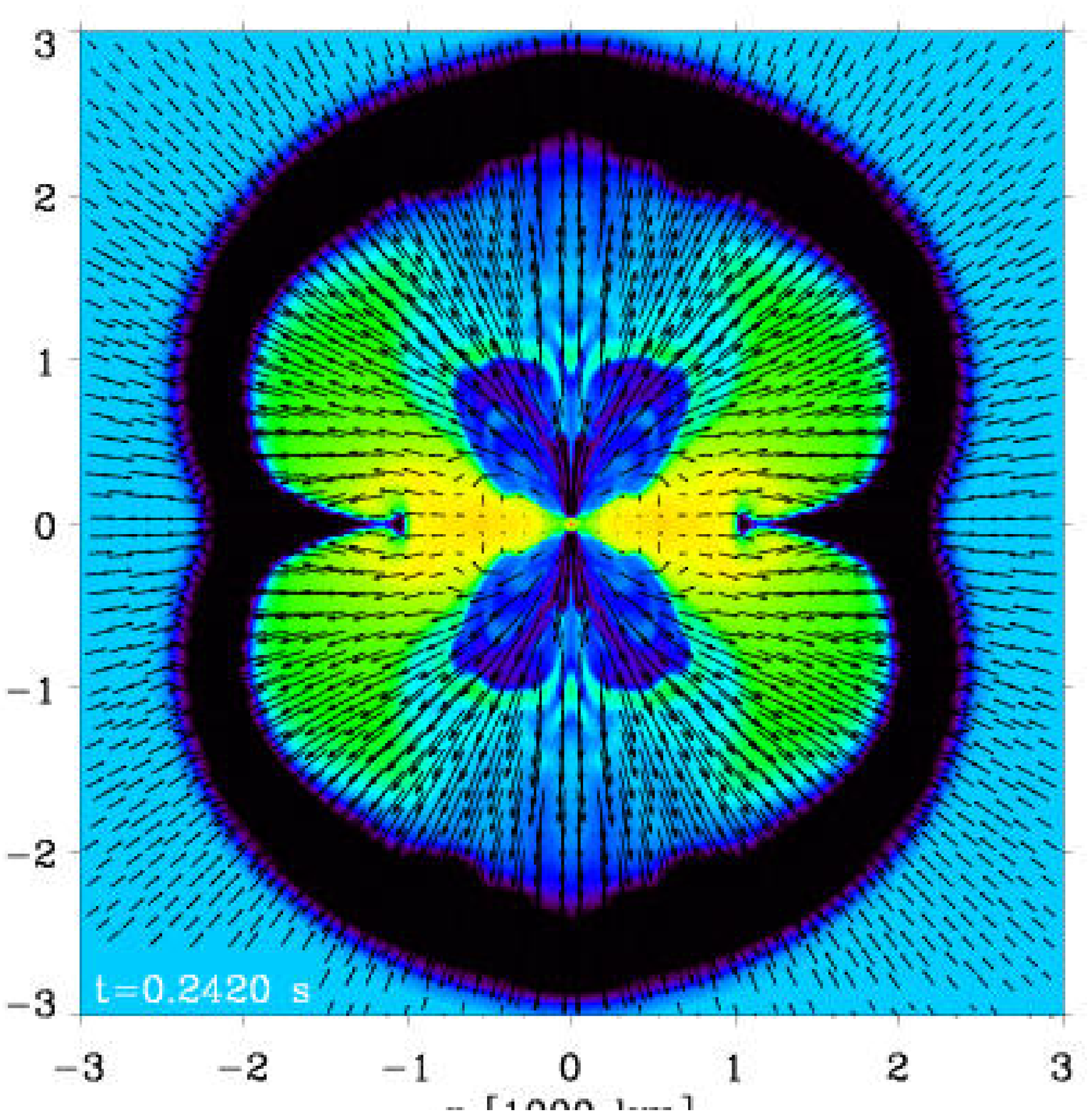}{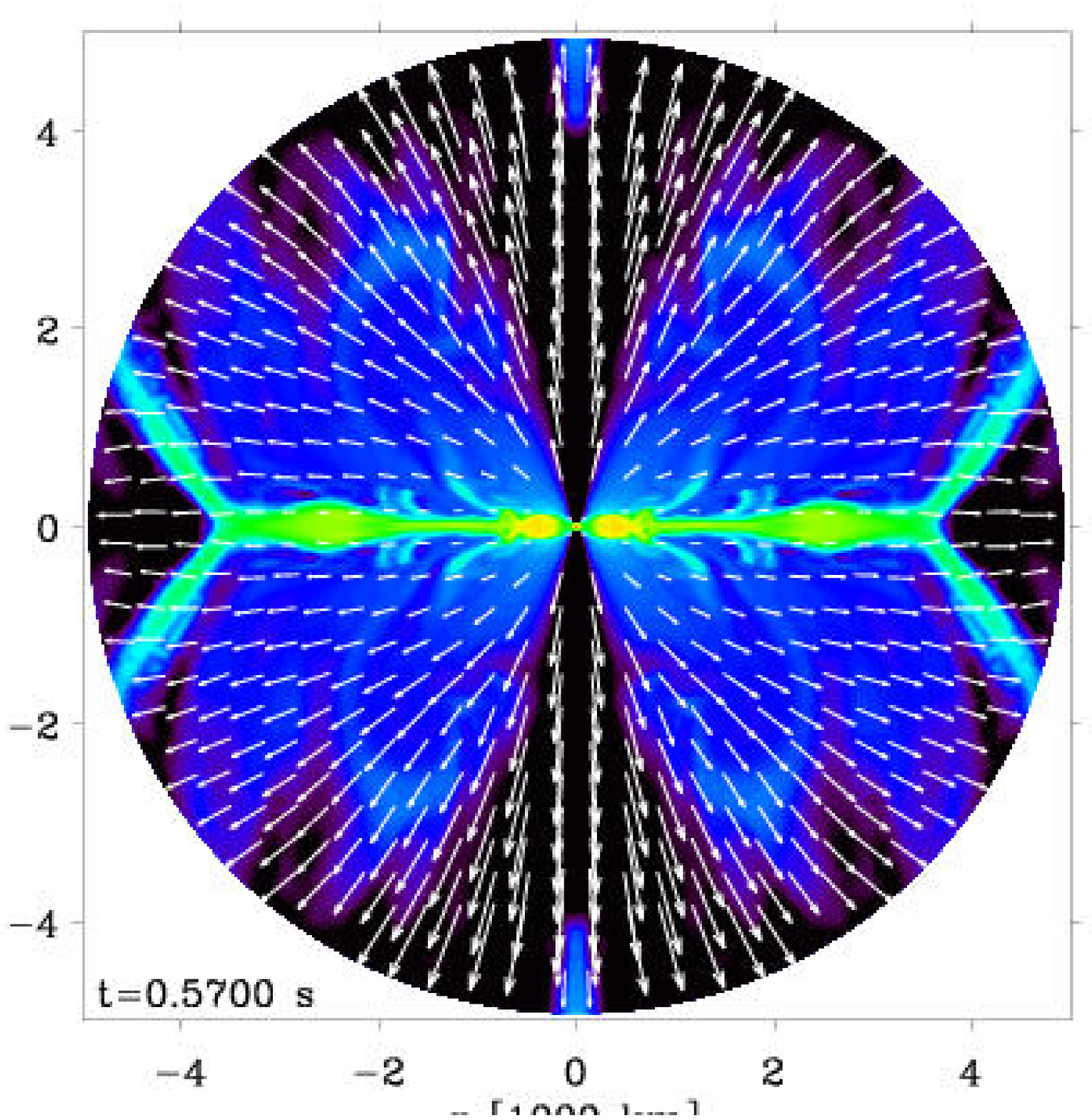}
%\clearpage
%\begin{figure*}
\caption{
Slide sequence of color maps of the entropy covering the post-bounce evolution of 
the 1.46-\mo progenitor, with velocity vectors overplotted as black or white arrows. 
The entropy, reaching up to 50\,$k_{\rm B}$/baryon at later times, has been saturated at 
20\,$k_B$/baryon in all panels to enhance the contrast.
From top to bottom, left to right, the vector of maximum length corresponds to a 
velocity magnitude of 52500\kms (infall), 28500\kms (infall), 24000\kms (infall), 
20500\kms (outflow), 16500\kms (outflow), and 34000\kms (outflow).
We indicate the time after bounce in the bottom-left corner of each panel.
Note also that the spatial scale varies between panels, with widths in the 
range 300\,km to 10000\,km. See text for details.
}
\label{fig_seq46}
\end{figure*}

\begin{figure*}
%\clearpage
\vspace*{-2.5cm}
% \plotone{colorbarS_aic_m1pt92_0_20.ps}
\plotone{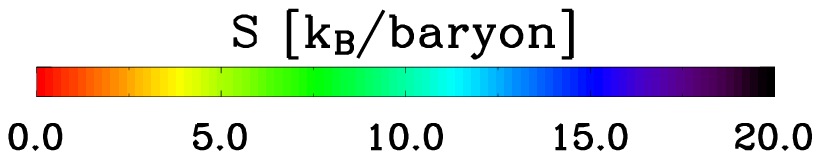}
%\vspace{-0.8cm}
% \plottwo{aic_m1pt92_S_t=0.0165s.ps}{aic_m1pt92_S_t=0.0590s.ps}
\plottwo{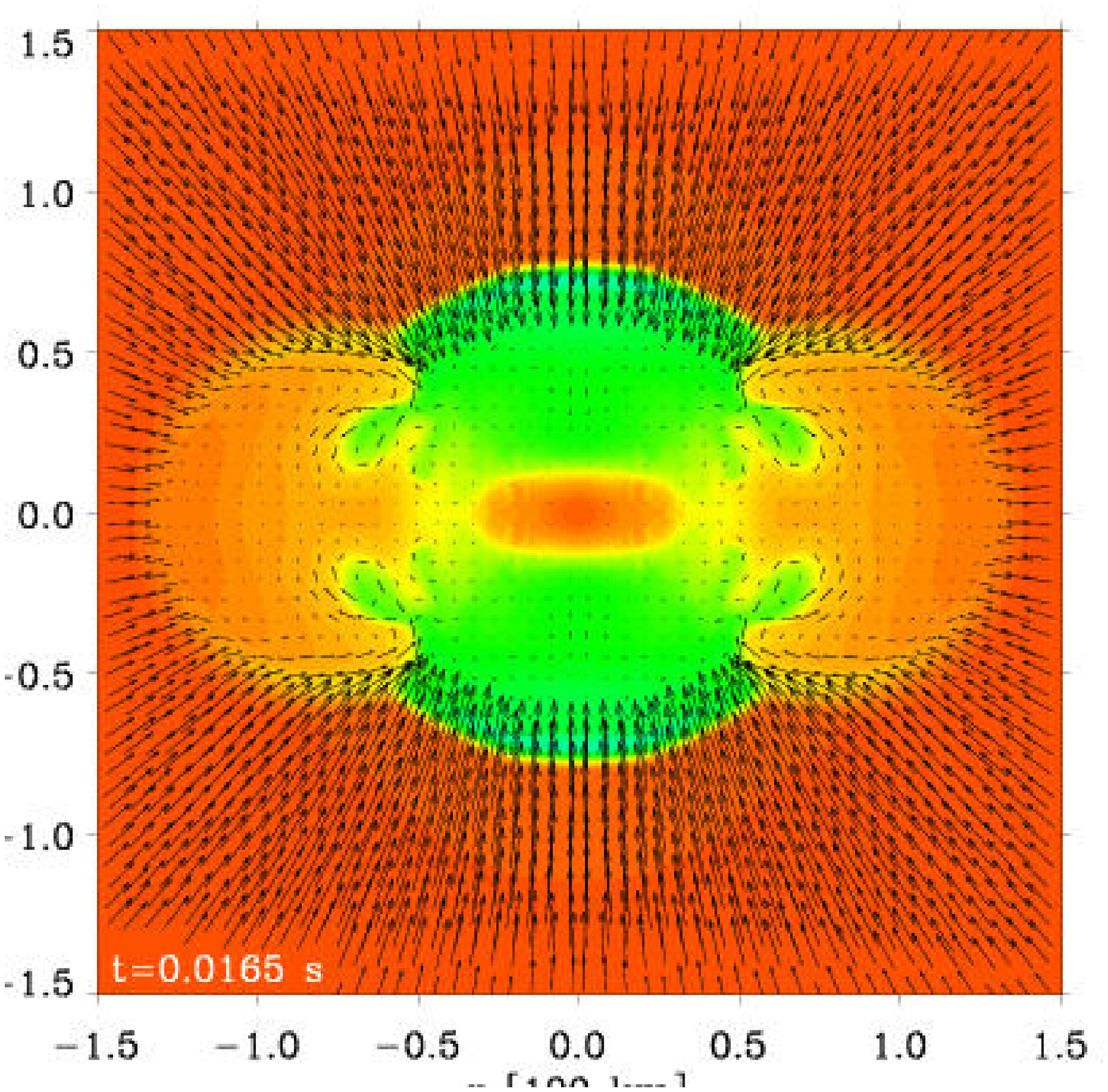}{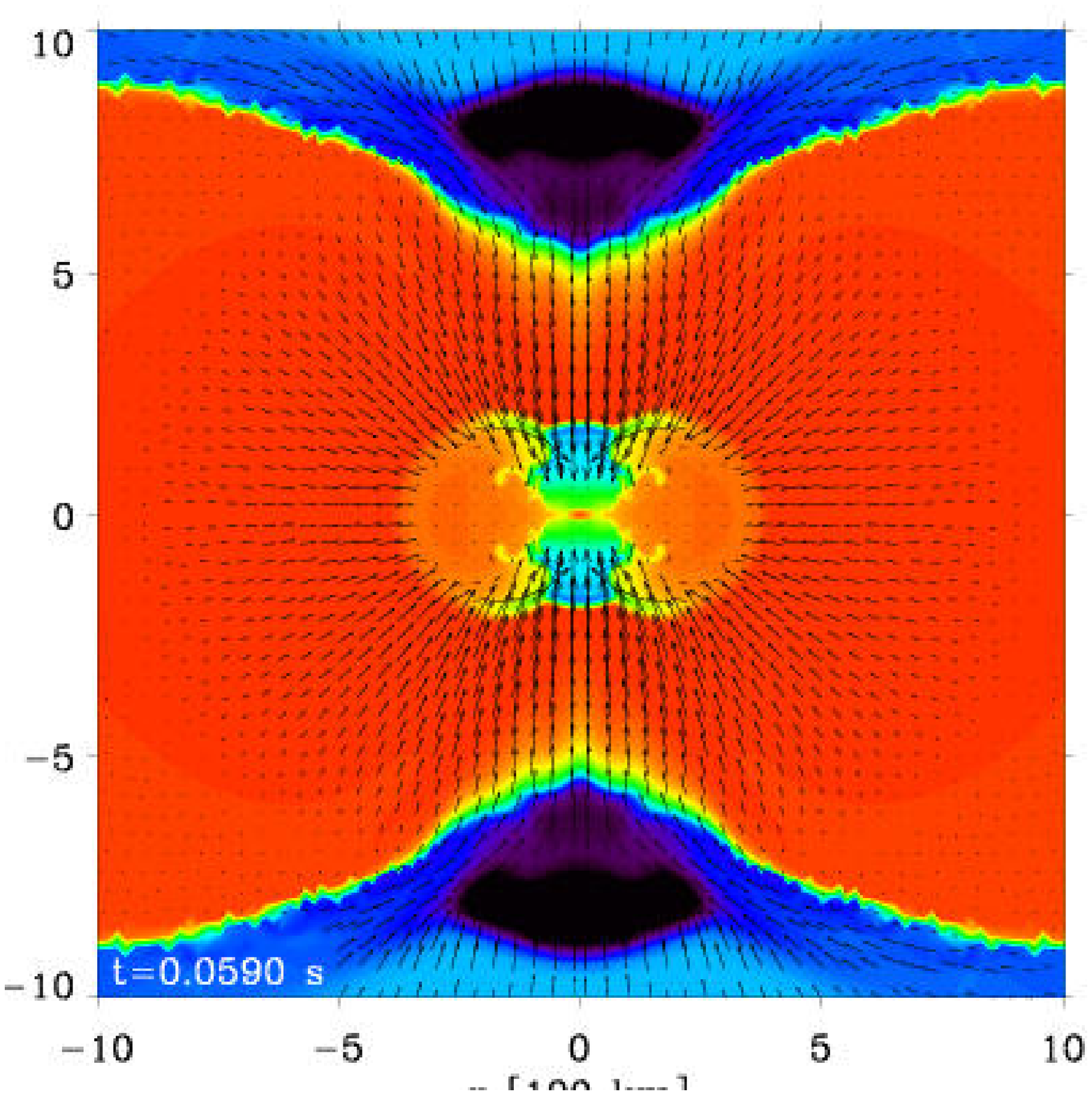}
\vspace{-0.3cm}
% \plottwo{aic_m1pt92_S_t=0.0700s.ps}{aic_m1pt92_S_t=0.0945s.ps}
\plottwo{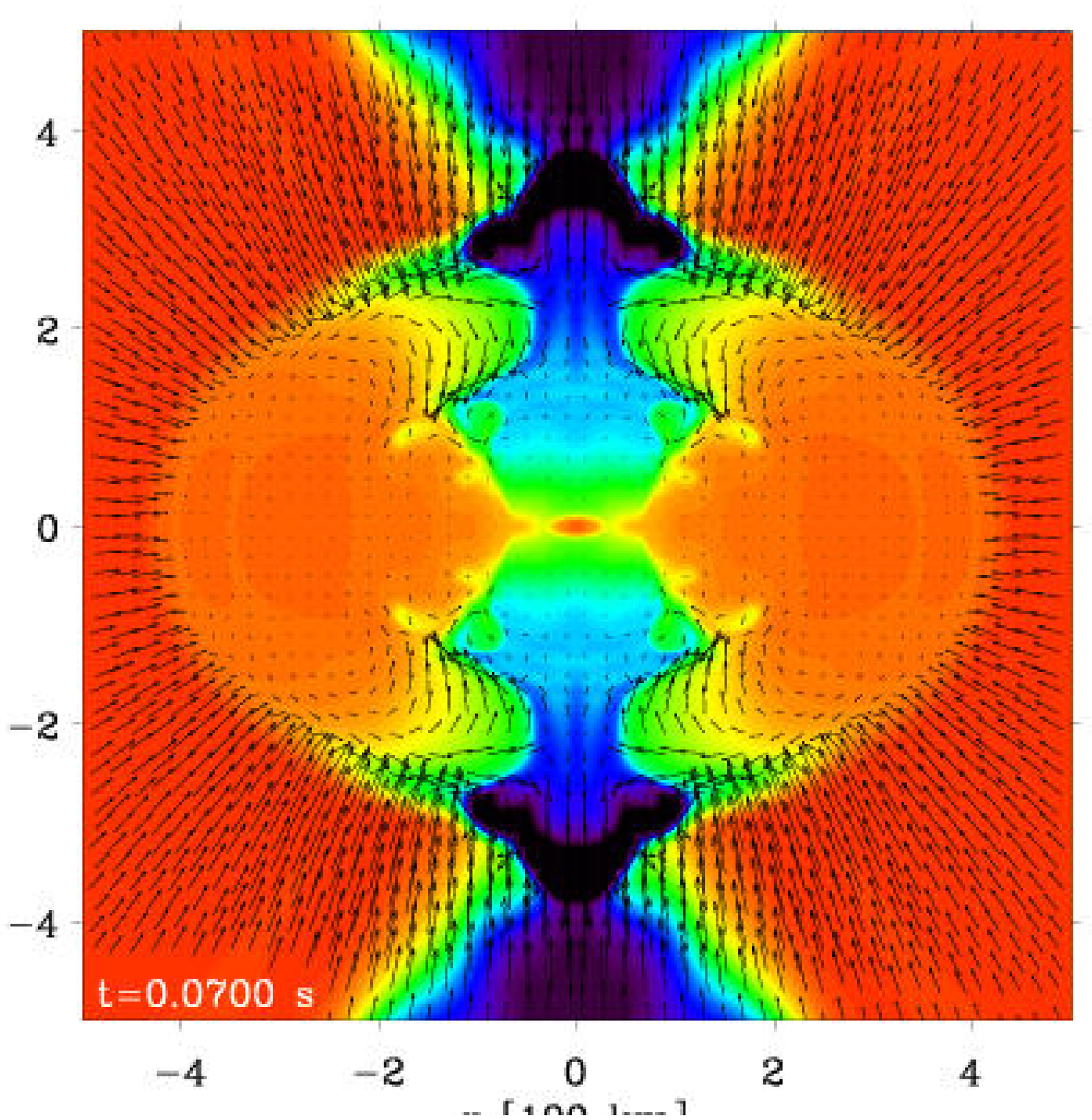}{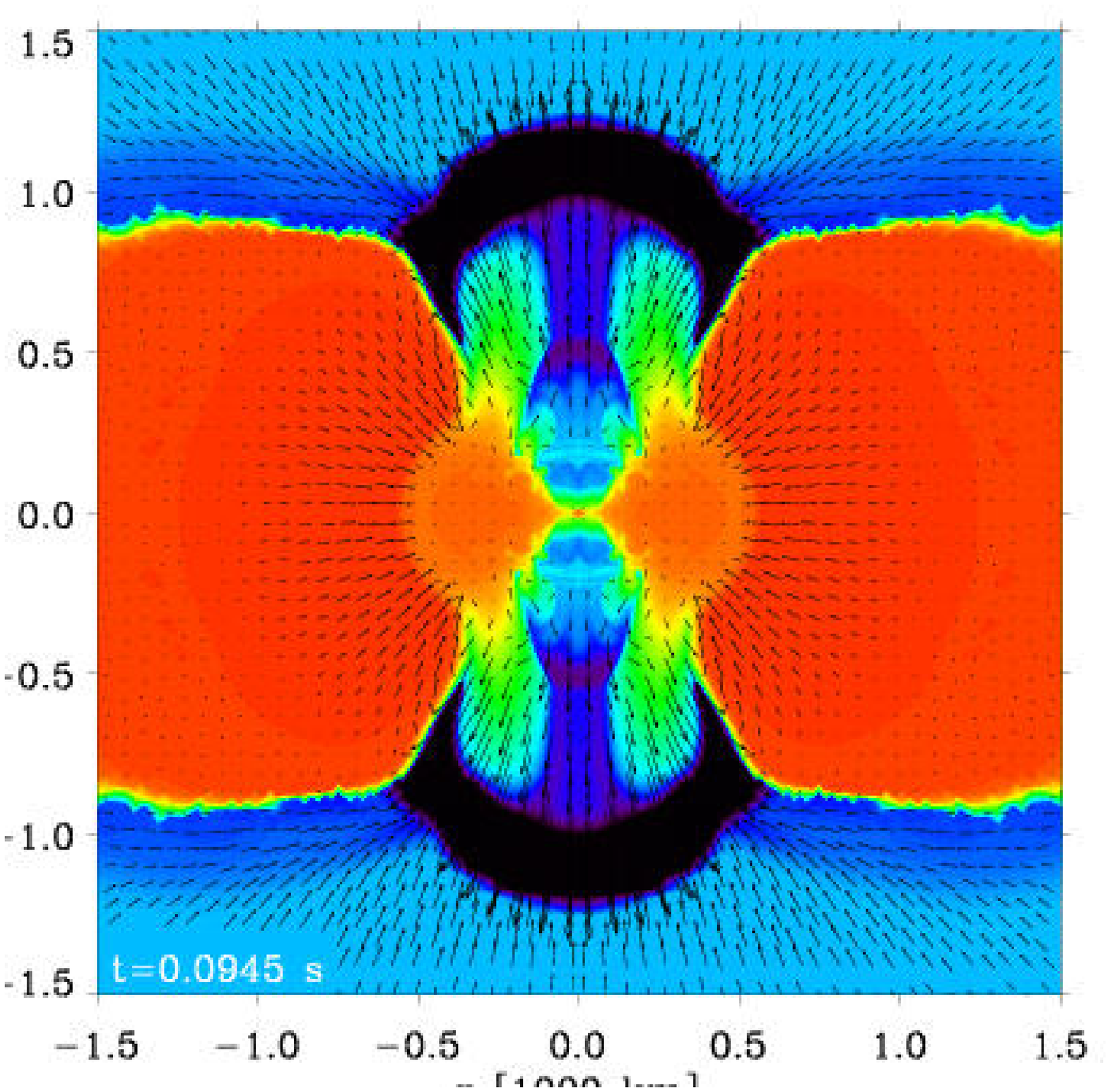}
\vspace{-0.3cm}
% \plottwo{aic_m1pt92_S_t=0.2185s.ps}{aic_m1pt92_S_t=0.7755s_white_vec.ps}
\plottwo{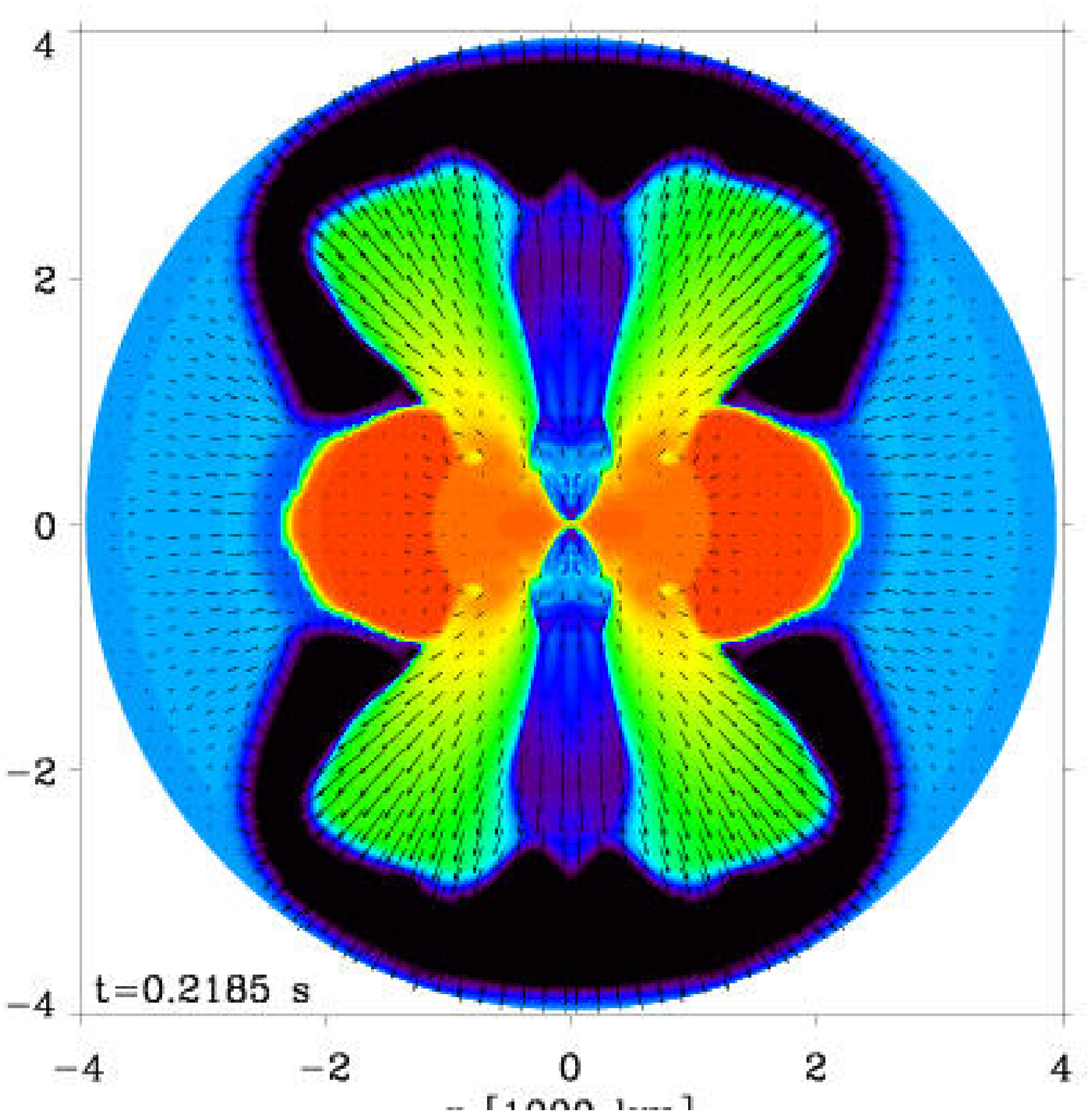}{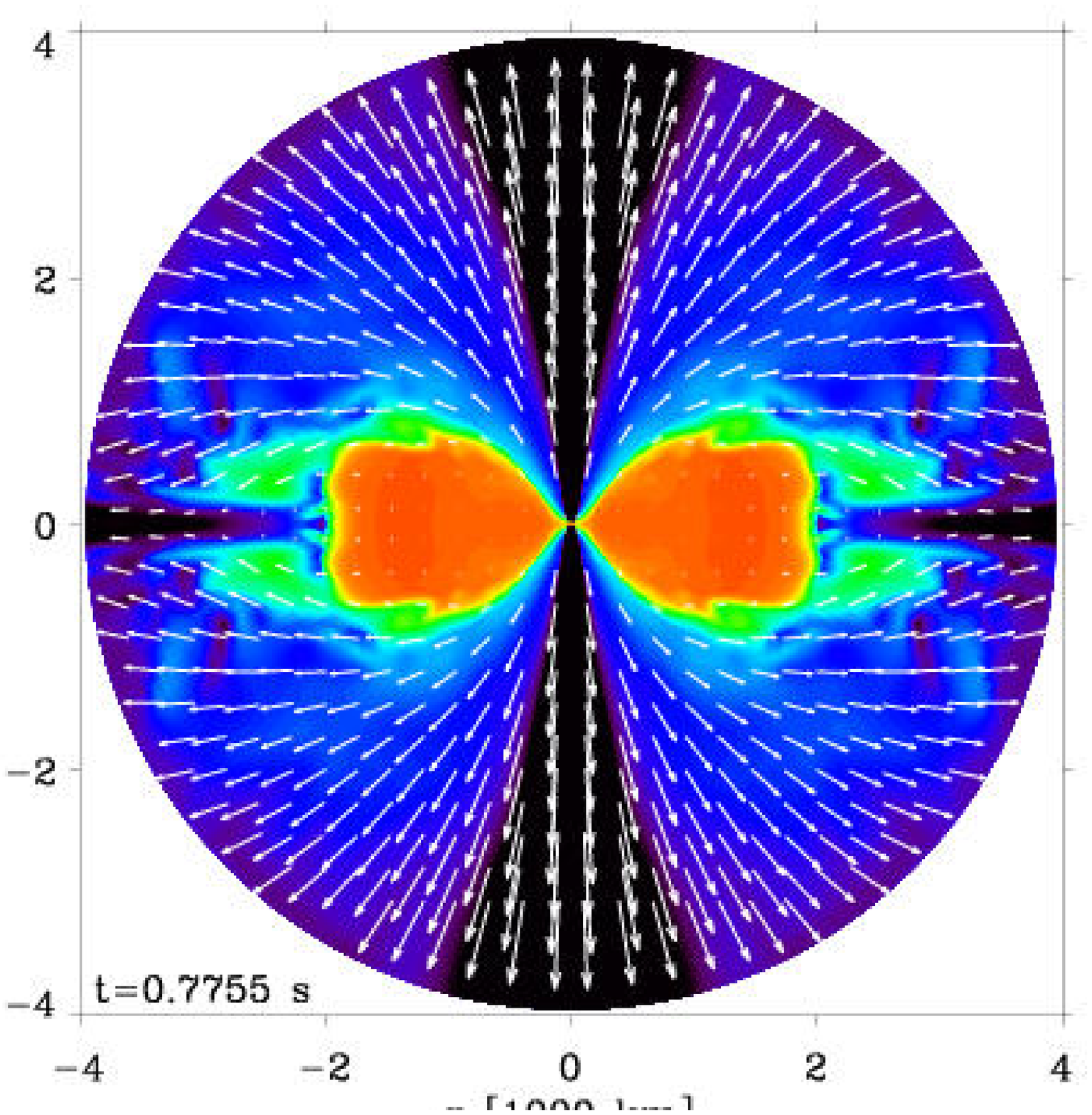}
%\clearpage
%\begin{figure*}
\caption{
Slide sequence of color maps of the entropy covering the post-bounce evolution of 
the 1.92-\mo progenitor, with velocity vectors overplotted as black or white arrows. 
The entropy, reaching up to 50\,$k_B$/baryon at late times, has been saturated at
20\,$k_B$/baryon to enhance the constrast.
From top to bottom, left to right, the vector of 
maximum length corresponds to a velocity magnitude of 53000\,\kms (infall), 
36500\,\kms (infall), 29000\,\kms (infall), 32000\,\kms (outflow),
32500\,\kms (outflow), and 34500\,\kms (outflow).
We indicate the time after bounce in the bottom-left corner of each panel.
Note also that the spatial scale varies between panels, with widths in the 
range 300\,km to 8000\,km. See text for details.
}
\label{fig_seq92}
\end{figure*}

\begin{figure*}
%\clearpage
\vspace*{-2.5cm}
% \plotone{colorbarye.ps}
\plotone{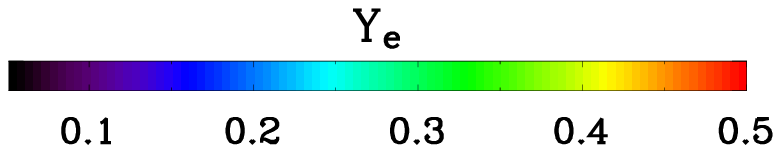}
%\vspace{-0.8cm}
% \plottwo{aic_m1pt46_ye_t=0.0590s.ps}{aic_m1pt92_ye_t=0.0590s.ps}
\plottwo{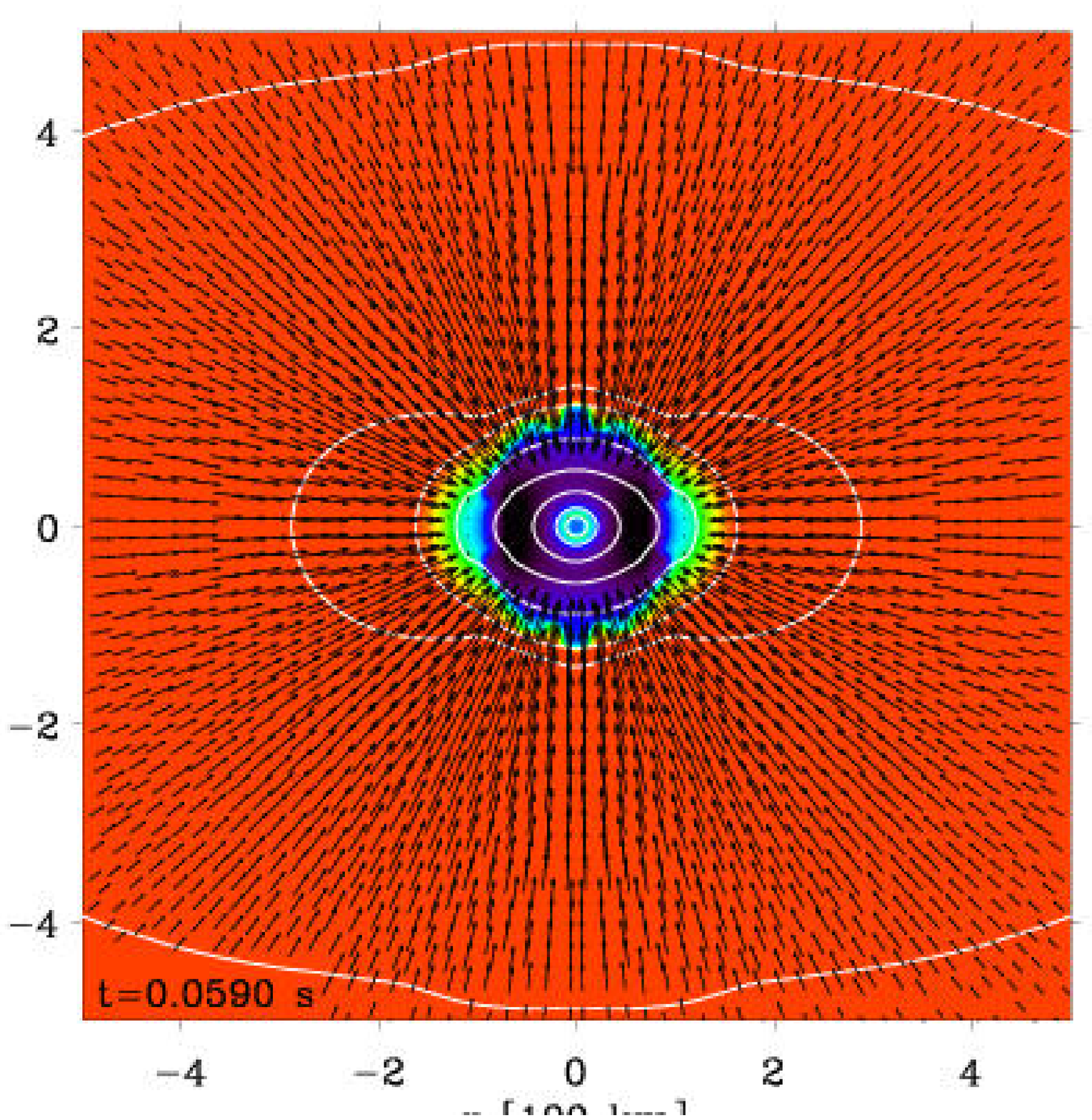}{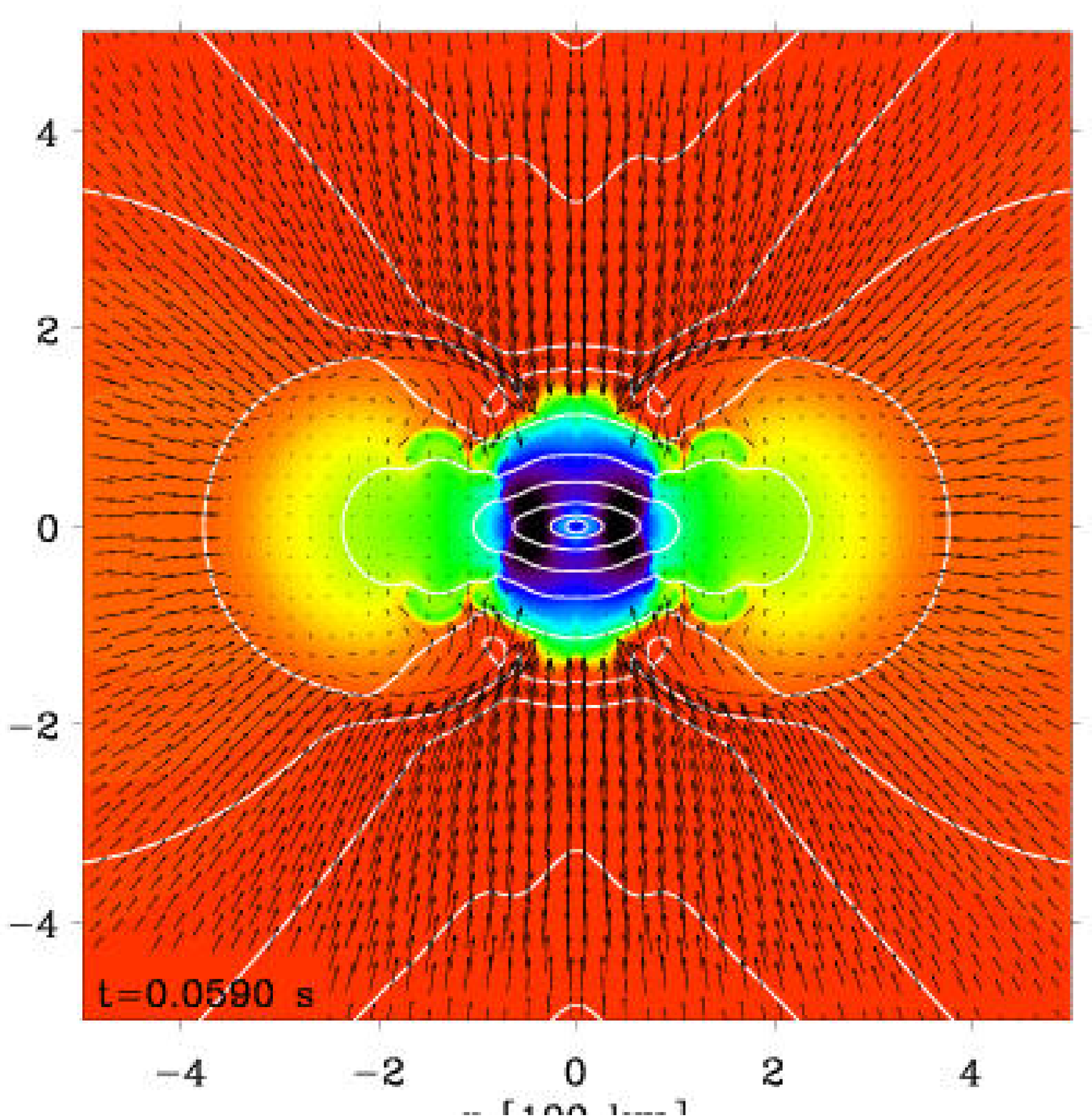}
\vspace{-0.3cm}
% \plottwo{aic_m1pt46_ye_t=0.3130s.ps}{aic_m1pt92_ye_t=0.3130s.ps}
\plottwo{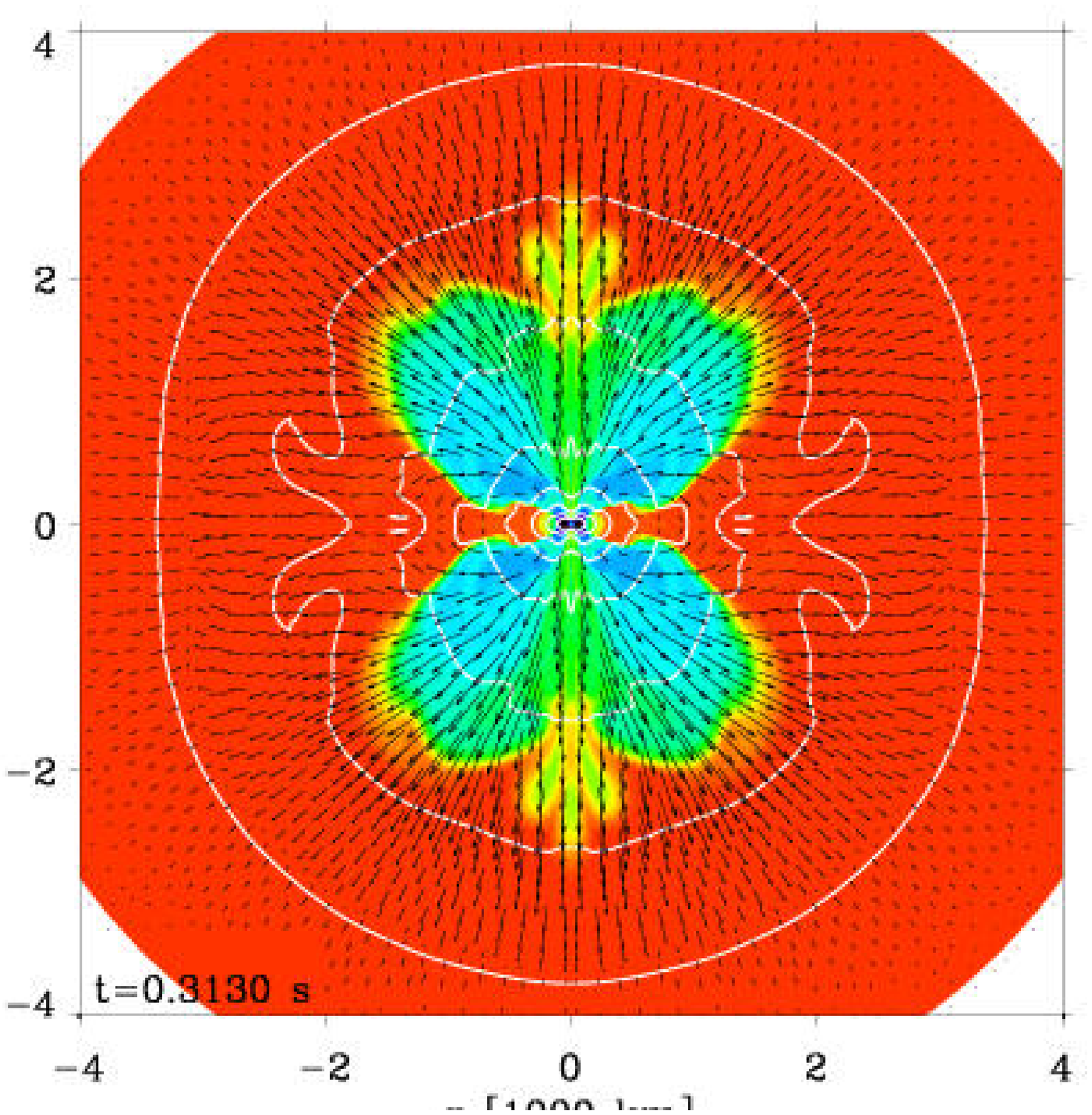}{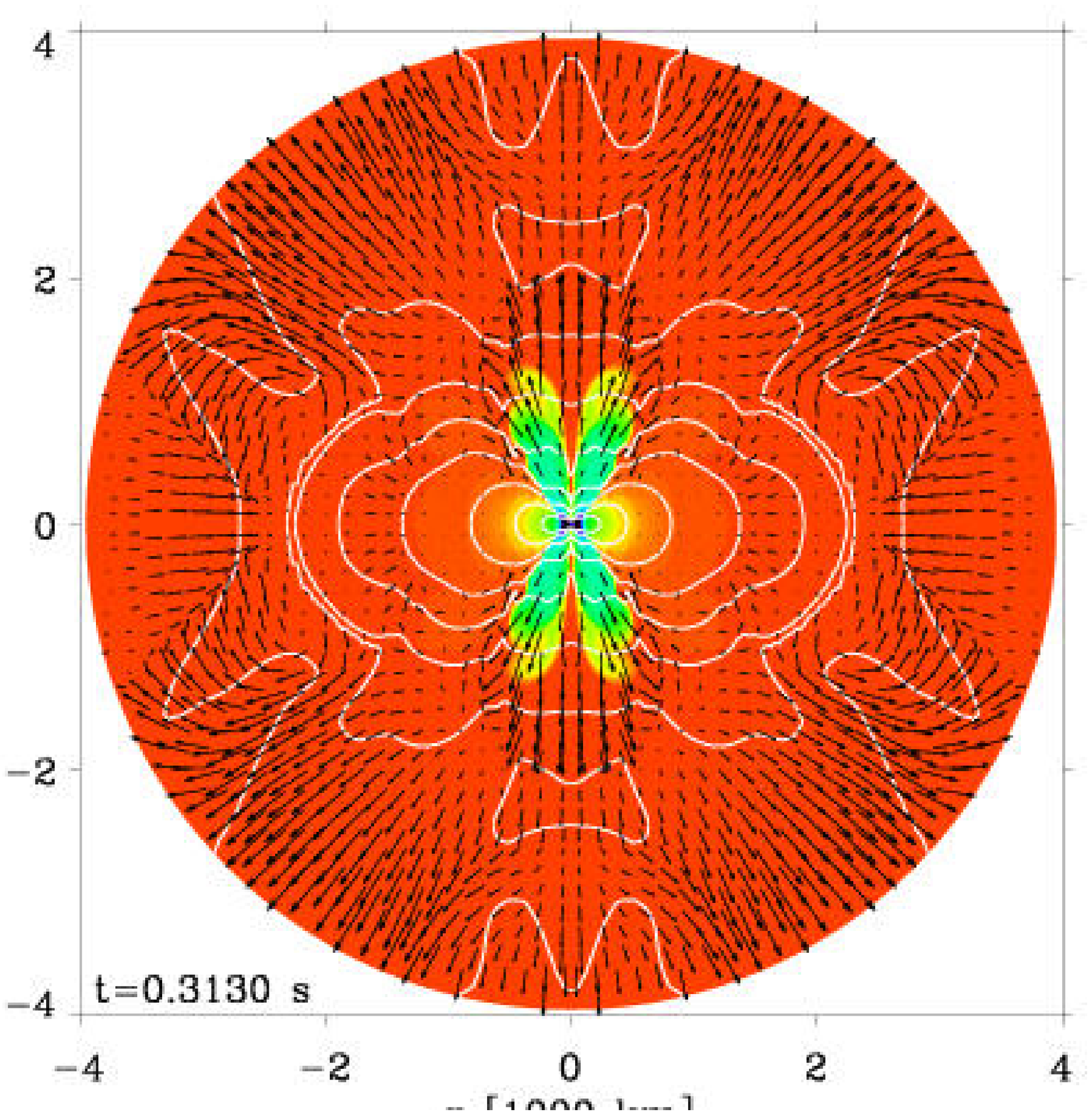}
\vspace{-0.3cm}
% \plottwo{aic_m1pt46_ye_t=0.5700s.ps}{aic_m1pt92_ye_t=0.7755s.ps}
\plottwo{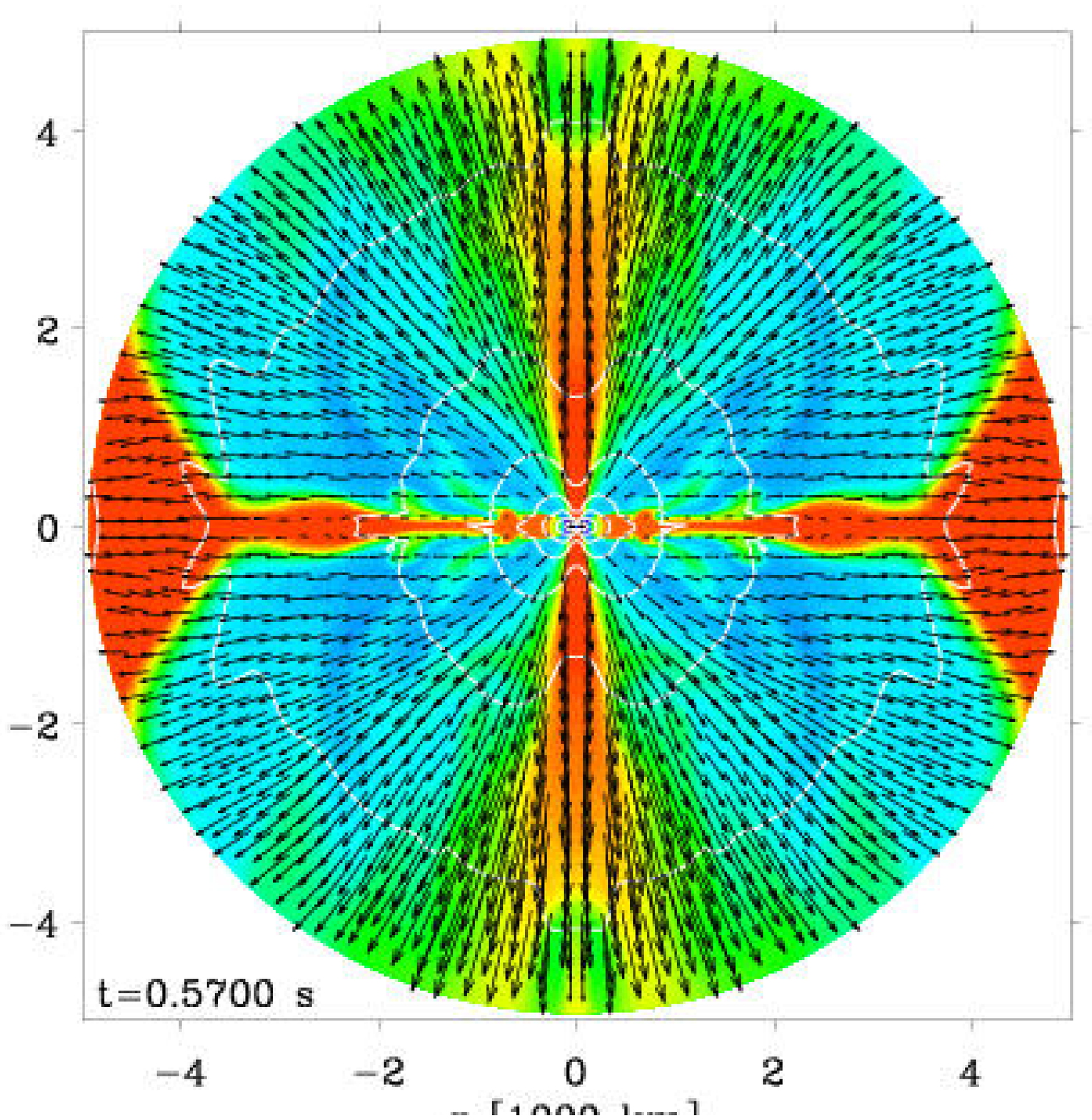}{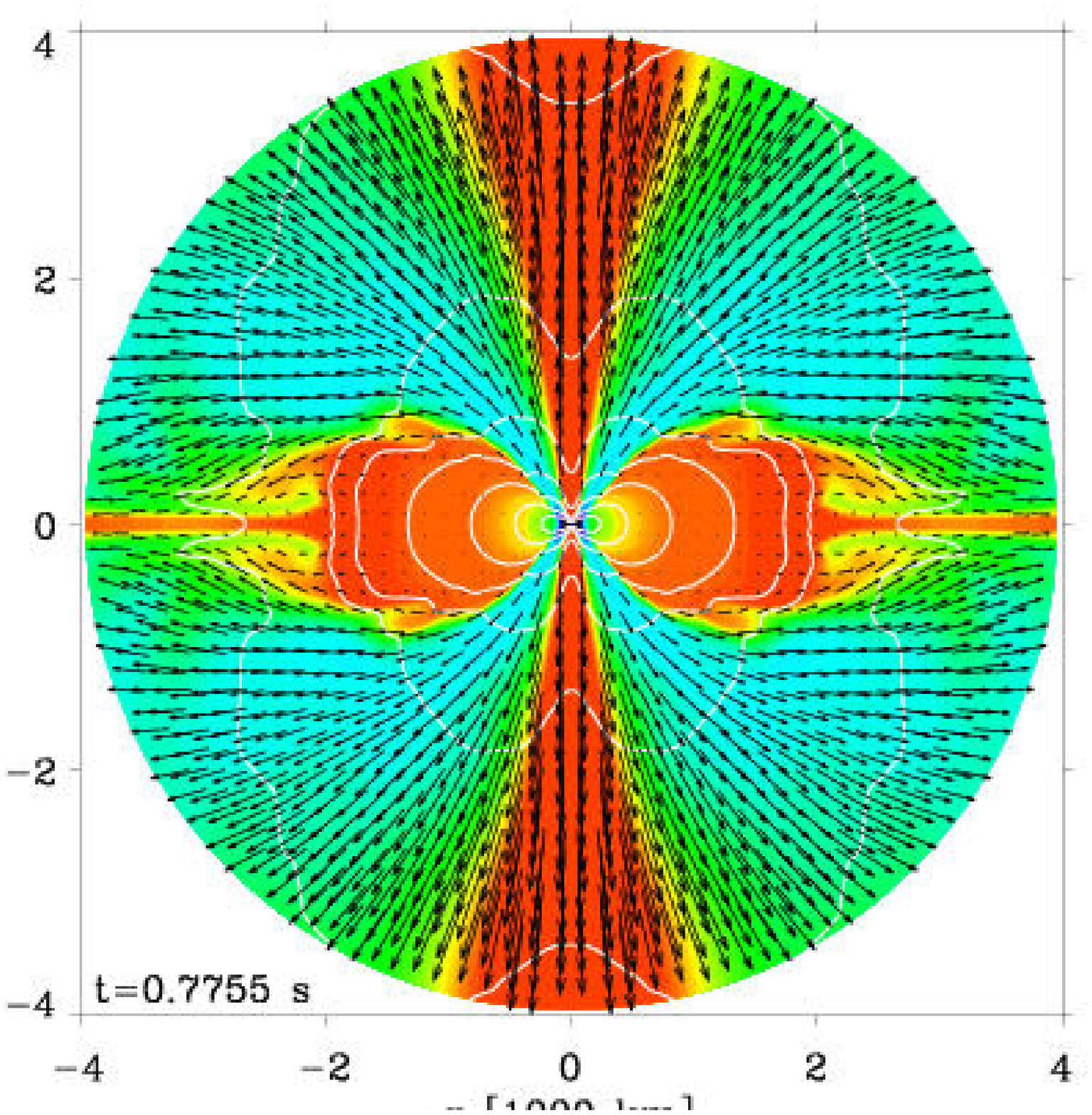}
%\clearpage
%\begin{figure*}
\caption{
Slide sequence of color maps of the electron fraction, $Y_{\rm e}$, covering 
the post-bounce evolution of the 1.46-\mo (left column) and 1.92-\mo (right column)
progenitors, overplotted with velocity vectors, as black arrows.
We also superpose isodensity contours as white lines, spaced every decade,
starting at 10$^{14}$\,g\,cm$^{-3}$ (top row) or 10$^{10}$\,g\,cm$^{-3}$ (bottom
two rows). 
From top to bottom, left to right, the vector of maximum length 
corresponds to a velocity magnitude of 43500\,\kms (infall), 
36500\,\kms (infall), 25500\,\kms (outflow), 19500\,\kms (outflow),
34000\,\kms (outflow), and 34500\,\kms (outflow).
We indicate the time after bounce in the bottom-left corner of each panel.
The spatial scale varies between panels, with widths in the 
range 500\,km to 10000\,km (8000\,km) for the 1.46-\mo (1.92-\mo) model. 
% !!!!LUC: the following cannot fit within the page
%
% Note, at late times (bottom row panels), the $Y_{\rm e}$ variations within the 
% ejecta (aspherically-distributed in the 1.92-\mo model), with lower values
% away from the poles, resulting from the lower electron-neutrino flux in the vicinity 
% of the neutrinosphere for the corresponding parcel trajectory (see \S\ref{sect_ye}).
See text for details.
}
\label{fig_seqye}
\end{figure*}

   In Figs.~\ref{fig_seq46}--\ref{fig_seq92}, we provide entropy (saturated at 
20\,k$_{\rm B}$/baryon, where k$_{\rm B}$ is Boltzmann's constant) color maps showing 
the key events in the evolution of both white dwarf models, with starting conditions discussed in 
the previous section and  displayed in Fig.~\ref{fig_init}. We complement these figures 
with Fig.~\ref{fig_seqye} for the electron fraction evolution at three reference 
times (left column: 1.46-\mo model; right column: 1.92-\mo model).
Note also that we overplot most figures with black arrows representing velocity vectors,
whose maximum length is set to 10\% of the width (or the height) of the image.
The corresponding maximum velocity is then given for each image in the figure captions - 
we also mention if we saturate the vector lengths. Having a large central density 
of 5$\times$10$^{10}$\,g\,cm$^{-3}$, both models achieve nuclear densities after the same time of $\sim$37\,ms. 
Differences in the bounce properties are attributable to the initial inner angular velocity 
distributions (Fig.~\ref{fig_init}). Compared to the 1.46-\mo model, the faster rotator 
has a lower maximum density at bounce (slightly shifted from the grid center to $r\sim$1\,km), 
i.e., 2.2 instead of 3.1$\times$10$^{14}$\,g\,cm$^{-3}$. It manifests an oblate, rather than 
a spherical, inner region (the inner tens of kilometers) of low entropy ($\sim$1$k_{\rm B}$/baryon;
visible in red). 
The deleptonized material in this inner region is, however, aspherically distributed
in both models, although more so in the 1.92-\mo model, with the lower $Y_{\rm e}$ material lying 
in a disk structure along the equatorial direction. The flatter density gradient
in this direction and the longer dwell time near the neutrinosphere 
conspire to produce this enhanced deleptonization. At core bounce, the low-density outer 
parts of the white dwarf progenitor have not yet started to infall and the progenitor still retains 
its original shape. 
The collapse of the inner regions, however, creates a rarefaction wave that triggers the infall
of the outer material, with a magnitude that is more pronounced along the poles due to the compact
structure of the white dwarf in these directions.

Although we do not observe a prompt explosion (i.e. occuring on a dynamical timescale of 
just a few milliseconds), the shock progresses slowly outwards without conspicuously 
stalling. This is slightly different from the core-collapse simulations recently published, 
whereby the shock {\it systematically} stalls (Burrows et al. 2005; Buras et al. 2005ab).
In the equatorial direction, facilitated by the centrifugal support of the infalling 
material, the shock progresses outwards steadily, and is faster in the faster rotating model, 
reaching a few hundred kilometers after $\sim$100\,ms.
Given the initial constant rotation rate on cylinders (see \S\ref{sect_progenitor}), 
the {\it radial} inflow of mass
brings angular momentum to the equatorial ($z \sim 0$) region, which becomes more strongly supported,
facilitating further the progress of the shock at low latitudes.
This exacerbates the non-spherical development of the shock structure in both models.
In its wake, we see a few large-scale whirls, resulting from the generation, at mid-latitudes
and due to shock passage, of large vortical motions. 

%\clearpage
\epsscale{1}
\begin{figure*}
%\plottwo{rho_radslice_m1.46_polar.ps}{rho_radslice_m1.92_polar.ps}
\plottwo{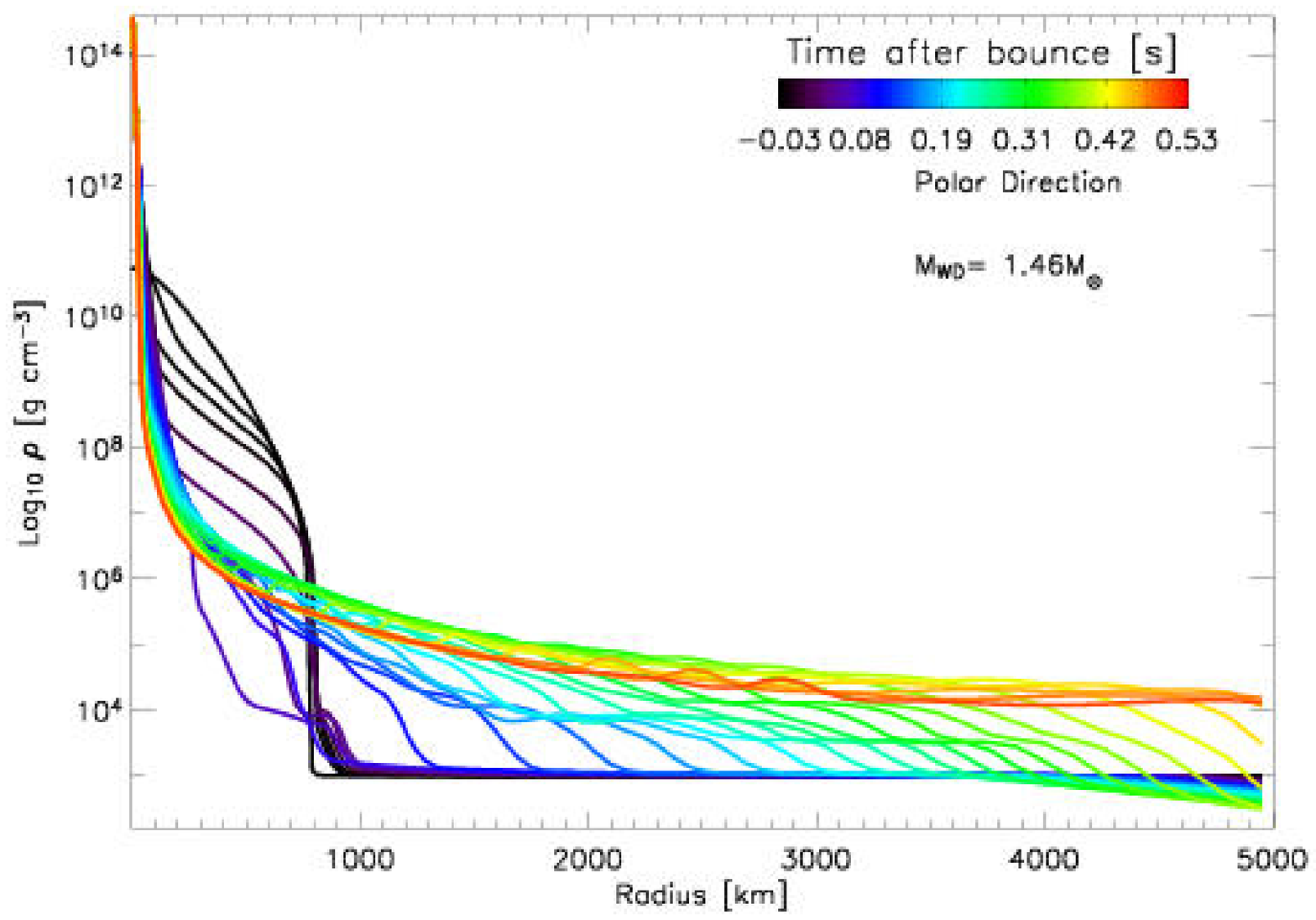}{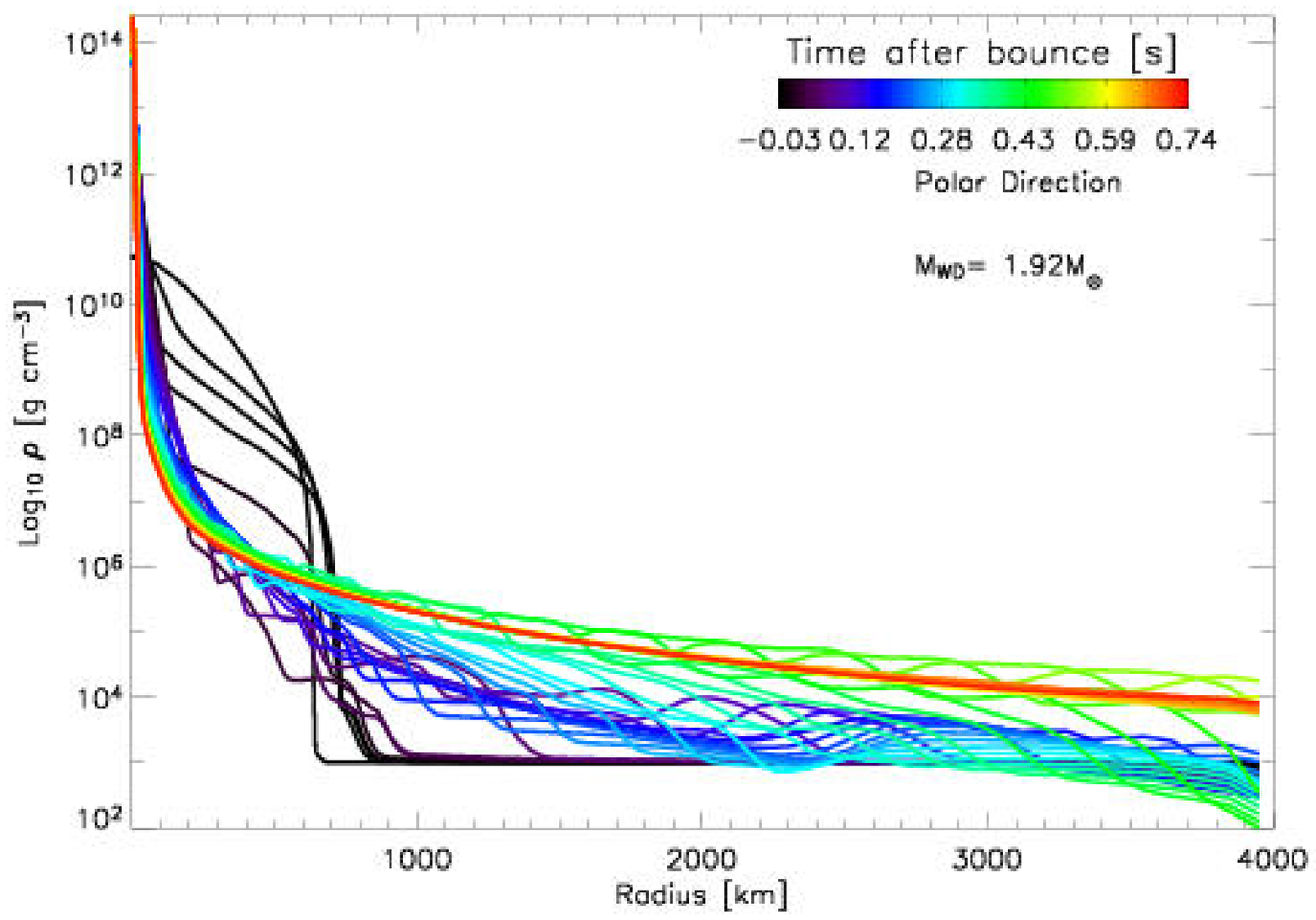}
%
%\plottwo{rho_radslice_m1.46_eq.ps}{rho_radslice_m1.92_eq.ps}
\plottwo{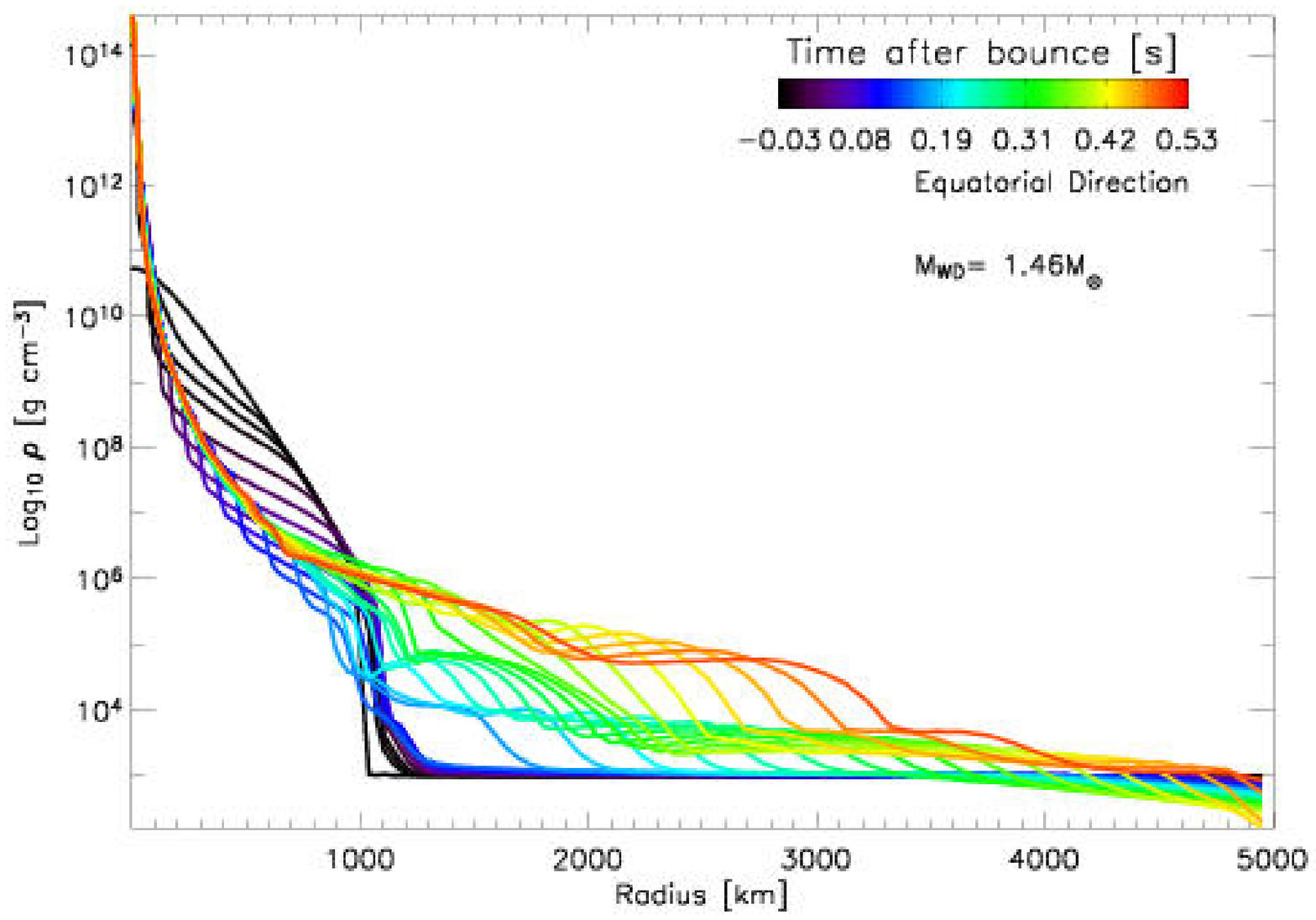}{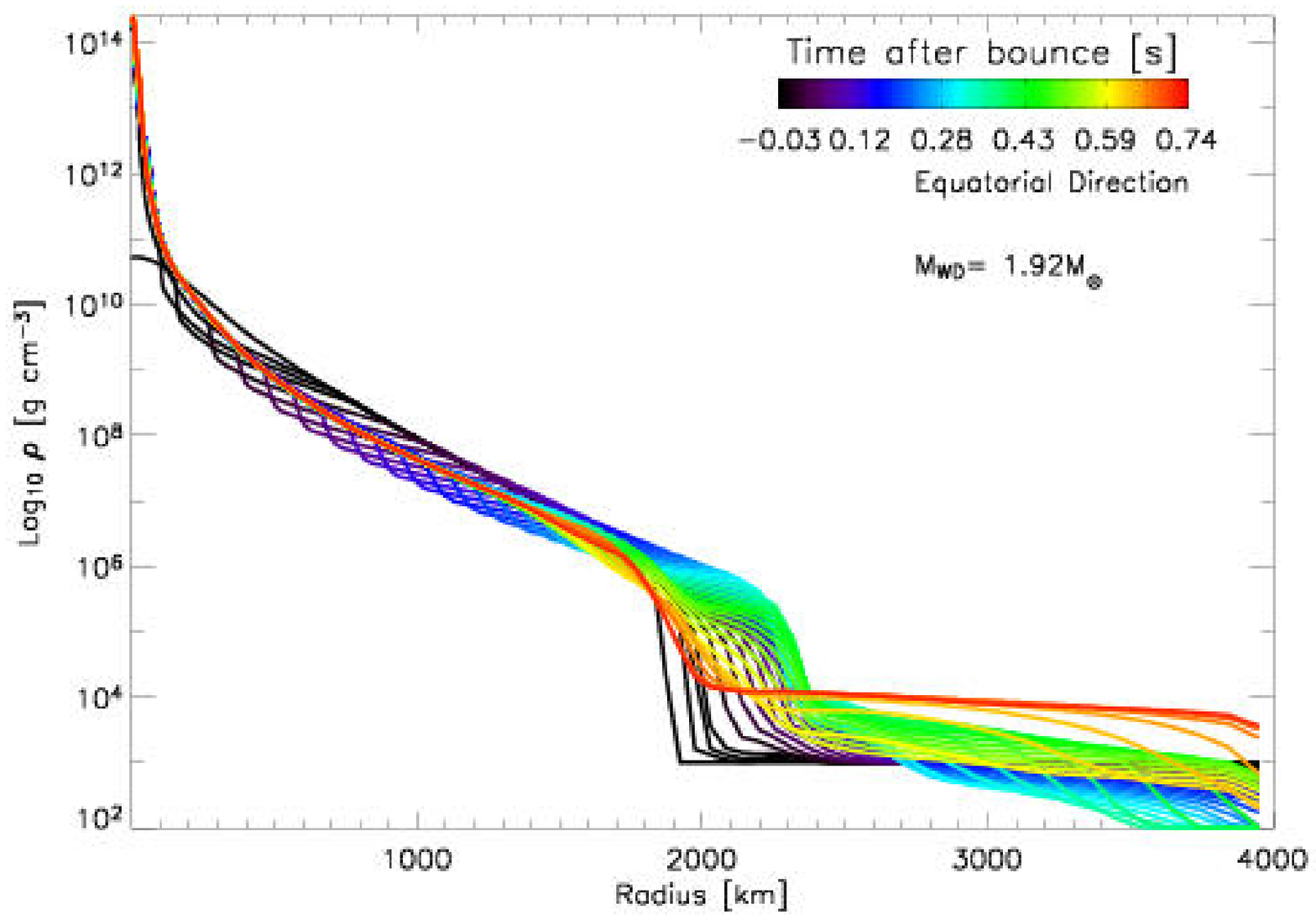}
\caption{
{\it Top}: Time evolution of the polar density profile for the 1.46-\mo
(left column) and 1.92-\mo (right column) progenitor models. 
{\it Bottom}: Same as above, but for the equatorial direction.
Note the steeper density profiles at all times along the pole compared to the equatorial
direction, highlighting the contrast between the wind (polar direction) and disk (equatorial 
direction; within a 2000\,km radius at 770\,ms for the 1.92-\mo model). 
See text for discussion.
}
\label{fig_rho_radslice}
\end{figure*}

\begin{figure*}
%\plottwo{vr_radslice_m1.46_polar.ps}{vr_radslice_m1.92_polar.ps}
\plottwo{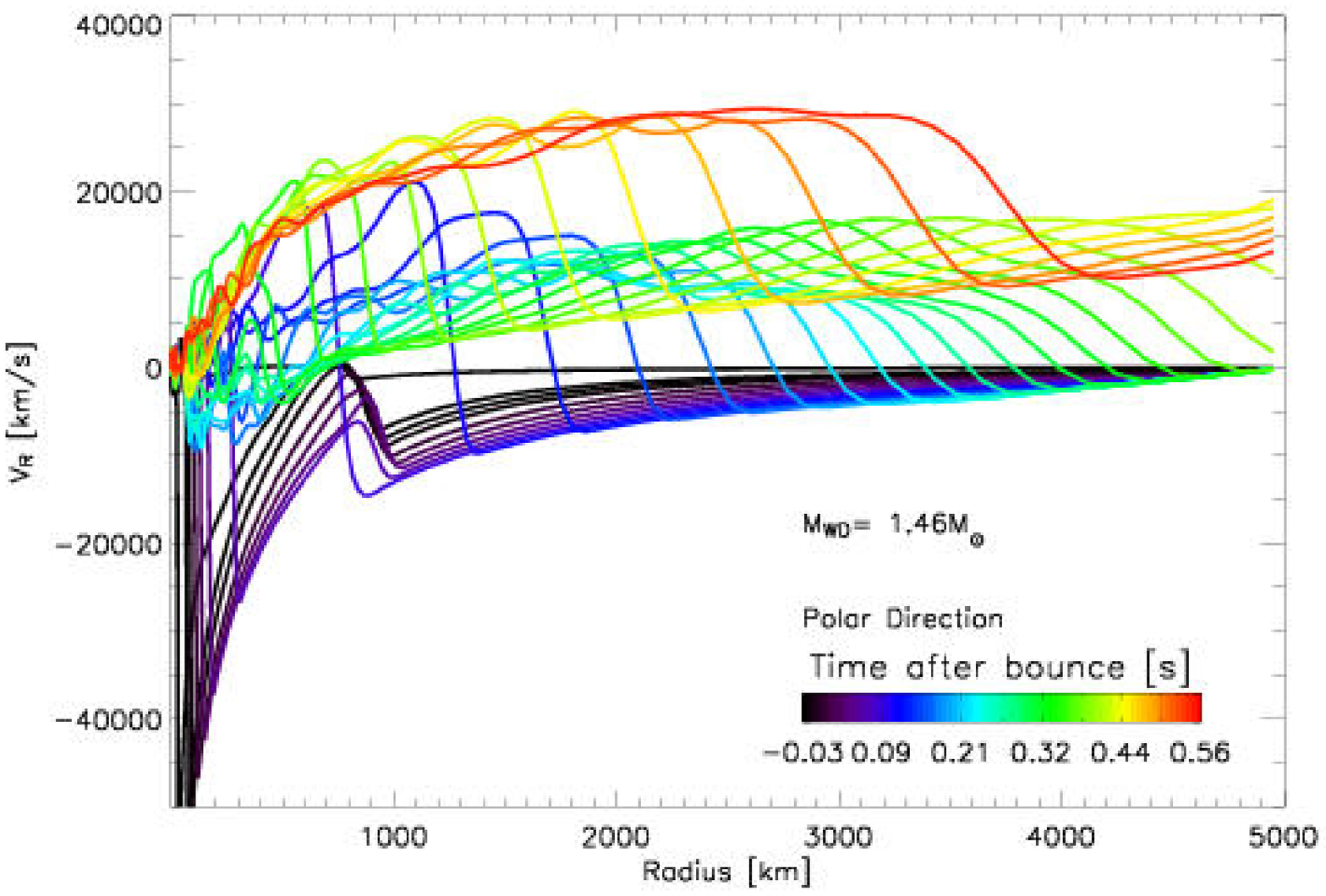}{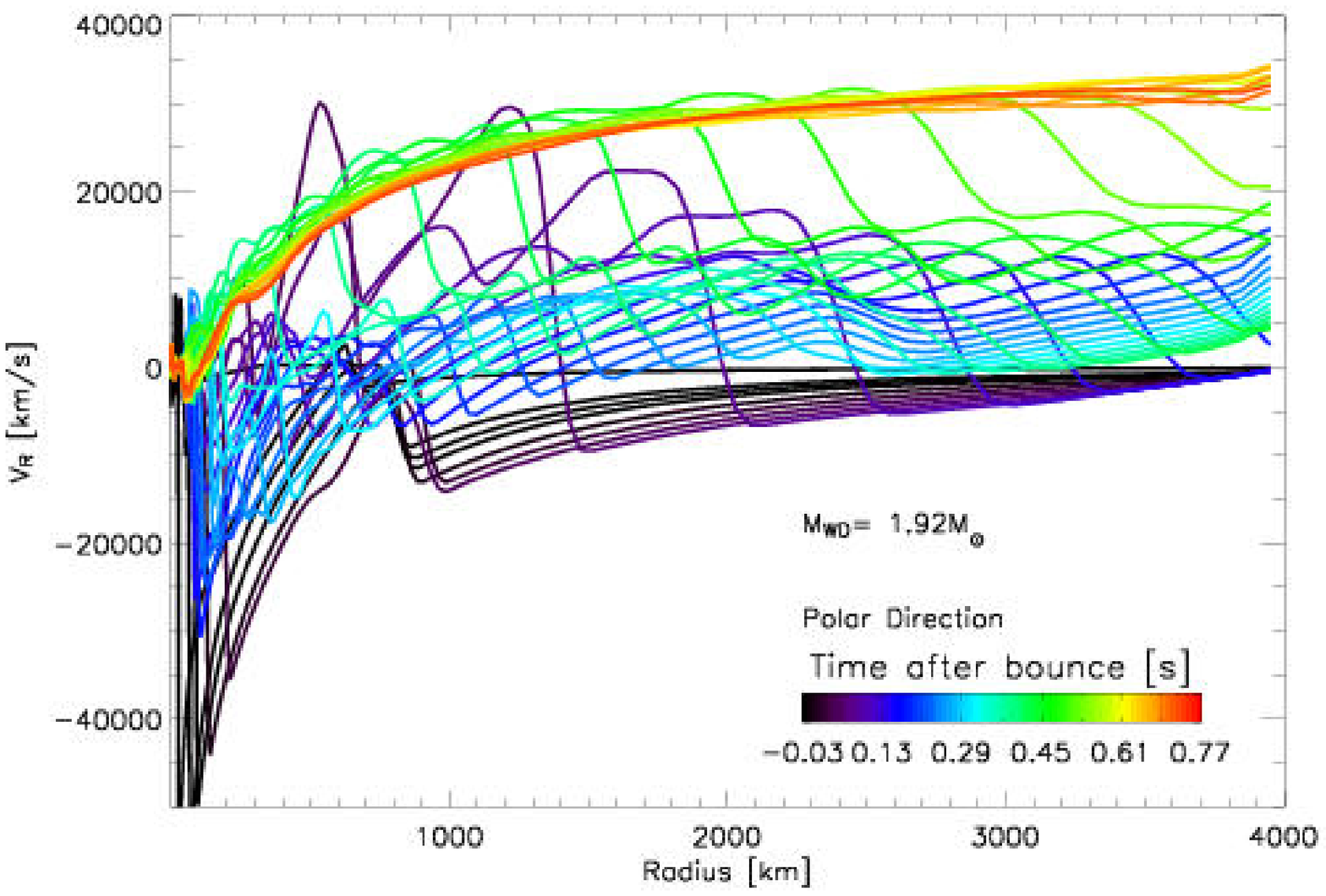}
%
%\plottwo{vr_radslice_m1.46_eq.ps}{vr_radslice_m1.92_eq.ps}
\plottwo{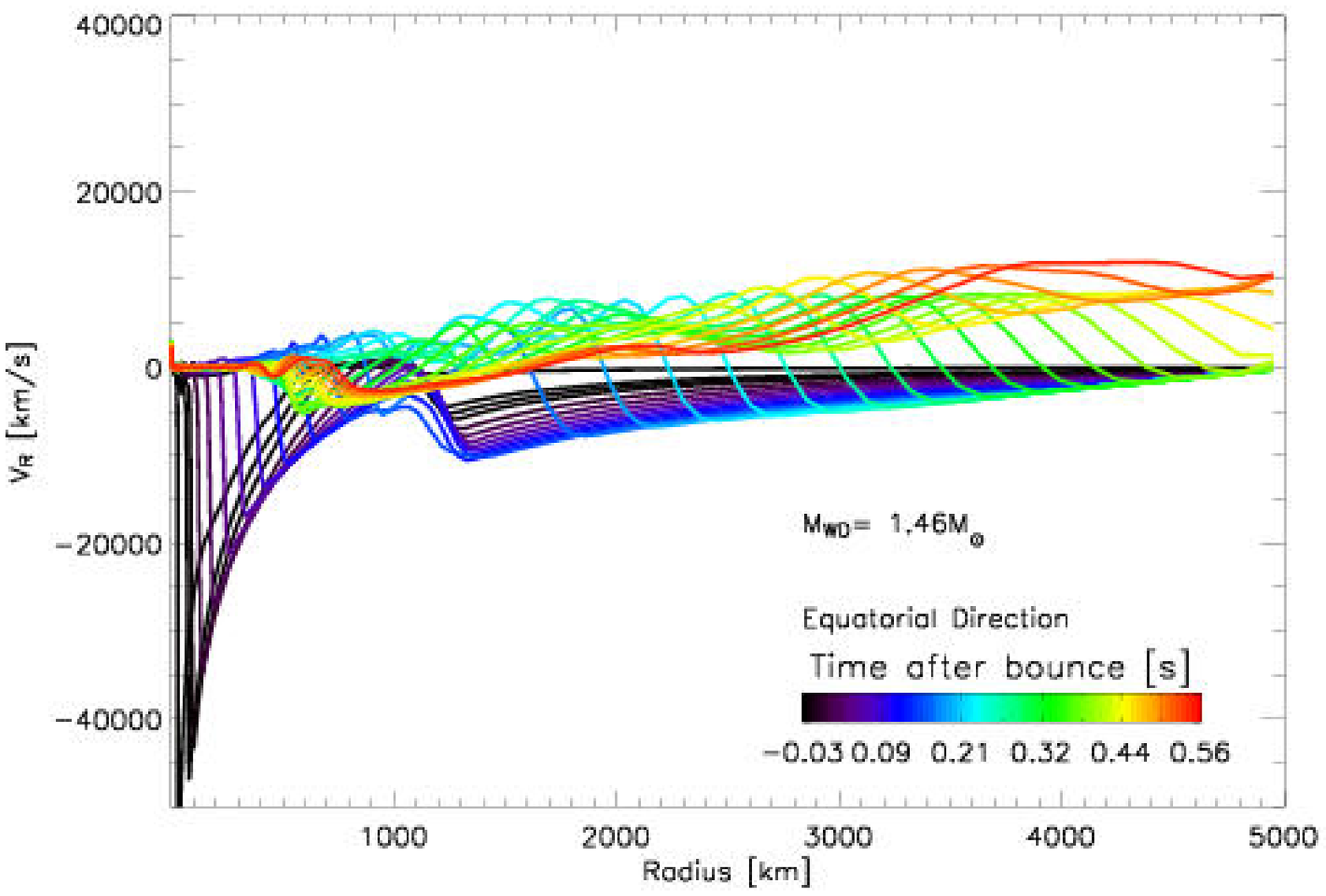}{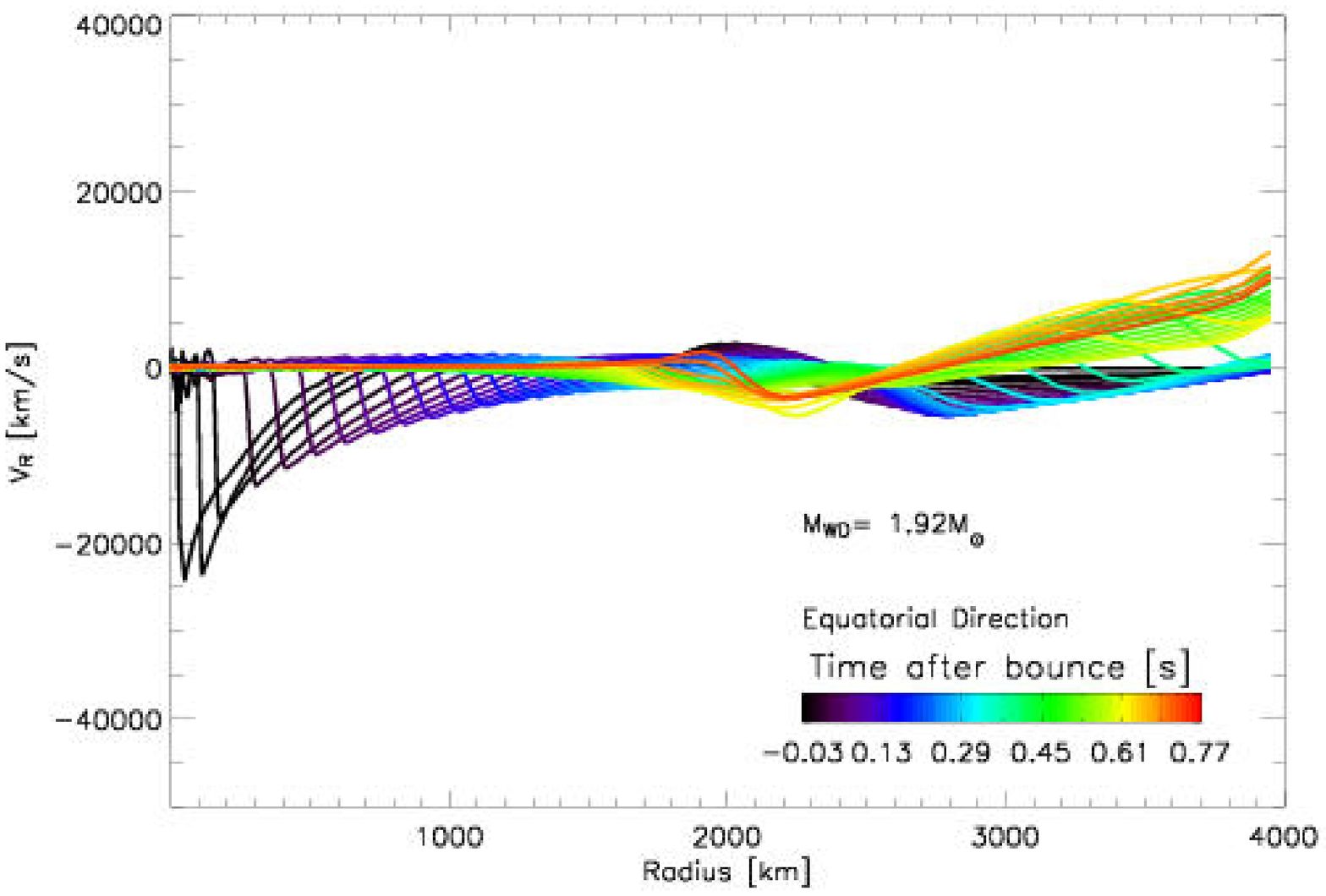}
\caption{
Same as for Fig.~\ref{fig_rho_radslice}, but this time for the radial velocity $V_R$.
Note the large contrast between the polar and equatorial pre-bounce (infall) velocities
for the 1.92-\mo model, the fast neutrino-driven wind at late times along the poles,
overtaking the slower expansion of the shock generated at core bounce.
Along the equator, material shows slow (1.46-\mo model) or absent (within a 2000\,km radius, 
1.92-\mo model) expansion, belonging to a quasi-Keplerian disk (\S\ref{sect_disk}).
See text for discussion.
}
\label{fig_vr_radslice}
\end{figure*}
%\clearpage
In the polar direction, centrifugal support is absent, but the stalling of the shock 
is prevented by the quickly decreasing accretion rate established by the steeper density gradient,
reduced polar radius, and smaller mass budget.
Hence, although the shock slowly migrates along the equatorial regions, it soon traverses the 
surface layer of the white dwarf along the poles and escapes outwards into the ambient medium.
Due to the strong asphericity of the progenitor, this occurs only 70\,ms after core bounce in the 
1.92-\mo model, $\sim$30\,ms earlier than for the 1.46-\mo model.
The ambient medium is then swept up by this blast, whose opening angle is constrained
by that of the uncollapsed disk of the progenitor. 
The outflow expansion rate is larger 30-40 degrees away from the poles than right along
the pole, coinciding in Fig.~\ref{fig_seqye} with the lower $Y_e$ material. The effect
is large for the 1.92-\mo model, giving a butterfly shape in cross section to the faster expanding portions of the
outflow. Off-axis material has more rotational kinetic energy available to convert to 2D planar ($r,z$) 
kinetic energy as it streams outward, reducing its rotational velocity while preserving 
angular momentum, thereby resulting in an enhanced acceleration compared to that of the material situated 
along the poles and lacking rotational energy.

As the shock expands, it wraps around
the disk, typically with a speed near that of the local sound speed. Nearer the pole, the outflow sweeps along
the pole-facing side of the pole-excavated white dwarf, entraining surface material and 
effectively loading the outflow with more mass, causing unsteady fallback onto the neutron star.
For the 1.46-\mo progenitor, the shock completely wraps around the low-latitude
regions of the white dwarf, and finally emerges from the outer equatorial regions, as witnessed 
by an outward-moving entropy jump. By 250\,ms after bounce, the 
shock reaches a few thousand kilometers, assumes a near spherical shape, and the entire 
white dwarf material outside the newly-formed neutron star flows nearly radially outward.
In the high-rotation (1.92-\mo) progenitor, the white dwarf possesses a lot more mass at 
near zero-latitude and this confines more drastically the emerging shock along the poles. 
As the shock migrates outwards, it opens up; it does wrap around the progenitor, but much 
later than when the shock escapes in the polar directions.
Despite the reduced ram pressure associated with centrifugal support, the shock stalls 
a few hundred milliseconds after bounce along all near-equatorial directions. 

Within 100-200\,ms, the newly-formed neutron star has a mass of $\sim$1.4\,\mo, similar in 
both models despite the 0.5\,\mo difference in progenitor mass.
Note that the large neutron star asphericity and the sizably lower density at the neutrinosphere for 
near-zero latitudes makes this mass definition ambiguous at such early times, especially
in the 1.92-\mo model.
Indeed, the neutron star is not clearly distinguishable from the 
surrounding equatorial material, so imposing either a density cut of 10$^{10-11}$\,g\,cm$^{-3}$ 
or a radius cut in defining the newly-born neutron star appears arbitrary when determining the residual mass.
In the 1.46-\mo model, about 0.06\,\mo remains outside of the neutron star, mostly in the
equatorial disk region; in the 1.92-\mo model, 0.6\,\mo is now lying in this disk-like structure. 
The rest of the initial mass is outflowing material, which, if selected according
to an outward radial velocity discriminant of 10000\,\kms (comparable to the escape velocity at
3000\,km), reaches 4$\times$10$^{-3}$\,\mo for the 1.46-\mo model and 3$\times$10$^{-3}$\,\mo 
for the 1.92-\mo model. These various components are documented in more detail in 
\S\S\ref{sect_disk}--\ref{sect_ye}.

The late-time evolution of both models is characterized by a strong neutrino-driven wind that 
sets in about 300\,ms after bounce, replenishing the grid with denser material (on average
10$^4$\,g\,cm$^{-3}$) and large velocities (with a maximum of 30000\,\kms along the poles).
The properties of the neutrino-driven wind are very angle-dependent, the density changing
by 30\% between the pole and the 40$^{\circ}$ latitude, while the radial outflow velocity
varies by a factor of 3 in the 1.92-\mo model. 
The latitudinal dependence of the mass flux per unit solid angle is therefore dominated by 
a variation in asymptotic velocity. We will discuss this result in more detail in \S\ref{sect_ener}.
By the time we stop the simulations, at 550\,ms and 780\,ms for the 1.46-\mo and 1.92-\mo models,
all the ambient medium originally placed around the white dwarf progenitor has been swept away by the
neutrino-driven wind, which occupies all the space outside the neutron star and the disk.
The electron fraction of the material ejected in the original blast is close to 0.45--0.5, 
while subsequently, in the neutrino-driven wind, the values are lower, with a pronounced
decrease towards lower latitudes (note, however, the high $Y_{\rm e}$ right along the pole; see 
\S\ref{sect_ye}).

In Fig.~\ref{fig_rho_radslice}, we recapitulate for both models the evolution described 
above by showing equatorial and polar slices of the density as a function of time.
Striking features are the distinct polar and equatorial surface radii, the fast infall of
the inner regions to nuclear densities, the slow plowing of the shock along the
equatorial directions, superseded in radial extent and velocity by the shock in the polar direction
as the surface mass shells collapse in, and finally the emergence of a sequence
of density kinks associated with the birth of the fast neutrino-driven wind that sweeps
away the previously shocked material that did not leave the grid.
Similarly, in Fig.~\ref{fig_vr_radslice}, we show slices of the radial velocity, $V_R$, 
along the polar (top row) 
and equatorial (bottom row) directions for the 1.46-\mo (left column) and 1.92-\mo (right
column) models. Notice the much larger infall velocities, similar along the poles
and the equator for the 1.46-\mo model, but with a strong latitudinal dependence
in the 1.92-\mo model. In that model, the speed contrast between the polar and equatorial 
directions is $\sim$30000\,\kms.
Overall, the evolution is more rapid along the poles than on the equator, with larger 
asymptotic velocities (30000\,\kms compared with 10000\,\kms), and with the establishment 
of a quasi-stationary outflow at late times along the pole.
These radial slices offer a means of better interpreting the fluid velocities, depicted
with vectors, in most color maps shown in this paper. We also show isodensity contours
in most color maps to provide some feeling for the density distribution.

Having described the general properties of the two simulations of the AIC of a 
1.46-\mo and 1.92-\mo white dwarf, we now address more specific issues, covering the
properties of the nascent neutron star (\S\ref{sect_pns}), the neutrino signatures (\S\ref{sect_nu}),
the properties of the residual disk and the angular momentum history (\S\ref{sect_disk}),
the neutrino-driven wind and the global energetics (\S\ref{sect_ener}), the electron fraction
of the ejected material (\S\ref{sect_ye}), and, finally, the gravitational wave signatures (\S\ref{sect_gw}).

\section{Neutron Star properties}
\label{sect_pns}
%\clearpage
\begin{figure*}
\vspace{-1.0cm}
% \plotone{colorbarnusphere.ps}
\plotone{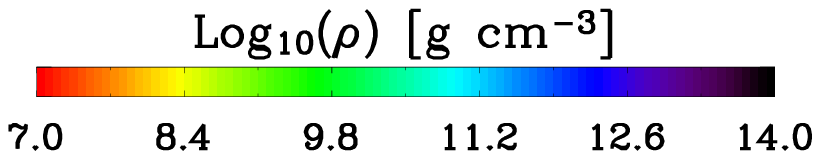}
%\vspace{-0.3cm}
% \plottwo{aic_m1pt46_nusphere_t=0.0590s_spec.ps}{aic_m1pt46_nusphere_t=0.0590s_flavor.ps}
\plottwo{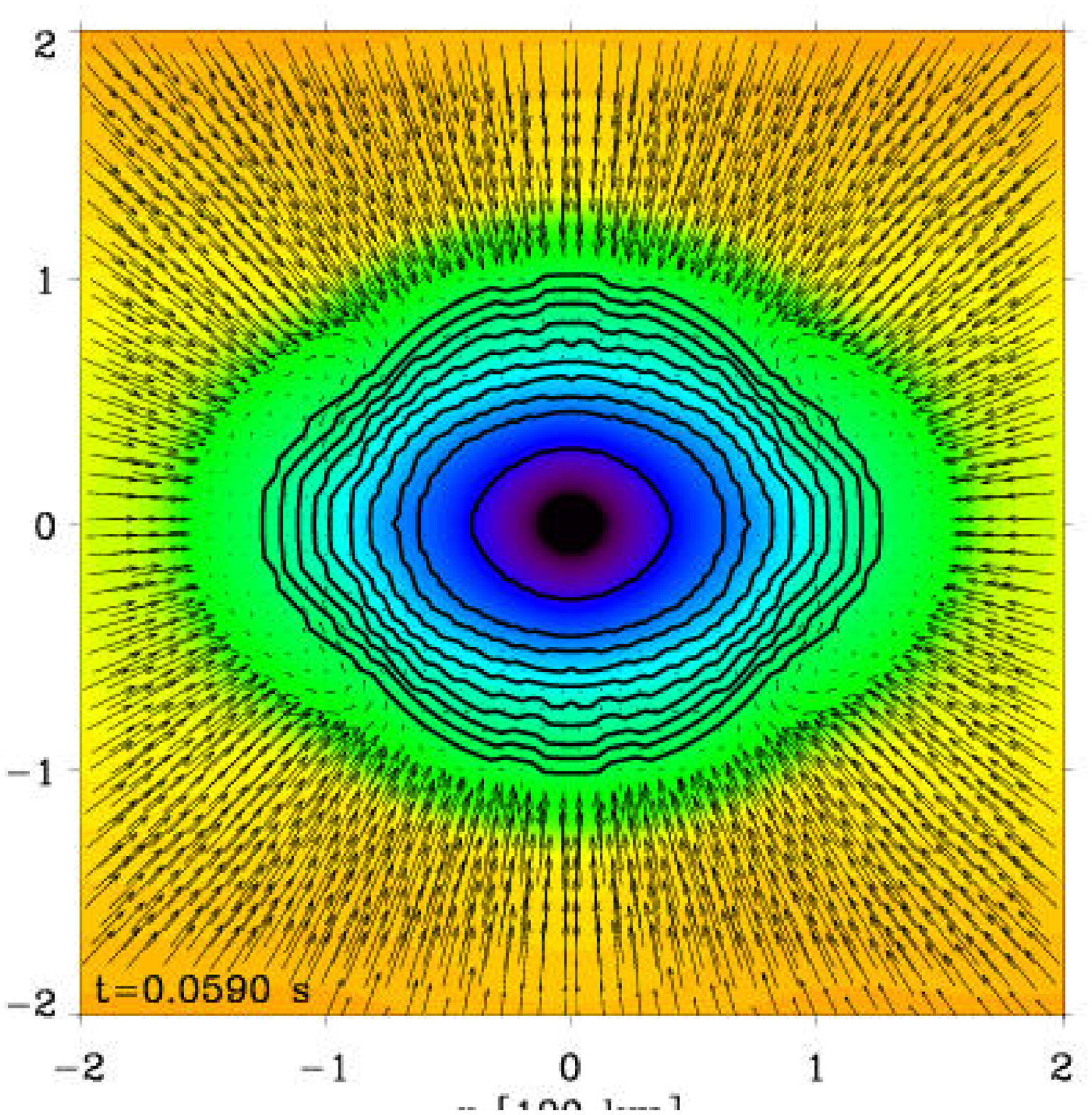}{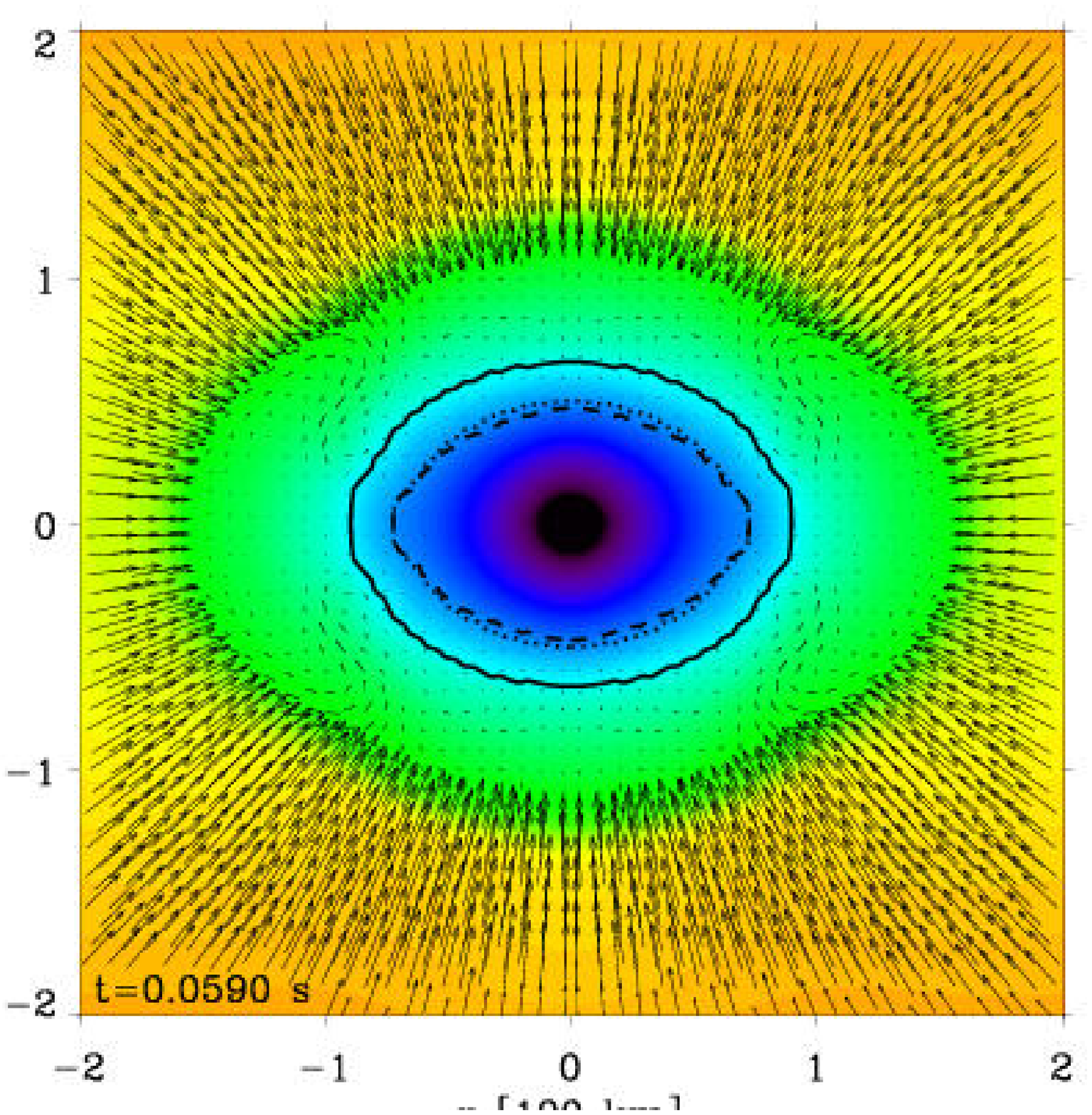}
\vspace{-0.3cm}
%\plottwo{aic_m1pt46_nusphere_t=0.5700s_spec.ps}{aic_m1pt46_nusphere_t=0.5700s_flavor.ps}
\plottwo{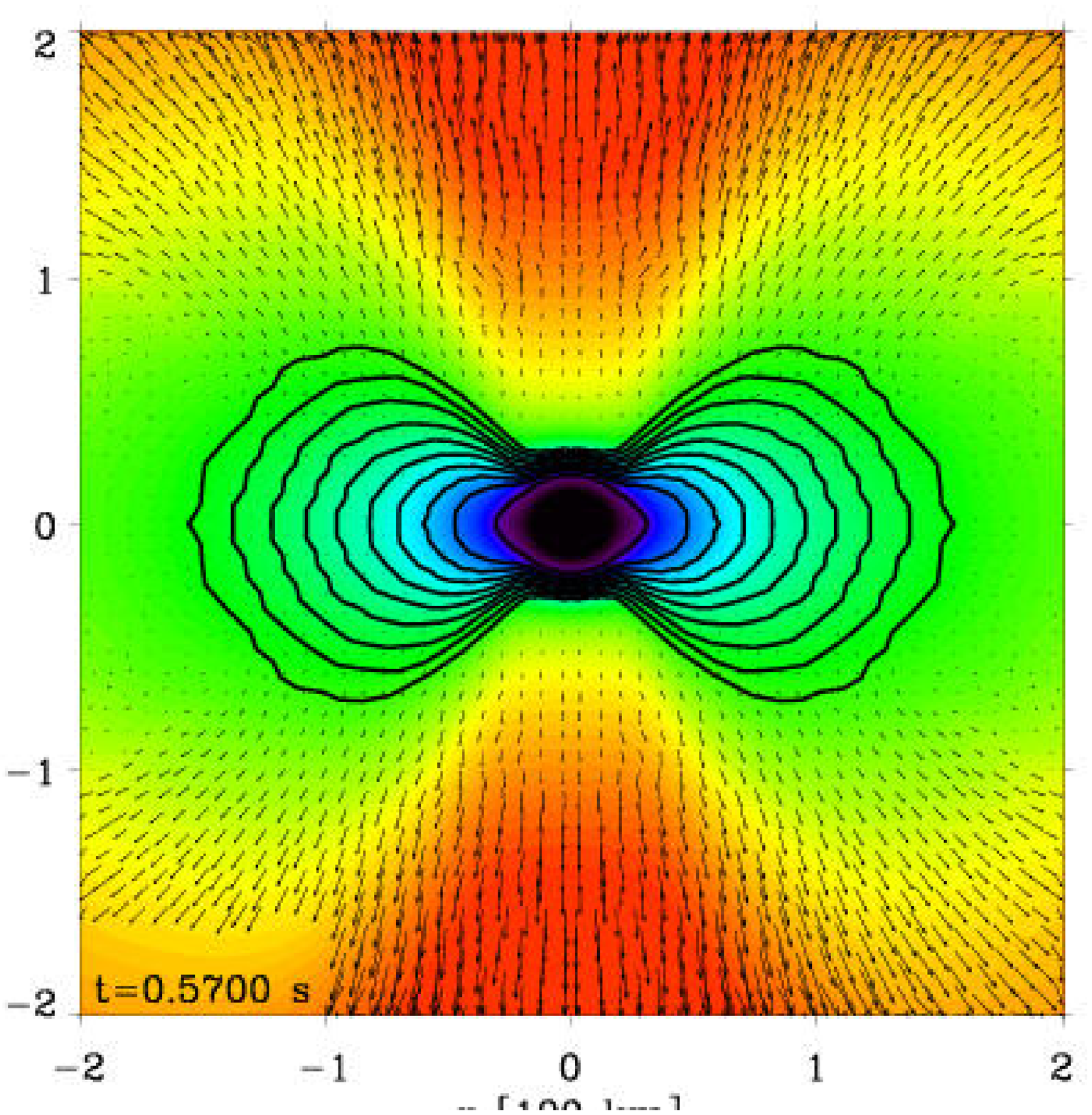}{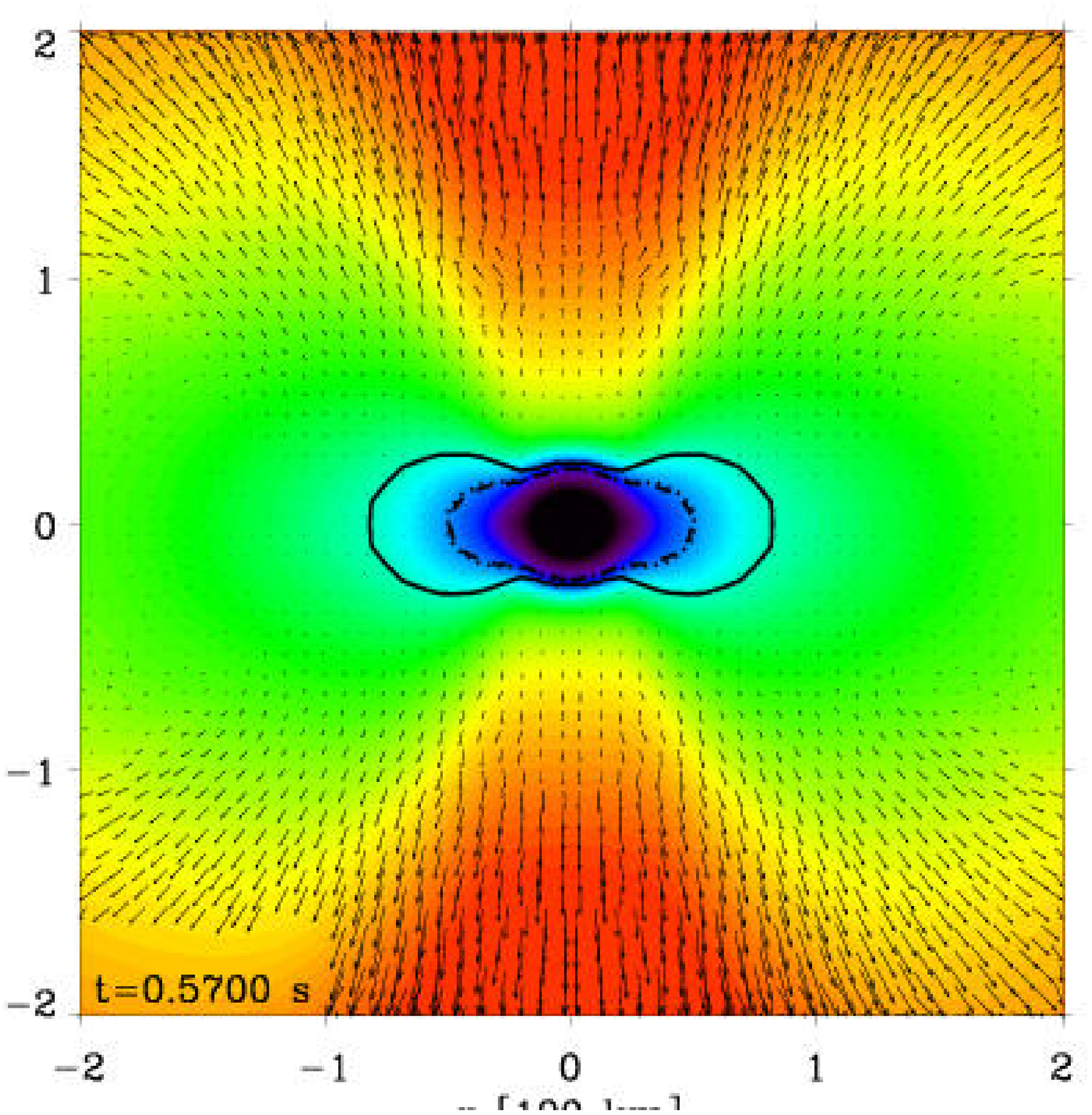}
\caption{
{\it Top}:
Color map of the density at 59\,ms after bounce for the 1.46-\mo model.
{\it Left}: Neutrinosphere radii for the $\nu_{\rm e}$ neutrino for ten energy groups
around the peak of the emergent energy distribution, i.e., at 2.5, 5.5, 
7.5, 9.5, 12.5, 16, 21, 27, 36, and 46\,MeV. 
{\it Right}: Neutrinosphere radii at 12.5\,MeV for the three neutrino flavors: 
$\nu_{\rm e}$ (solid lines), ${\bar{\nu}}_{\rm e}$ (dotted line), and ``$\nu_{\mu}$'' 
(dash-dotted line, which overlaps with the dotted line).
{\it Bottom}: Same as above, but for the last computed time at 570\,ms after bounce.
The vector with the maximum length corresponds to a velocity magnitude of 46300\,\kms 
(infall; top panels) and 9440\,\kms (outflow; bottom panels). 
The fixed spatial coverage of the display is 200$\times$200\,km$^2$.
}
\label{fig_nusphere_46}
\end{figure*}

\begin{figure*}
% \plotone{colorbarnusphere.ps}
\plotone{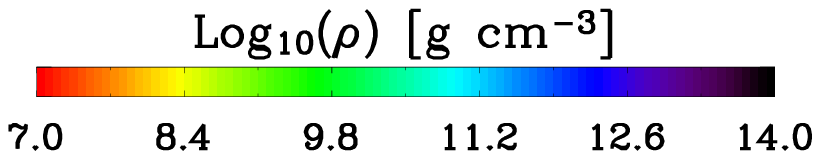}
%\vspace{-0.3cm}
% \plottwo{aic_m1pt92_nusphere_t=0.0590s_spec.ps}{aic_m1pt92_nusphere_t=0.0590s_flavor.ps}
\plottwo{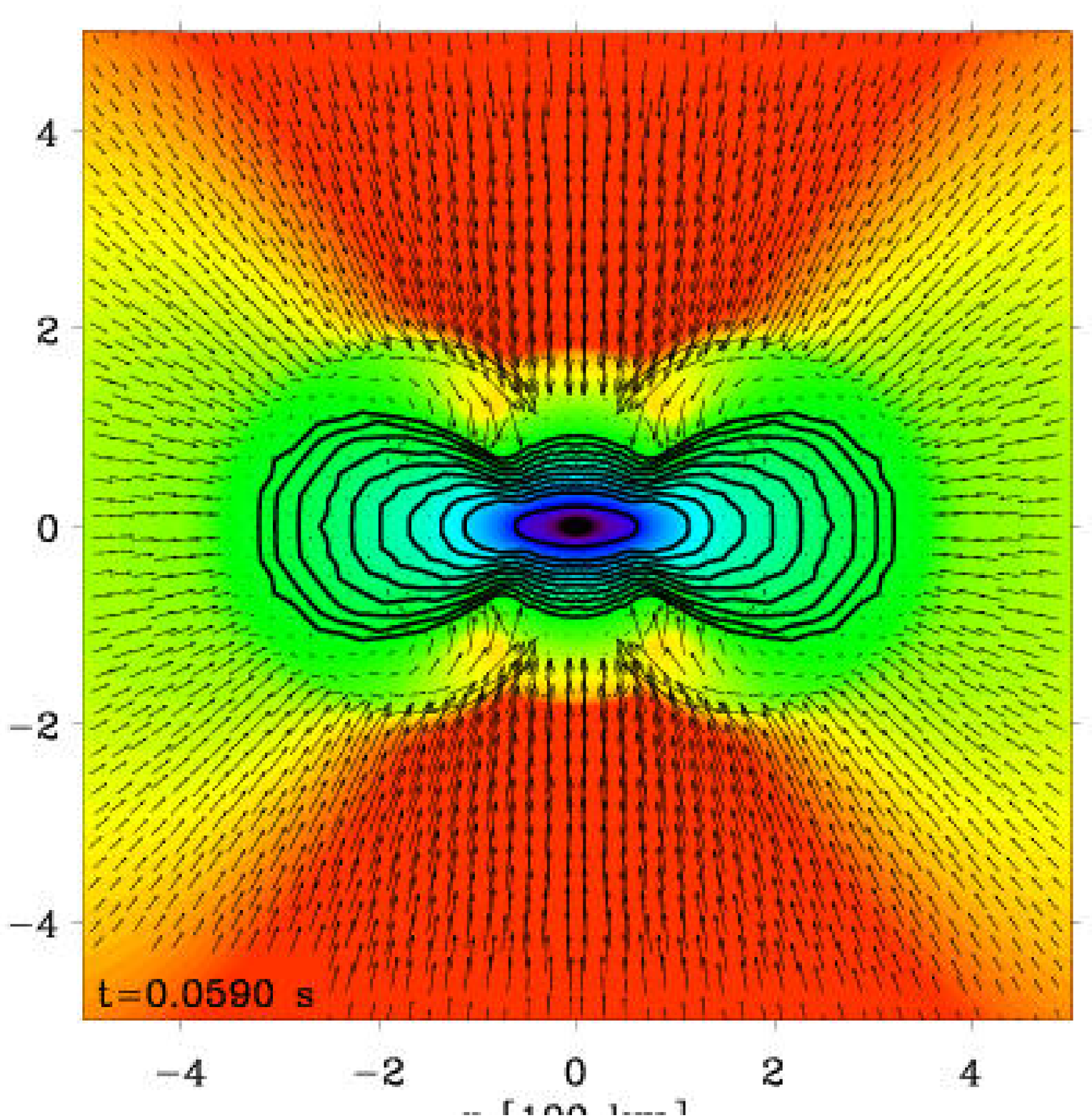}{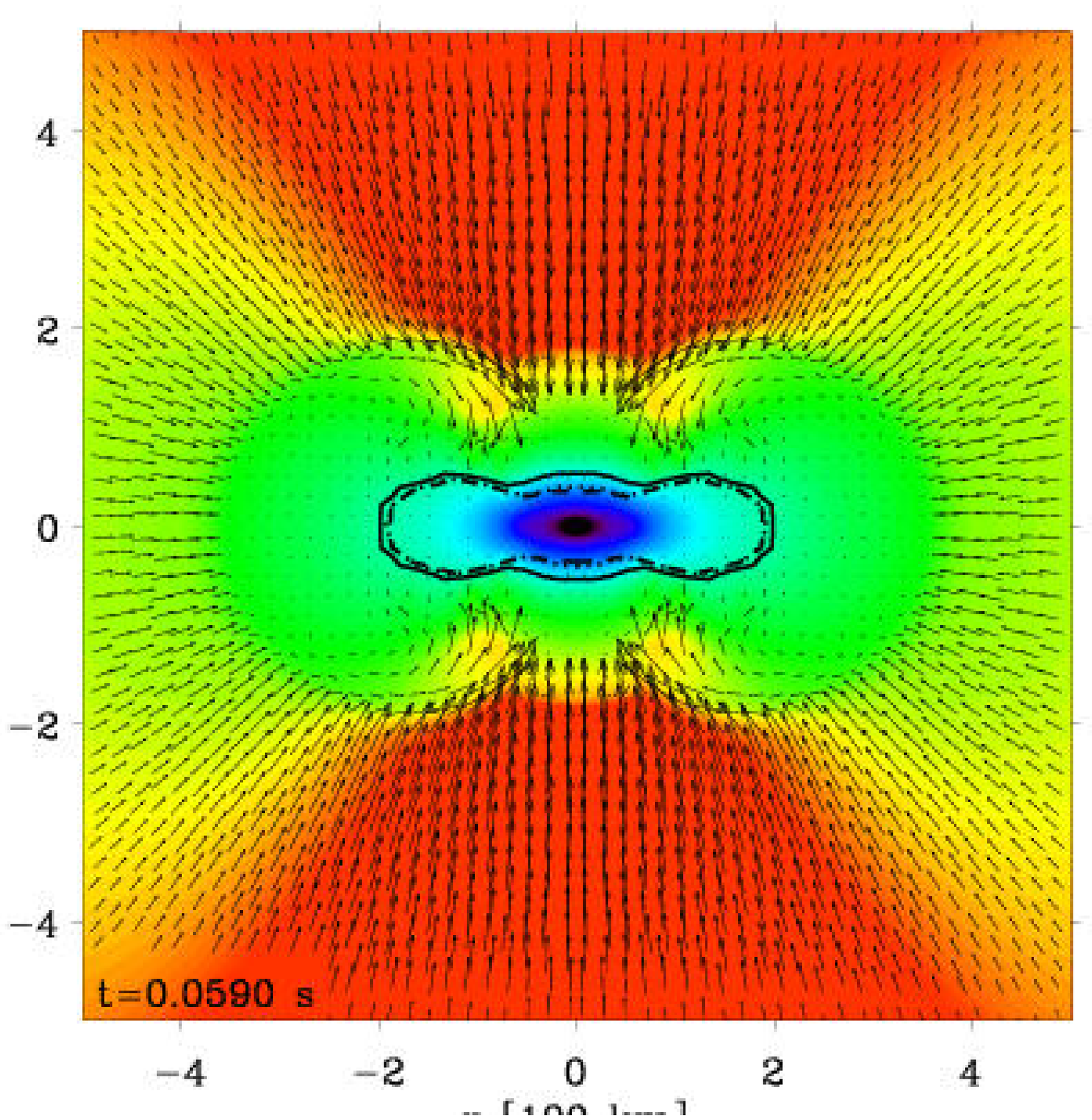}
\vspace{-0.3cm}
% \plottwo{aic_m1pt92_nusphere_t=0.7755s_spec.ps}{aic_m1pt92_nusphere_t=0.7755s_flavor.ps}
\plottwo{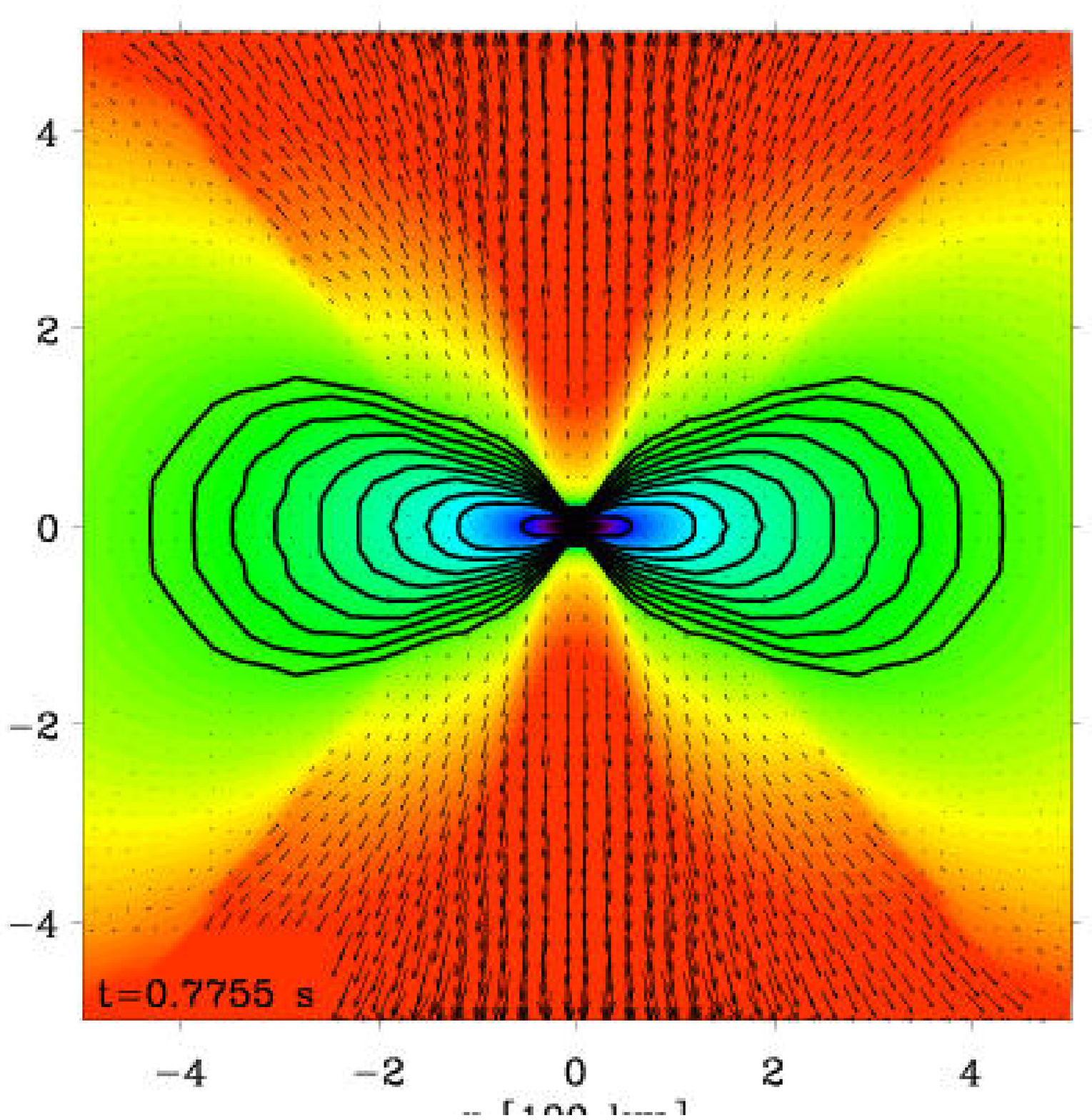}{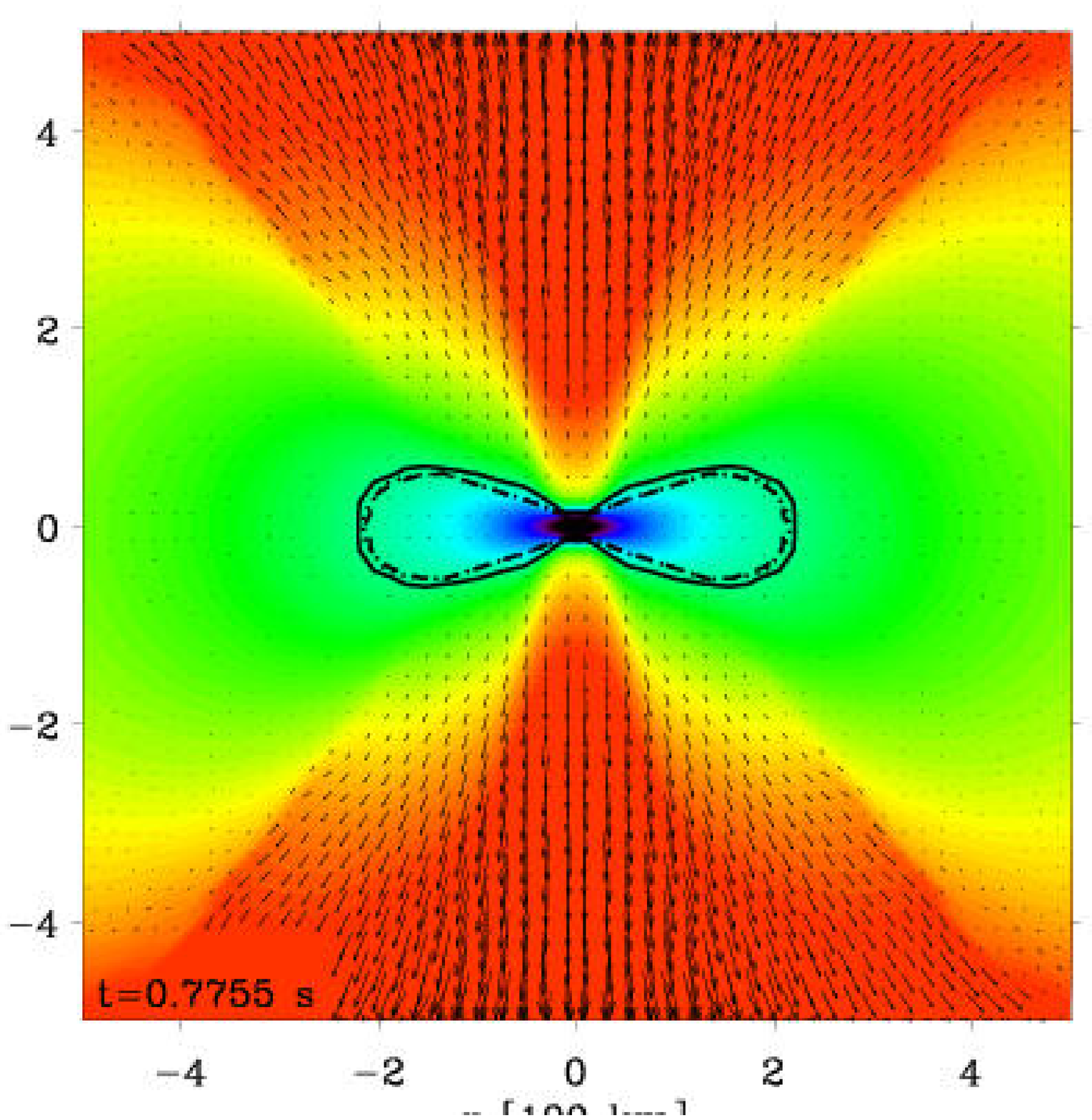}
\caption{
{\it Top}:
Color map of the density at 59\,ms after bounce for the 1.92-\mo model.
{\it Left}: Neutrinosphere radii for the $\nu_{\rm e}$ neutrino for ten energy groups
around the peak of the emergent energy distribution, i.e., at 2.5, 5.5, 
7.5, 9.5, 12.5, 16, 21, 27, 36, and 46\,MeV. 
{\it Right}: Neutrinosphere radii at 12.5\,MeV for the three neutrino flavors: 
$\nu_{\rm e}$ (solid lines), ${\bar{\nu}}_{\rm e}$ (dotted line), and ``$\nu_{\mu}$'' 
(dash-dotted line, which overlaps with the dotted line).
{\it Bottom}: Same as above, but for the last computed time at 775.5\,ms after bounce.
The vector of maximum length corresponds to a velocity magnitude of 36500\,\kms 
(infall; top panels) and 15500\,\kms (outflow; bottom panels). 
The fixed spatial coverage of the display is 500$\times$500\,km$^2$.
}
\label{fig_nusphere_92}
\end{figure*}
%\clearpage
  The white dwarf progenitors discussed in this paper, due to their evolution to high central
densities, high rotational kinetic energies, and high mass, are distinctive in that their 
cores always collapse to form neutron stars rather than being disrupted by the explosive 
burning of carbon and oxygen.

   We find that neutron stars formed from the AIC of the progenitor white dwarfs used in this work 
are very aspherical (see, for example, Walder et al. 2005; Liu \& Lindblom 2001; Janka \& M\"onchmeyer 1989ab), 
although there are significant differences in evolution after bounce between 
the two models. In Fig.~\ref{fig_nusphere_46},  we show color maps of the density field 59\,ms 
after bounce (top row), as well as for the last time computed ($t=570$\,ms after bounce; bottom row) 
in the 1.46-\mo model. 
To render more striking the level of asphericity of the neutron star ``surface,'' 
we overplot the neutrinospheres $R_{\nu}(\varepsilon_{\nu},r,z)$, adopting the definition 
 $$ \tau(R_{\nu}(\varepsilon_{\nu},r,z)) = 
 \int_{R_{\nu}(\varepsilon_{\nu},r,z)}^\infty \kappa_{\nu}(\rho,T,Y_{\rm e}) \rho(r',z') dR' = 2/3\,,$$ 
%$$ \tau(R_{\nu}(r,z)) = \int_{R_{\nu}(r,z)}^\infty \kappa_{\nu}(\rho,T,Y_{\rm e}) \rho(r',z') dR' = 2/3\,,$$ 
where $\kappa_{\nu}(\rho,T,Y_{\rm e})$ is the combined material absorption and scattering opacity 
to neutrinos, and the integration is carried out along radial rays, with $R' = \sqrt{r'^2 + z'^2}$, 
using 30 equally spaced latitudinal directions per quadrant. 
Strictly, this definition is most appropriate for the photosphere/neutrinosphere in a plane-parallel
atmosphere, but it gives a sense of the asphericity of the collapsed core.
Note that the above expression contains a dependence both on the neutrino flavor and the
neutrino energy $\varepsilon_{\nu}$. 
In the left column of Fig.~\ref{fig_nusphere_46}, line contours correspond to such neutrinosphere radii as 
a function of energy group, bracketing the peak of the neutrino energy distribution at the 
neutrinosphere, i.e., between 2.5 and 46\,MeV. Material opacity to neutrinos increases with
the square of the energy, so that higher-energy neutrinos have larger neutrinospheres.
Here, for the 1.46-\mo model, these radii vary from $\sim$30 to $\sim$120\,km along the equatorial 
direction, with little departure from sphericity (30\% lower values are obtained in the polar direction).
In the right panel, line contours correspond to the neutrinosphere for the three different flavors at
a neutrino energy ($\varepsilon_{\nu}$) of 12.5\,MeV, associated with the peak of the energy distribution
at infinity.
Note that this is approximate, since the neutrino energy distribution hardens
with time and manifests a latitudinal dependence (see \S\ref{sect_nu}).  
As in standard 1D and 2D core-collapse computations, we find that the 
electron neutrinos decouple from matter
at larger radii than the $\bar{\nu}_e$ and ``$\nu_{\mu}$'' neutrinos. Here, the former decouple 
at 90\,km (60\,km) along the equator (pole), the latter two further in but at a similar radius of
70\,km (50\,km).
% the latter two nearly overlapping and situated inwards at 70\,km (50\,km). 
The neutrinospheres show a similar shape for all three
flavors (and all energy groups), reflecting the corresponding asphericity in the density field.

In the bottom-row panels, we reproduce the above for the last time in the 1.46-\mo 
simulation. The departure from sphericity is now considerable, with both an oblateness 
and a strong pinching of the neutrinospheres along the polar directions.
Along the equatorial direction, the radial spacing between neutrinospheres of consecutive 
and higher energy groups has increased, and the lower (higher) energy groups decouple further 
in (out) than in the previous snapshot, with neutrinospheres between 20 and 150\,km 
(from 2.5 to 50\,MeV), 80 and 50\,km (for the $\nu_e$ neutrino, and 
$\bar{\nu}_e$/``$\nu_{\mu}$'' neutrinos, respectively).  
Along the polar direction, neutrinospheres of all neutrino energy groups (and all neutrino
flavors at 12.5\,MeV) shown reside in a narrow range of radii between 20 and 30\,km (22 to 25\,km).
Here again, the neutrinospheres depicted follow very closely the contours of density, which is the primary
factor controlling the neutrino optical depth.

In Fig.~\ref{fig_nusphere_92},  we duplicate Fig.~\ref{fig_nusphere_46} (note the different spatial 
scale) for the 1.92-\mo model, showing the same 
quantities both early after bounce (59\,ms) and  
significantly later at 775.5\,ms after bounce. There are numerous differences with the 1.46-\mo model.
First, the neutrinospheres are aspherical even right after bounce (top row), with equatorial (polar) radii
larger (smaller) by a factor of 2-3 compared with those in the 1.46-\mo model. 
All flavors reveal similar neutrinosphere locations.

  In the 1.46-\mo model, the later ratio of the equatorial and polar radii is 2.5:1, irrespective of energy 
group and flavor. This becomes 15:1 in the 1.92-\mo model, and is thus a considerable departure from
sphericity; the faster rotating model has a neutrinosphere radius of just 14\,km along the pole, but
215\,km along the equator.
This is the most conspicuous difference with the essentially spherical neutron stars seen in non-rotating
simulations of the more standard core collapse of massive stars (Keil et al. 1996; 
Swesty \& Myra 2005; Buras et al. 2005a; Dessart et al. 2005), with neutrinosphere radii of the 
order of 20-30\,km at comparable times after core bounce.

   The very aspherical neutron stars formed through the AIC of a white dwarf make the determination
of the neutron star mass somewhat ambiguous. Rather than taking the enclosed mass within a given spherical
radius, we compute the total mass from all regions above a given mass density.
For a density cut of 10$^{10}$\,g\,cm$^{-3}$, we obtain neutron star masses of 1.42\,\mo 
for the 1.46-\mo model, and 1.5\,\mo for the 1.92-\mo model. 
However, if we adopt a density cut of 10$^{11}$\,g\,cm$^{-3}$, the neutron 
star masses are, respectively, 1.39\,\mo and 1.30\,\mo.
The higher-mass progenitor model has now a smaller neutron star mass, reflecting 
the strong asphericity of the density field.
While we might associate the neutrinosphere with the neutron star surface and with a standard 
mass density of 10$^{11}$\,g\,cm$^{-3}$, such an association
in the present fast rotating neutron star is inappropriate, since the neutrinospheres extend 
well into regions where the density is $\sles$10$^{10}$\,g\,cm$^{-3}$.
These neutron star mass values are reached at 100\,ms after core bounce and remain essentially constant.
The neutrino-driven wind that appears after a few 100\,ms decreases the neutron star mass at 
a rate of just a few 10$^{-3}$\,\mo\,s$^{-1}$ (see \S\ref{sect_ener}).
Note also that the enhanced centrifugal support in the faster-rotating, higher-mass model
leads to bounce at a 30\% lower maximum density compared with the 1.46-\mo model,
and both a reduced and a delayed mass accretion rate along the equatorial direction compared with 
what would prevail in the absence of rotation. At the end of each simulation, the total 
neutron star angular momentum is 
1.35$\times$10$^{49}$ erg$\cdot$s (1.13$\times$10$^{49}$ erg$\cdot$s) for the 1.46-\mo model, 
using the density cut at 10$^{10}$\,g\,cm$^{-3}$ (10$^{11}$\,g\,cm$^{-3}$) , and 
4.57$\times$10$^{49}$\,erg$\cdot$s (2.79$\times$10$^{49}$\,erg$\cdot$s) for the 1.92-\mo model.
However, the accretion rate at a given Eulerian radius is higher and longer-lived
at smaller latitudes, because of the larger amount of mass available and the flatter 
density profile in those regions.
Overall, the presence of a massive accretion disk in the fast rotating model complicates
the definition of the neutron star mass at such early times. Evolution over minutes/hours/days 
will likely lead to significant accretion onto the neutron star, resulting in a much higher 
final mass.

The final rotational to gravitational energy ratio $T/|W|$ is 0.059 for the 1.46-\mo
model and 0.262 for the 1.92-\mo model. These values are large and for the latter model, large enough
to cause the growth of secular and perhaps even dynamical instabilities (Tassoul 2000).
Using realistic post-bounce configurations for a rotating massive star progenitor,
Ott et al. (2005a) find a dynamically unstable spiral mode for $T/|W|$ as low as $\sim$0.08.
Thus, it is likely that the PNS structures found here, especially for the 1.92-\mo model, would
develop some non-axisymmetric instability that would cause, among other things, outward angular momentum transport.

These results are significantly different from those of Fryer et al. (1999), who obtained 
a successful explosion $\sles$100\,ms after bounce, a PNS mass of $\sim$1.2\,\mo, and a $\sim$1\,s 
period at $\sim$200\,ms after core bounce.  Such different conclusions stem 
from their adoption of slow, solid-body progenitor rotation, with most of 
the angular momentum stored in the outer mantle and blown away by the explosion,
rather than being accreted by the PNS.

\section{Neutrino signatures}
\label{sect_nu}
%\clearpage
\begin{figure*}
% \plottwo{aic_m1pt46_flux_r250km.ps}{aic_m1pt92_flux_r400km.ps}
\plottwo{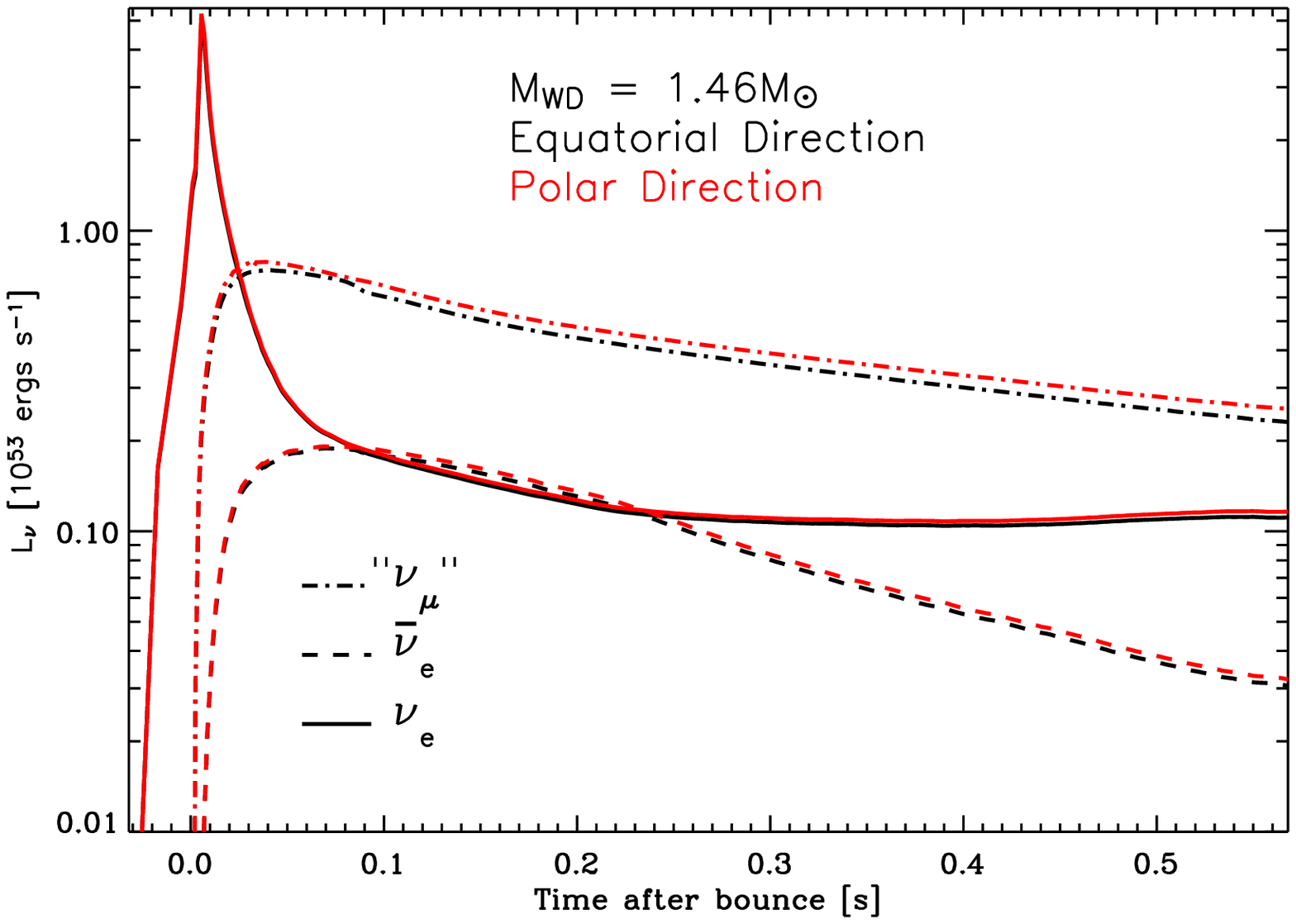}{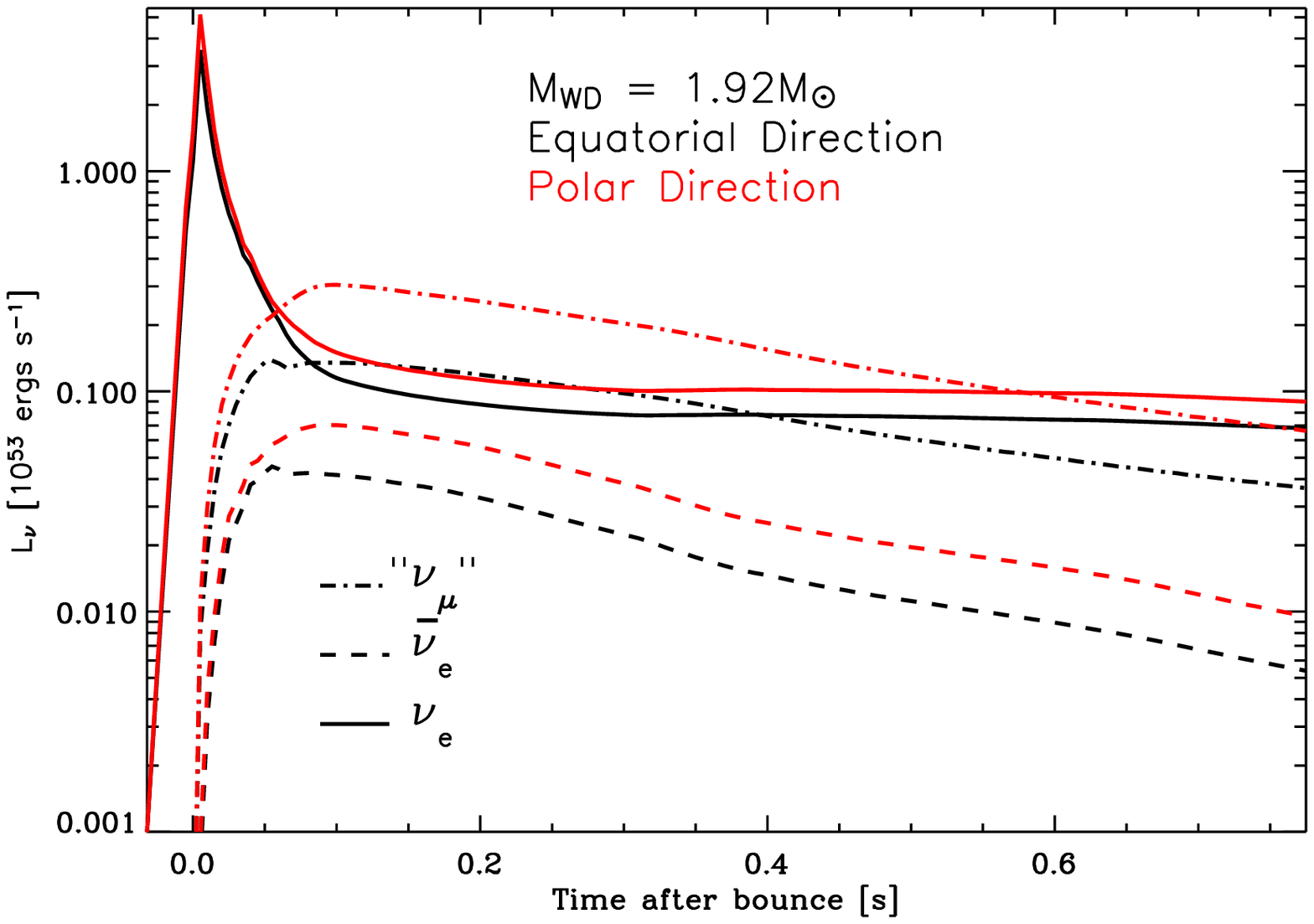}
\caption{
Flavor- and angle-dependent neutrino luminosities for the 1.46-\mo (left) and 1.92-\mo (right) 
progenitors, for an adopted radius of $R=250$\,km for the former and $R=400$\,km for the latter.
Luminosities shown as solid, dash-dotted, and dashed lines correspond to 
$\int_{d\Omega} d\Omega R^2 F_{\nu}(R,\theta,\varepsilon_{\nu})$ for the
$\nu_e$, ${\bar{\nu}_e}$, and ``$\nu_{\mu}$'' neutrinos, repectively, and in black for the equatorial
direction ($\theta=\pi/2$) and in red for the polar direction ($\theta=0$).
Of particular relevance to the asymptotic electron fraction of the ejecta is the
higher electron-neutrino luminosity compared to that of the anti-electron neutrinos,
by a factor of at least ten in the 1.92-\mo model (right).
Note that the vertical scale extends down to 10$^{51}$\,erg\,s$^{-1}$ for the 1.46-\mo model
and to 10$^{50}$\,erg\,s$^{-1}$ for the 1.92-\mo model.
}
\label{fig_nuflux}
\end{figure*}

\begin{figure*}
% \plottwo{aic_m1pt92_T_t=0.5700s_with_vel_Rmax30km.ps}{aic_m1pt92_T_t=0.5700s_Rmax250km.ps}
\plottwo{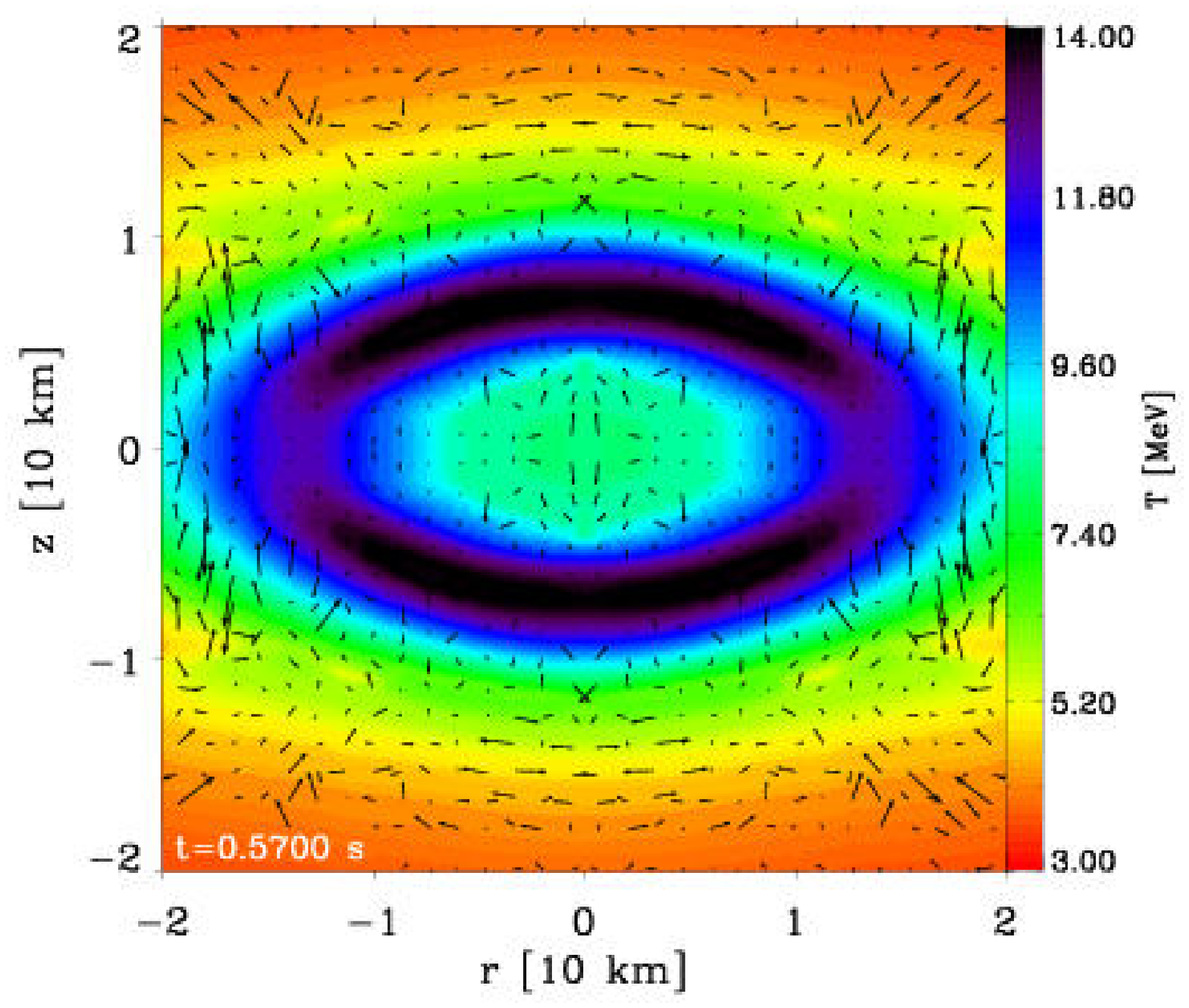}{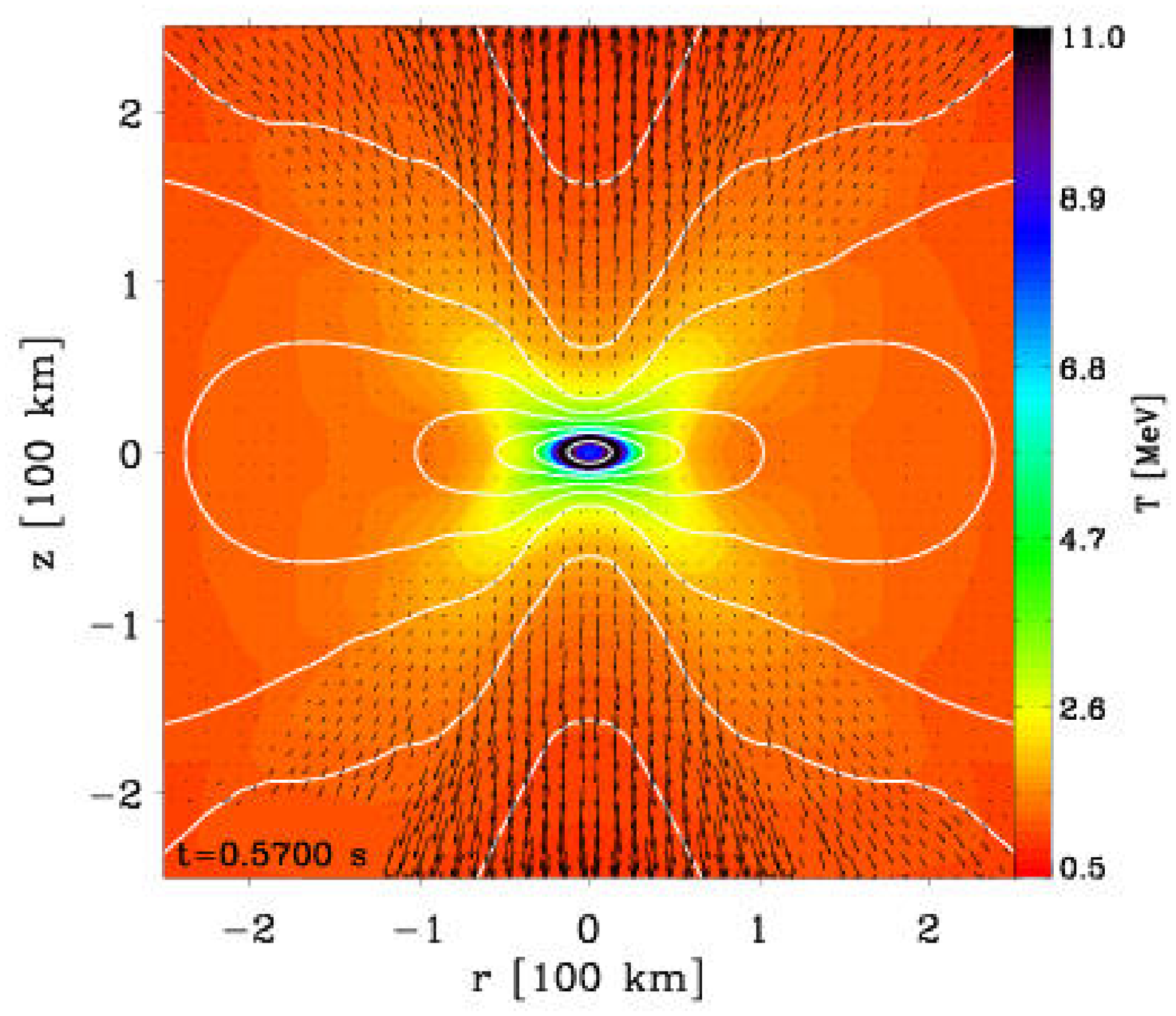}
\caption{
Color map of the matter temperature for the 1.92-\mo model at the last snapshot. We 
overplot the contours of the density every decade, starting with 10$^{14}$\,g\,cm$^{-3}$ 
for the innermost line.
{\it Left}: View of the innermost region ($\sim$20\,km), showing the latitudinal 
variation of the temperature even on the innermost density contour. The velocity field 
reveals the presence of fluid motions along cylinders due to rotation (see \S\ref{sect_disk}), 
thereby preventing convection.
A ten-fold magnification of the velocity-vector length, saturated at a value of 
9130\,\kms, has been applied to better render the fluid motions.
{\it Right}: Zoom out of the left slide, showing the large temperature variation
along the 10$^{10-11}$\,g\,cm$^{-3}$ density contour, which corresponds reasonably well 
to the neutrinosphere location at the peak of the neutrino energy distribution 
(see Fig.~\ref{fig_nusphere_92}). 
The vector of maximum length corresponds to a velocity magnitude of 11300\,\kms. 
}
\label{fig_temp}
\end{figure*}

\begin{figure*}
%\epsscale{1.3}
\plottwo{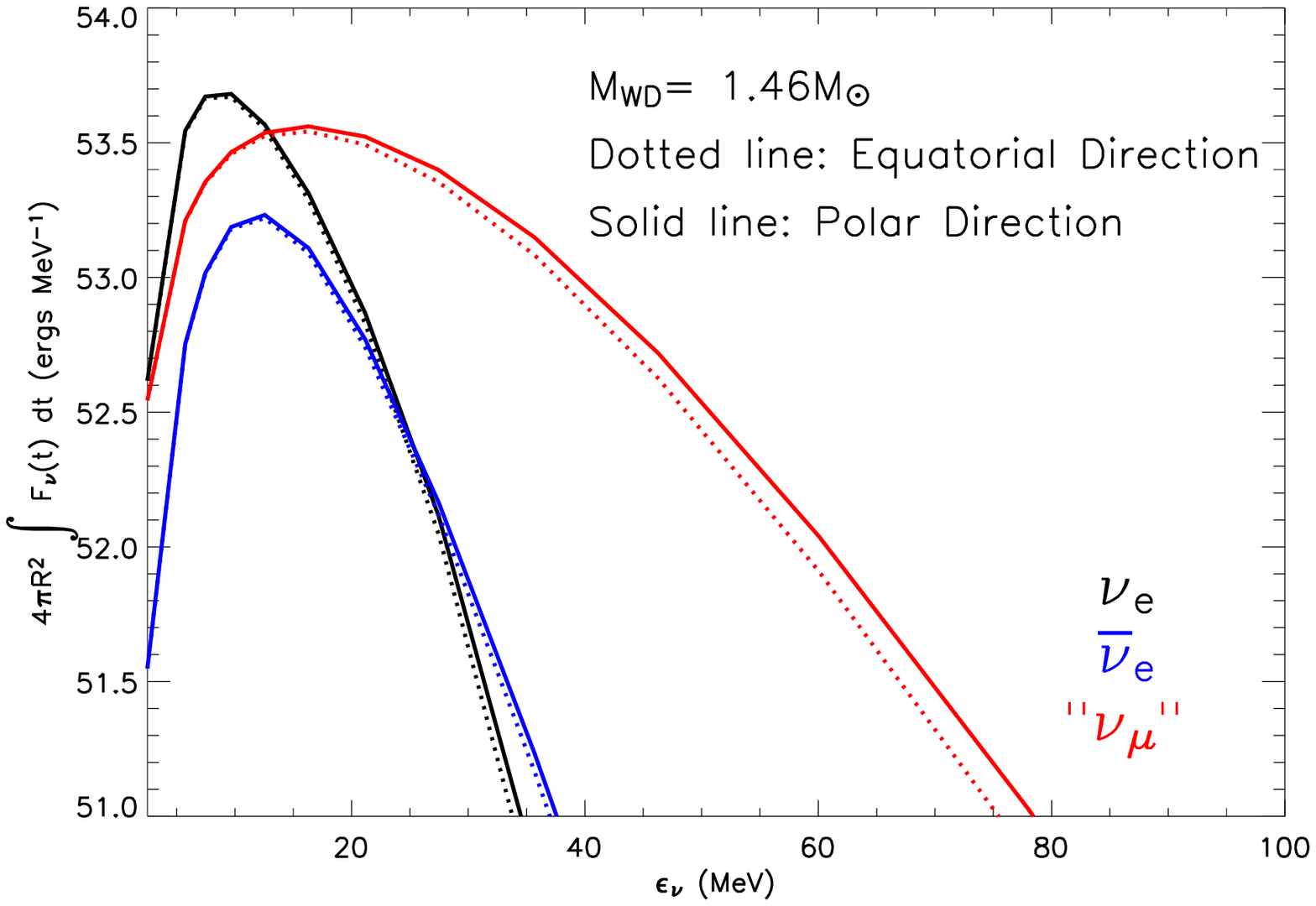}{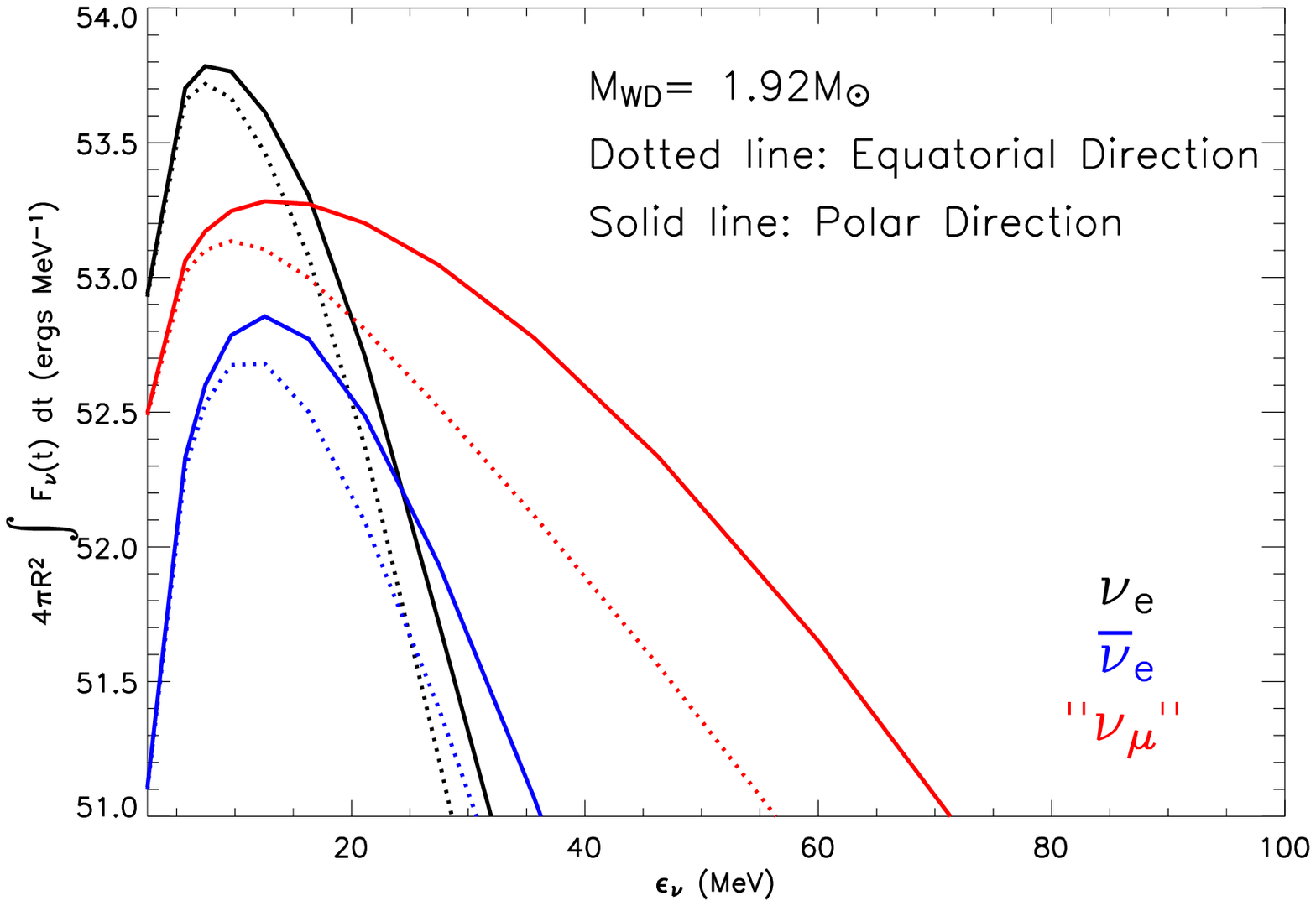}
\plottwo{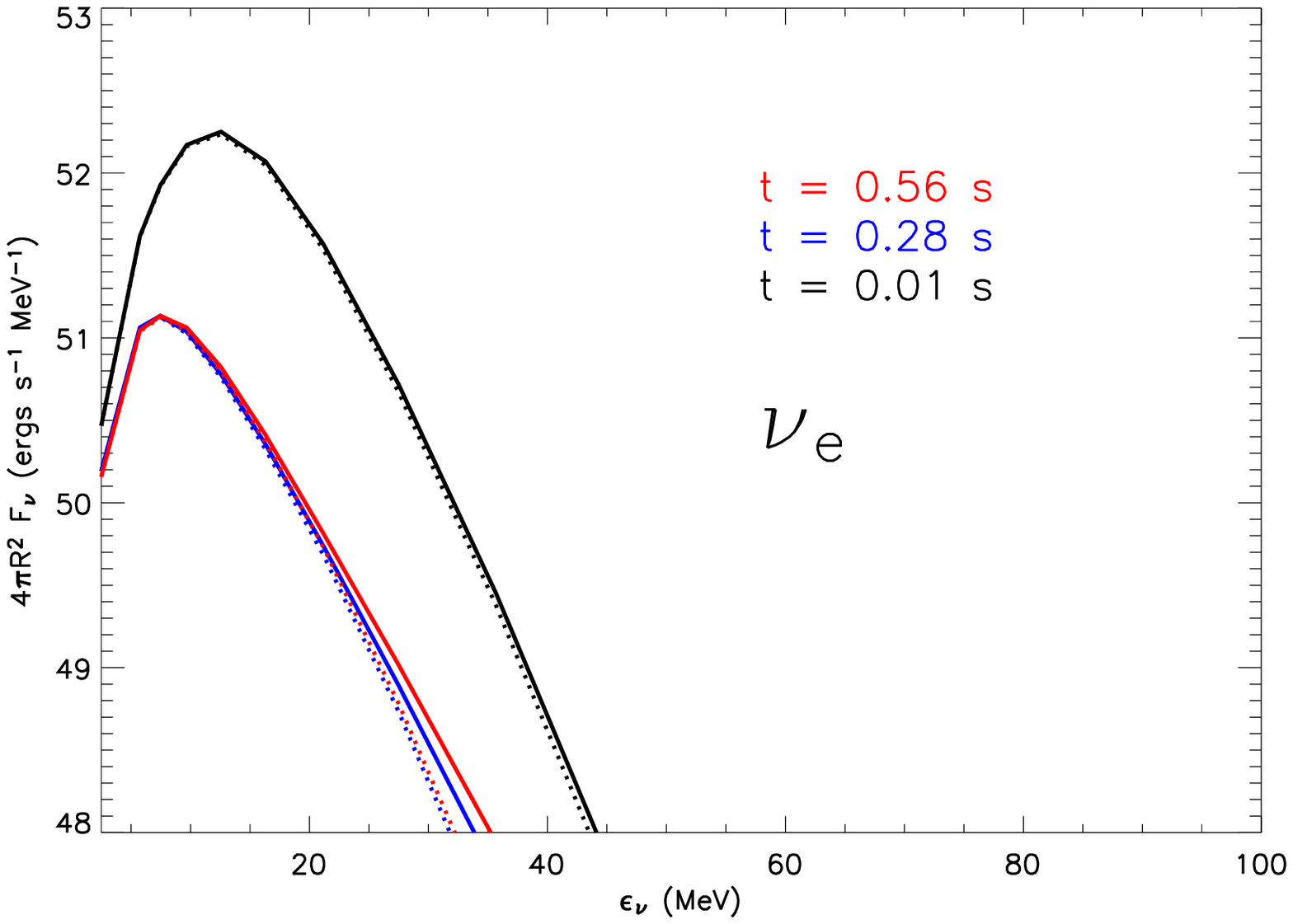}{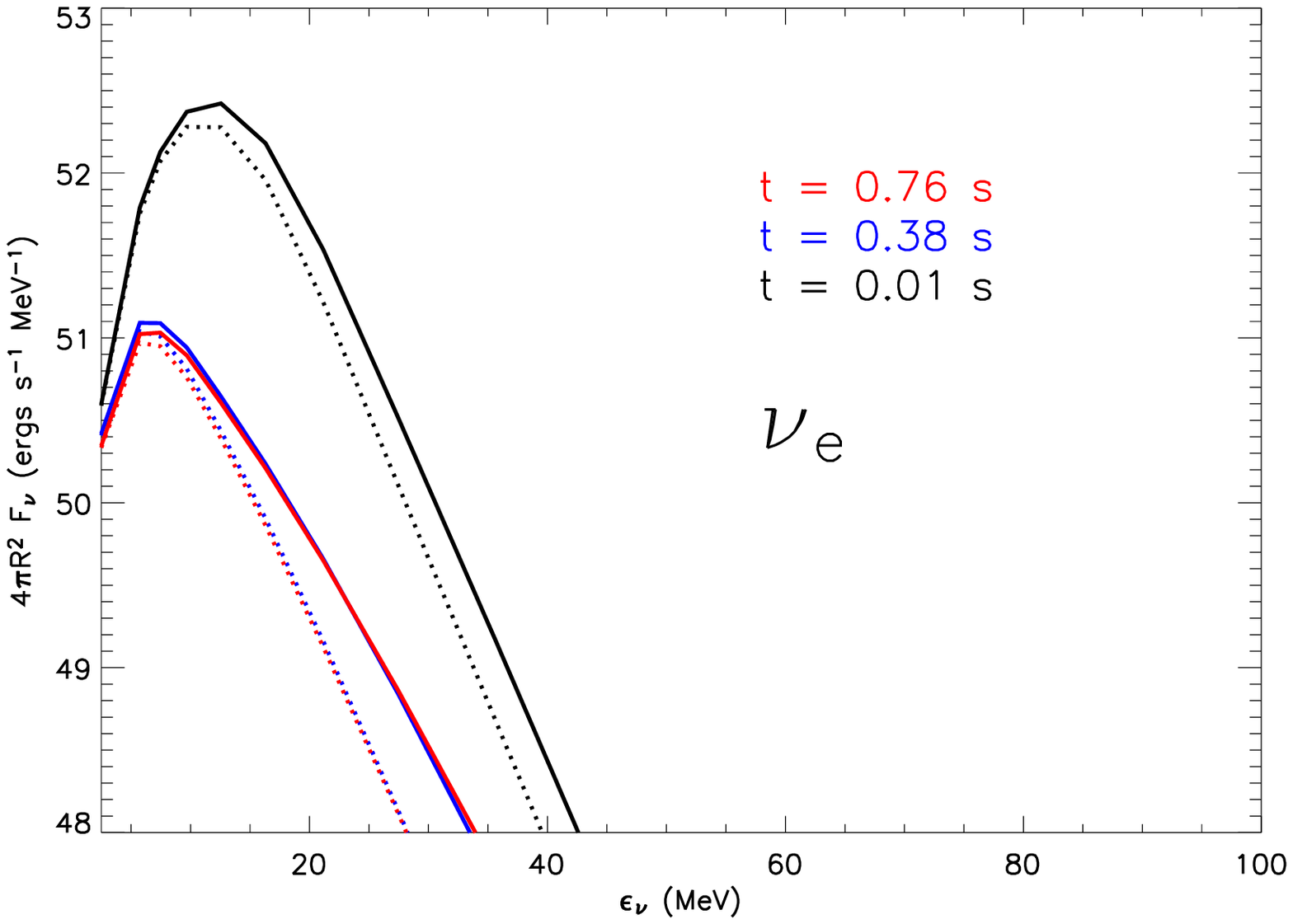}
\plottwo{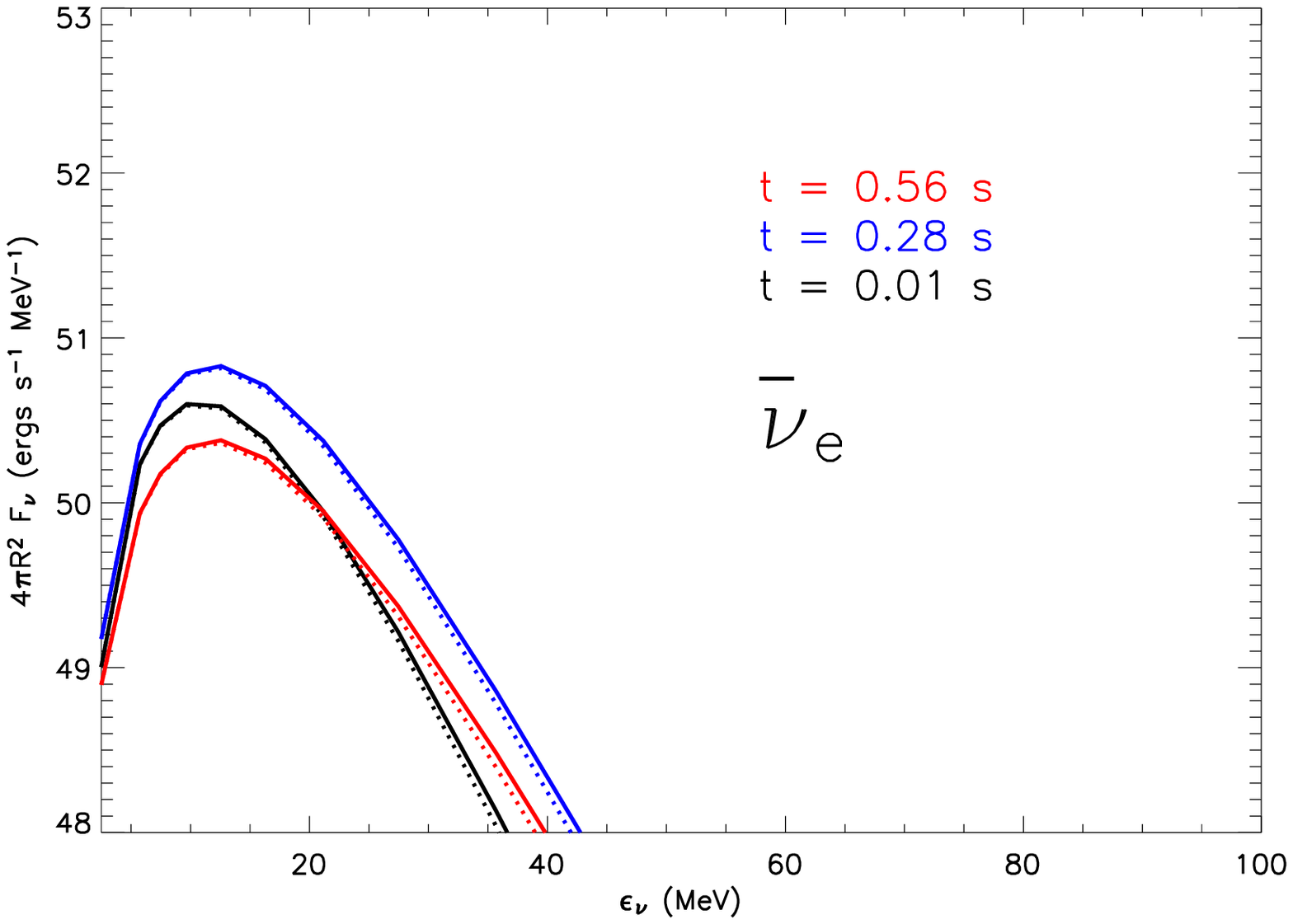}{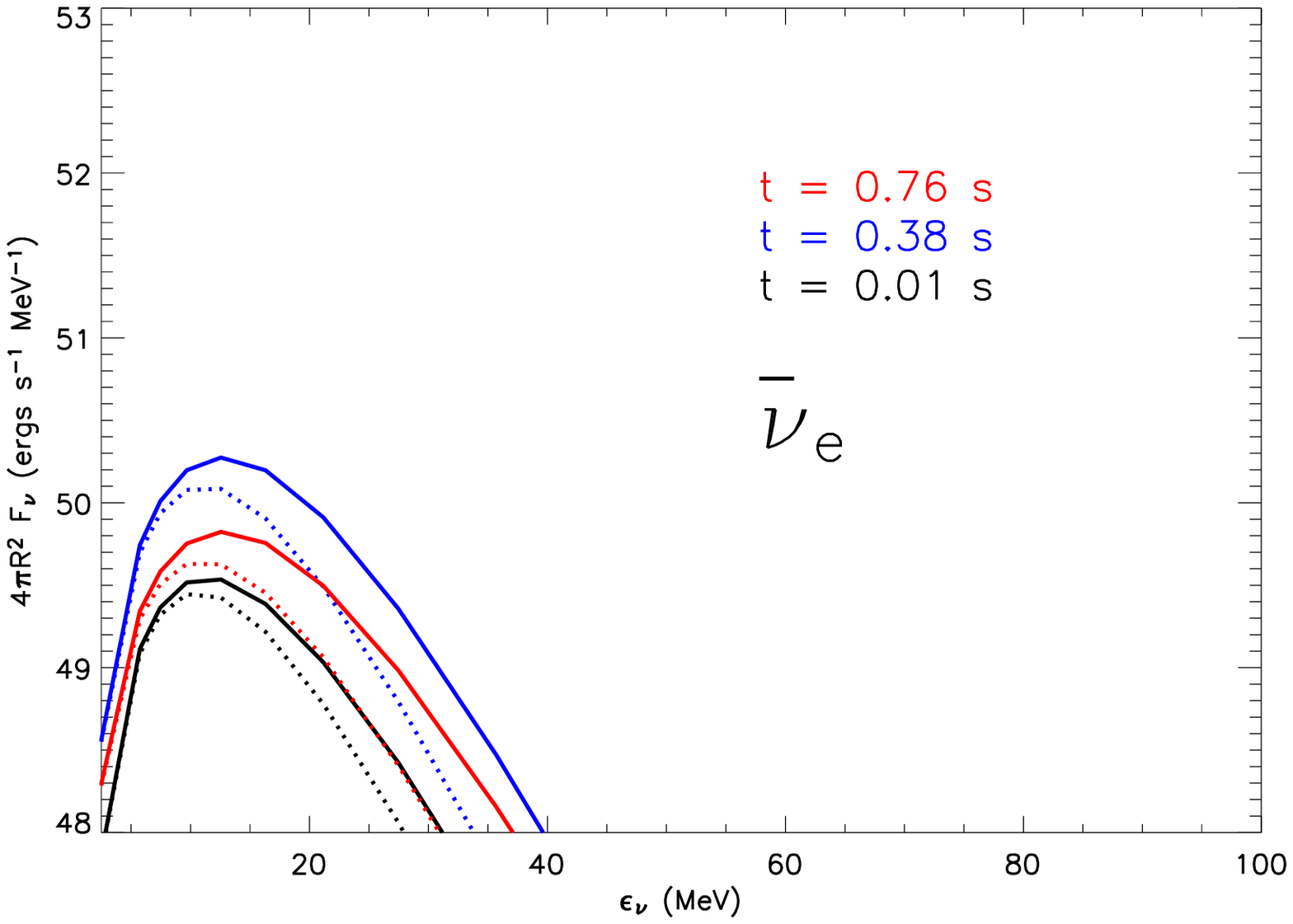}
\plottwo{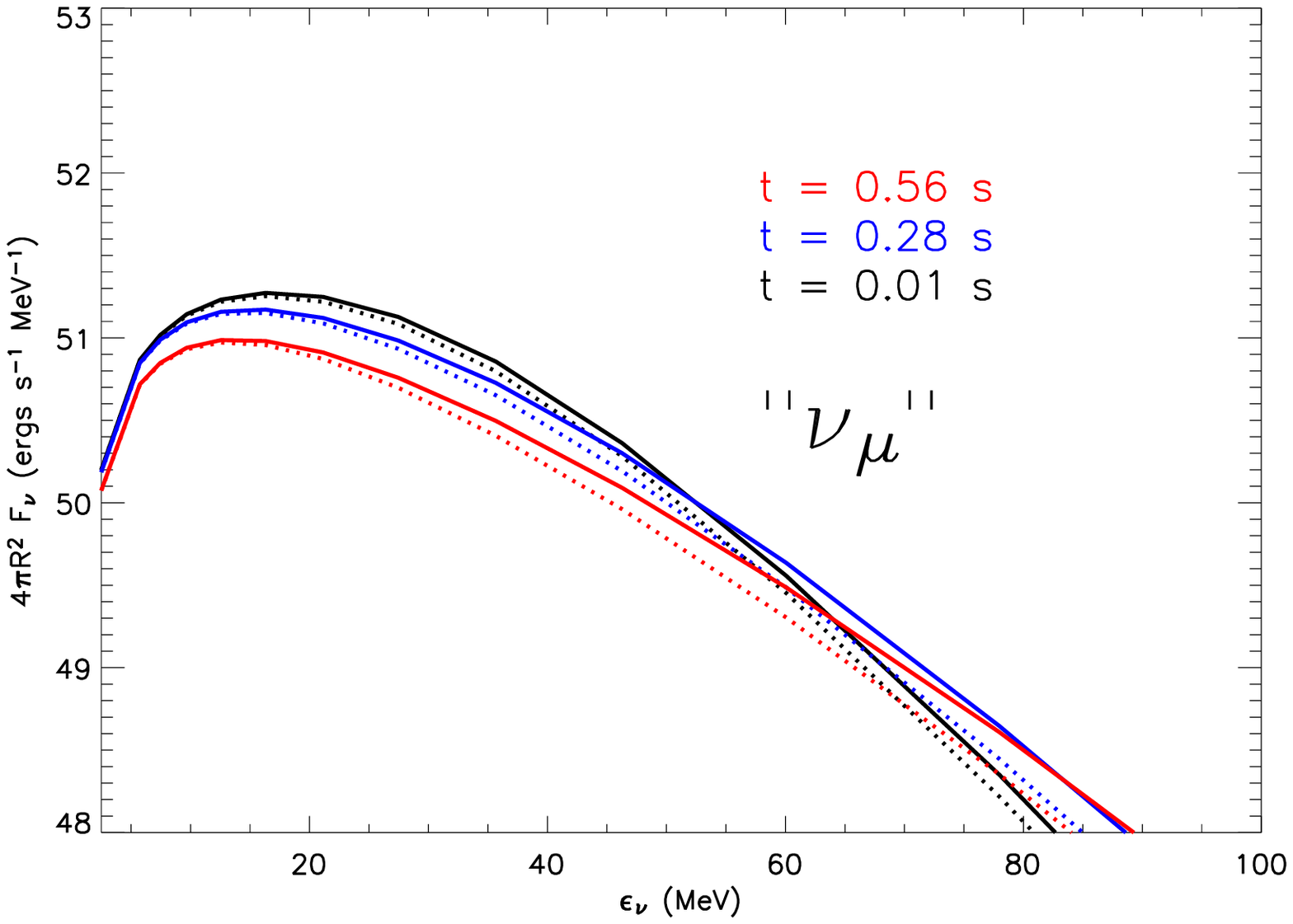}{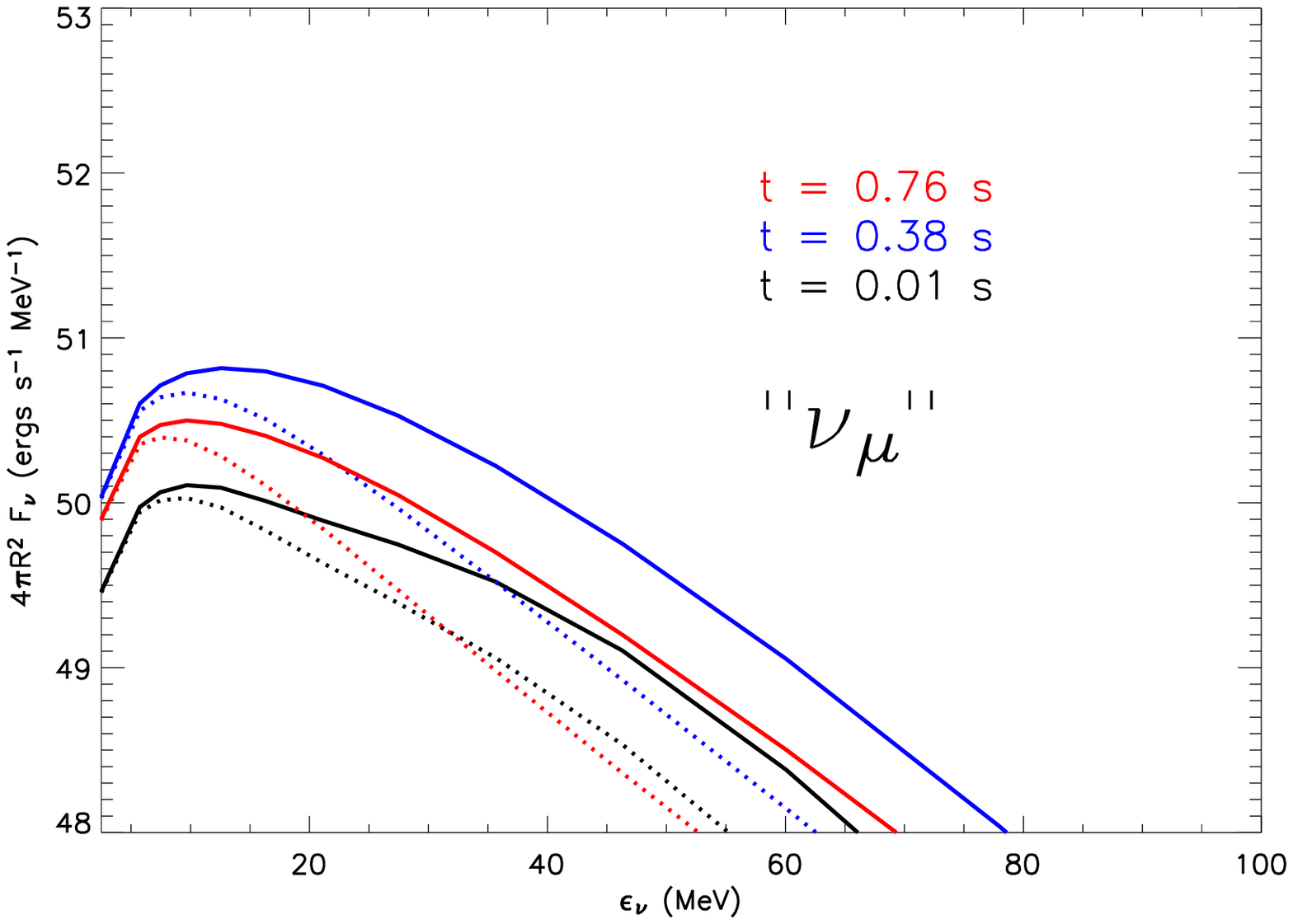}
\caption{
{\it Top Left}: Time integral, until 570\,ms after bounce, of the neutrino luminosity,
set equal to $\int_\Omega R^2 F_{\nu}(R,\theta) d\Omega$ ($\theta$ is the polar angle,
$R$ is spherical radius, and $\Omega$ is the solid angle; see \S\ref{sect_nu}),
at $R=250$\,km, for each neutrino flavor (black: $\nu_e$; blue: ${\bar{\nu}_e}$; 
red: ``$\nu_{\mu}$''), and along the polar (solid line) and equatorial (dotted line)
directions.
{\it Bottom left panels}: Neutrino energy distribution, for each neutrino flavor, 
at 10\,ms (black), 280\,ms (blue), and 560\,ms (red) post-bounce times, for both 
equatorial (dotted) and polar (solid) directions.
{\it Right column}: Same as left column, but this time for the 1.92-\mo model,
using a radius $R=600$\,km.
In the bottom three panels, the times selected are now 10\,ms (black), 380\,ms (blue), 
and 760\,ms (red) after bounce. Note the much larger contrast between the equatorial
and polar fluxes and energy distributions.   
}
\label{fig_nufluxspec}
\end{figure*}
%\clearpage
The first observational signature of an AIC explosion would be the copious emission of 
neutrinos immediately after core bounce.
As discussed above, the properties of the bounce of the core and the nascent neutron star
are close enough to those obtained in simulations of the core collapse of massive progenitors 
that one expects a neutrino signal with a somewhat similar evolution and character (see, for example,
the predictions for the 11-\mo model of Woosley \& Weaver 1995 
in Dessart et al. 2005).

In Fig.~\ref{fig_nuflux}, we show the neutrino luminosities (on a log scale) for the 
1.46-\mo (left panel) and 1.92-\mo (right panel) models, with distinct curves for the 
different neutrino flavors (solid line: $\nu_e$; dashed line: ${\bar{\nu}_e}$; 
dash-dotted line: ``$\nu_{\mu}$''), as well as different colors for the 
equatorial (black) and polar (red) directions.
These are luminosities in the sense that the flux in each direction is scaled by 4$\pi R^2$
where $R$ is a spherical radius, of 250\,km for the 1.46-\mo model and 400\,km for the 
1.92-\mo model (chosen to be well above the neutrinosphere for the energy at the peak
of the neutrino distribution). 
They correspond to the total luminosity that would have obtained had
the selected directional flux been the same in all directions.
% We also draw the sum of the three neutrino, angle-integrated, luminosity (black curve), i.e., 
% $\sum_{\nu} \int_\Omega R^2 F_{\nu}(R,\theta) d\Omega$, where $\theta$ is the polar angle.
The temporal evolution of the various fluxes for both models is comparable.
The total neutrino luminosity reaches a maximum of 5.2$\times$10$^{53}$\,erg\,s$^{-1}$, 
mostly due to the $\nu_e$ neutrino contribution, and decreases to $\sim$4$\times$10$^{52}$\,erg\,s$^{-1}$ 
at 500\,ms after bounce in the 1.46-\mo model, with a further 30\% decrease for the 1.92-\mo model.
At later times, the main reason for this difference is the much lower ``$\nu_{\mu}$'' 
neutrino luminosity in the 1.92-\mo model.
This reduction has been seen and discussed by Fryer \& Heger (2000) in the context 
of the collapse of rotating cores of massive progenitors. The smaller core densities 
(weaker bounce) achieved in models with fast rotation lead to smaller temperatures 
and, consequently, smaller neutrino emission, with a larger effect for the 
$\nu_{\mu}$ and $\nu_{\tau}$ neutrinos (grouped under the name ``$\nu_{\mu}$'' here).
So, while the globally lower neutron star densities in the fast-rotating model induce a reduction 
in neutrino luminosity compared to the 1.46-\mo model, the same effect introduces a 
latitudinal variation of neutrino fluxes in the faster rotating model, with fluxes, 
irrespective of neutrino flavor, larger by a factor of about two along the pole than 
along the equator (note that at a radius of 250\,km, the difference is higher and on
the order of three).
This variation, not discussed by Fryer \& Heger (2000), results from the further 
variation of the neutrinosphere temperatures with latitude within a given rotating model.
In both models, but more so in the 1.92-\mo model, the temperature gradient and the 
temperatures are reduced at the neutrinosphere (for a given energy group) 
along the equator compared to the poles, 
irrespective of energy group and flavor, as is clearly visible in Fig.~\ref{fig_temp}.
For example, we see that the temperature on the 10$^{10}$\,g\,cm$^{-3}$ contour
is 4\,MeV in the polar direction and 0.5\,MeV in the equatorial direction.
Within the neutrinosphere region, we find fluid velocities that are oriented 
preferentially in the $z$-direction, along cylinders, illustrating the weak or absent 
convection that results from the stabilizing specific angular 
momentum profile (Fryer \& Heger 2000; Heger et al. 2000; Ott et al. 2005b; 
see also \S\ref{sect_disk}).
Along a given angular slice, the temperature has a maximum at mid-latitudes, caused 
by enhanced ($\nu_e$) neutrino energy deposition in this direction.
These regions offer a tradeoff, since the flux is still higher at such latitudes than 
along the equator, while the density is relatively higher than along the poles 
(see \S\ref{sect_ener}). 
The latitudinal variations seen in the collapsed models of AIC 
progenitors are extreme, and, indeed, for the slower rotation rates typically obtained for
massive-star core-collapse progenitors (Heger et al. 2000), a modest anisotropy is found 
instead (Walder et al. 2005). 

We document further the neutrino signatures of the AIC of white dwarfs by showing in 
the top row of Fig.~\ref{fig_nufluxspec}, for the 1.46-\mo (left) and 1.92-\mo (right) 
models, the time-integrated neutrino emission at infinity 
for each flavor, as a function of neutrino energy. In the bottom three panels of each column, 
we show the individual neutrino distributions at three representative times of the simulations 
(at bounce, halfway through the simulation, and at the last simulated time).
The overall flux level and hardness of the energy distribution are higher along 
the polar direction, the variation at a given time towards higher latitude mimicking the time 
evolution seen for non-rotating core collapse simulations of massive star progenitors.
We compute the average neutrino energies, here defined as
$$ \sqrt{\langle\varepsilon_{\nu}^2\rangle} \equiv 
\left[ 
    \frac{\int d\varepsilon_{\nu} \varepsilon_{\nu}^2 F_{\nu}(\varepsilon_{\nu},R)} 
         {\int d\varepsilon_{\nu} F_{\nu}(\varepsilon_{\nu},R)} 
\right]^{\frac{1}{2}}\,.$$ 
For the 1.46-\mo model ($R=250$\,km), we obtain similar values to within 
2-3\% along the pole and along the equator with $\langle\varepsilon_{\nu_e}\rangle=10$\,MeV,
$\langle\varepsilon_{{\bar\nu}_e}\rangle=15$\,MeV, 
and $\langle\varepsilon_{\nu_{\mu}}\rangle=24$\,MeV.
For the 1.92-\mo model ($R=600$\,km), we obtain systematically lower values than for the 1.46-\mo 
model, and for lower latitudes. Along the equator (pole), we obtain 
$\langle\varepsilon_{\nu_e}\rangle=9$\,MeV (10\,MeV), 
$\langle\varepsilon_{{\bar\nu}_e}\rangle=14$\,MeV (16\,MeV), and 
$\langle\varepsilon_{\nu_{\mu}}\rangle=16$\,MeV (21\,MeV).
Note that this definition of the neutrino ``average'' energy is similar to that found in 
Thompson et al. (2003), who used the mean intensity $J_{\nu}$ in place of the flux 
$F_{\nu}$. Since we are 
close to free-streaming regimes at the adopted radii for the peak of the neutrino energy 
distribution, the two are equivalent.

In the 1.46-\mo model, and at late times, we see that the neutrino 
display is more dramatic (total flux is twice as high) and
is characterised by a harder spectrum than in the 1.92-\mo 
model. This is due to the more compact and, thus, hotter neutrinospheres of the neutron star 
formed by the lower-mass white dwarf progenitor.

\section{Rotation and the remnant disk}
\label{sect_disk}
%\clearpage
\begin{figure*}
% \plottwo{w_radslice_m1.46.ps}{w_radslice_m1.92.ps}
% \plottwo{j_radslice_m1.46.ps}{j_radslice_m1.92.ps}
\plottwo{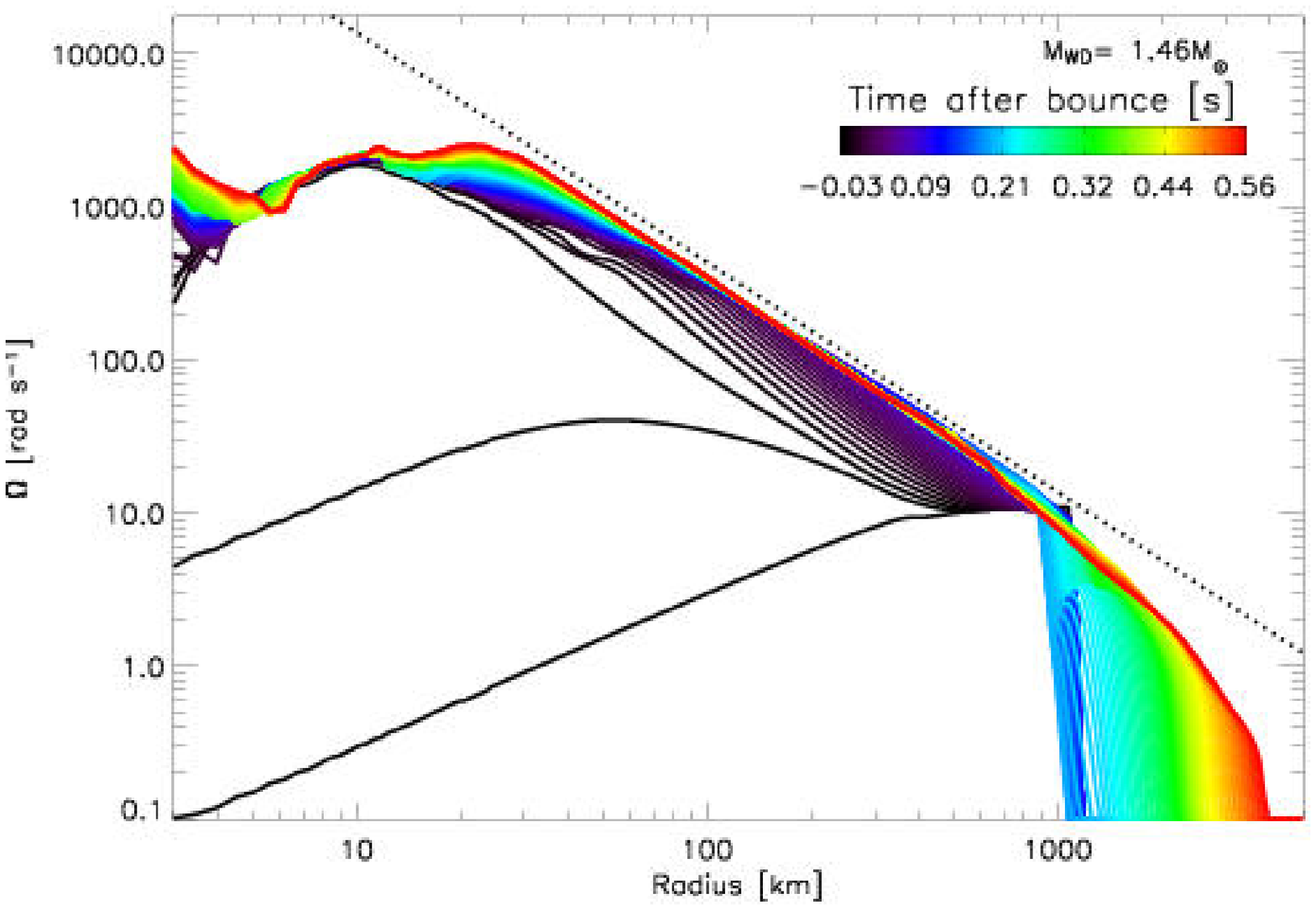}{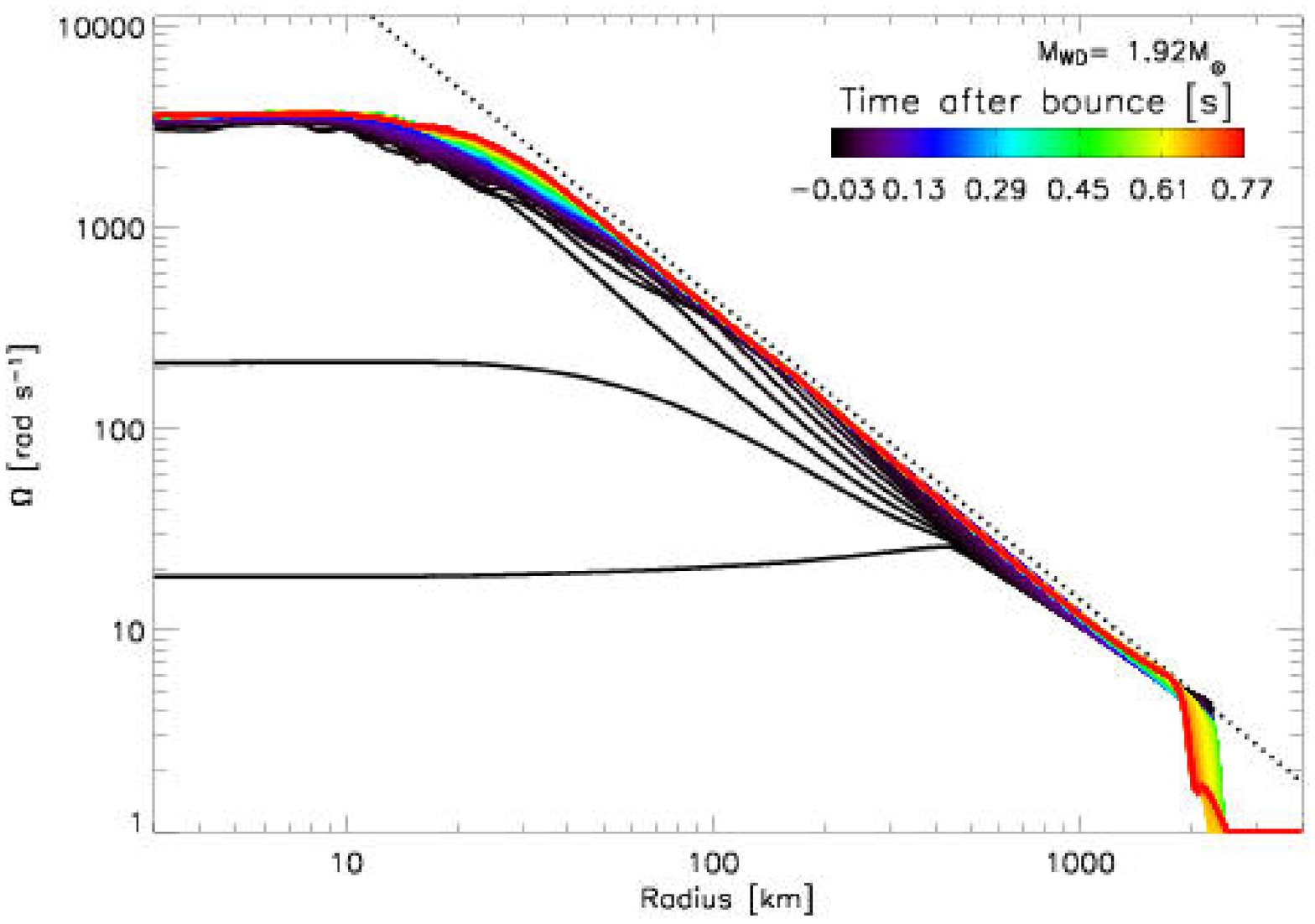}
\plottwo{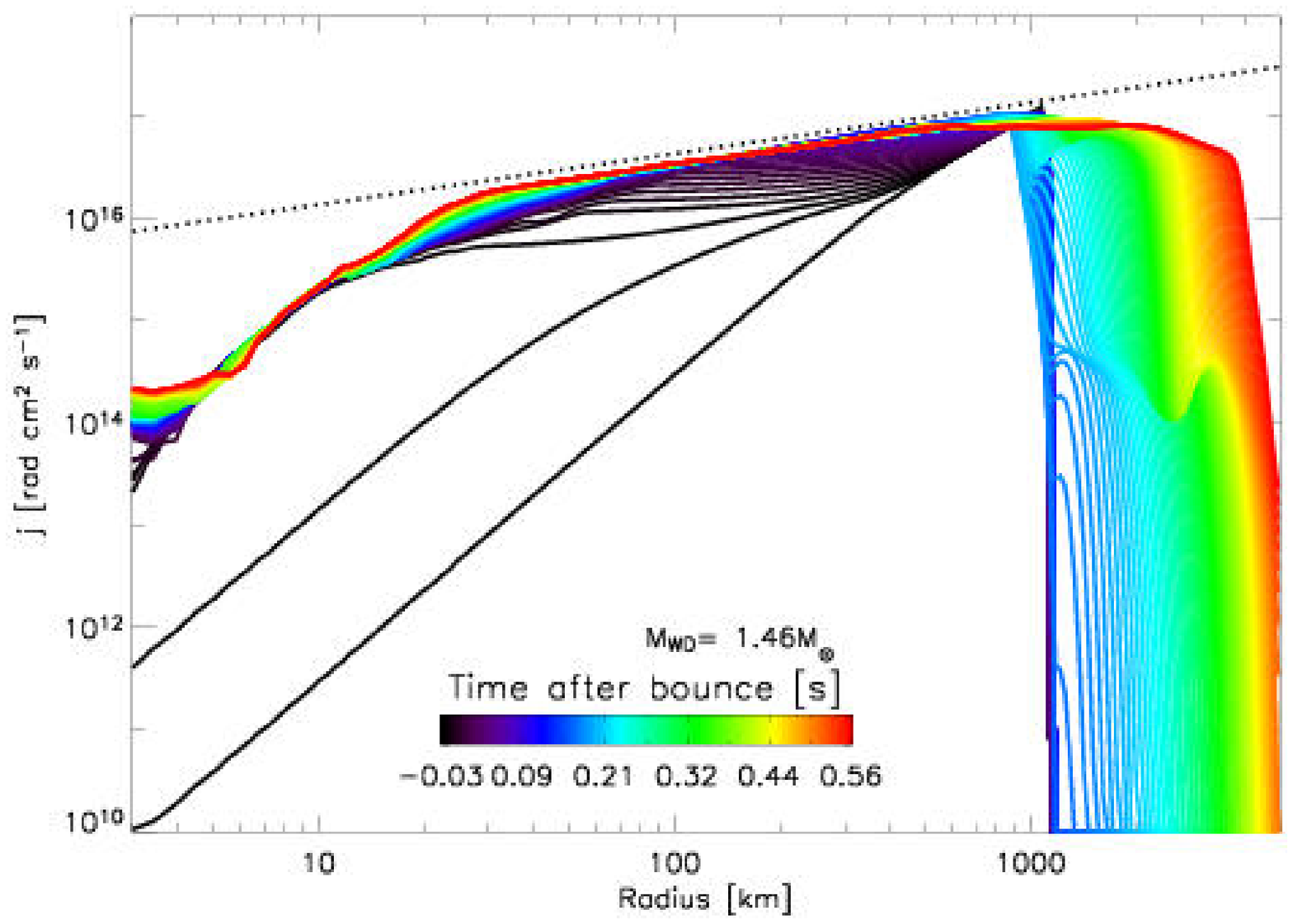}{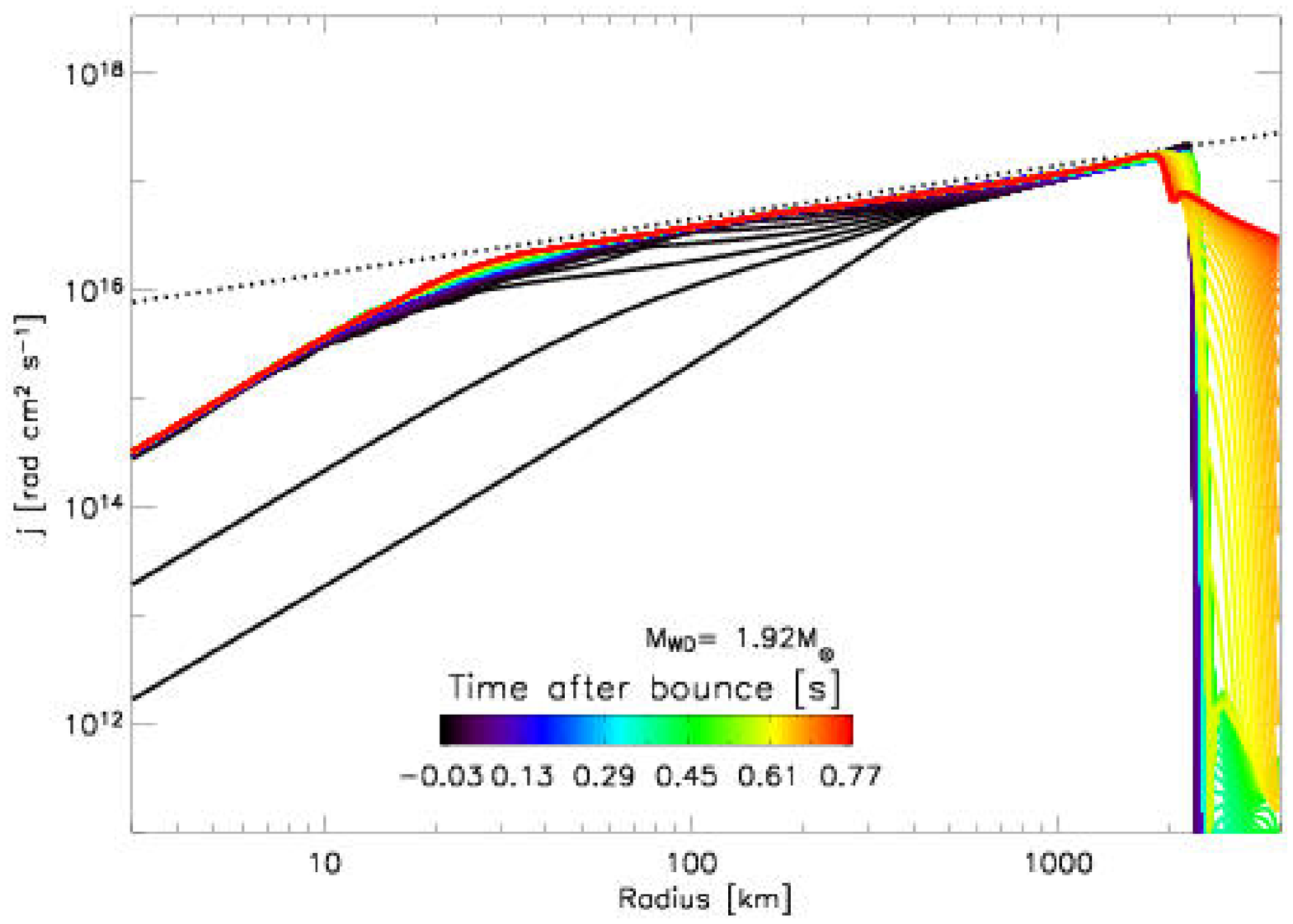}
\caption{
{\it Top}: Time evolution of the equatorial angular velocity profile for the 1.46-\mo
(left) and 1.92-\mo (right) progenitor models.
{\it Bottom}: Same as above, but this time for the specific angular momentum $j$.
Note its outward increase, which acts as a stabilizing influence on convective 
instabilities (Heger et al. 2000), as well as the rise, from zero, to larger
values after $\sim$200\,ms as the neutrino-driven wind wraps around the white dwarf
and brings rotating material to low latitudes.
For all panels, we overplot as a dotted black line the corresponding Keplerian value,
i.e., $\Omega_{\rm Keplerian}(r) = \sqrt{GM/r^3}$.  See text for details.
}
\label{fig_w_j}
\end{figure*}

\begin{figure*}
%\plotone{colorbarw.ps}
\plotone{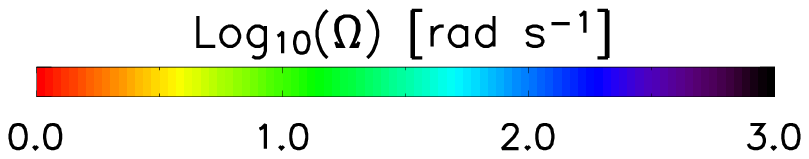}
%\vspace{-0.8cm}
% \plottwo{aic_m1pt46_w_t=0.5700s.ps}{aic_m1pt92_w_t=0.7755s.ps}
\plottwo{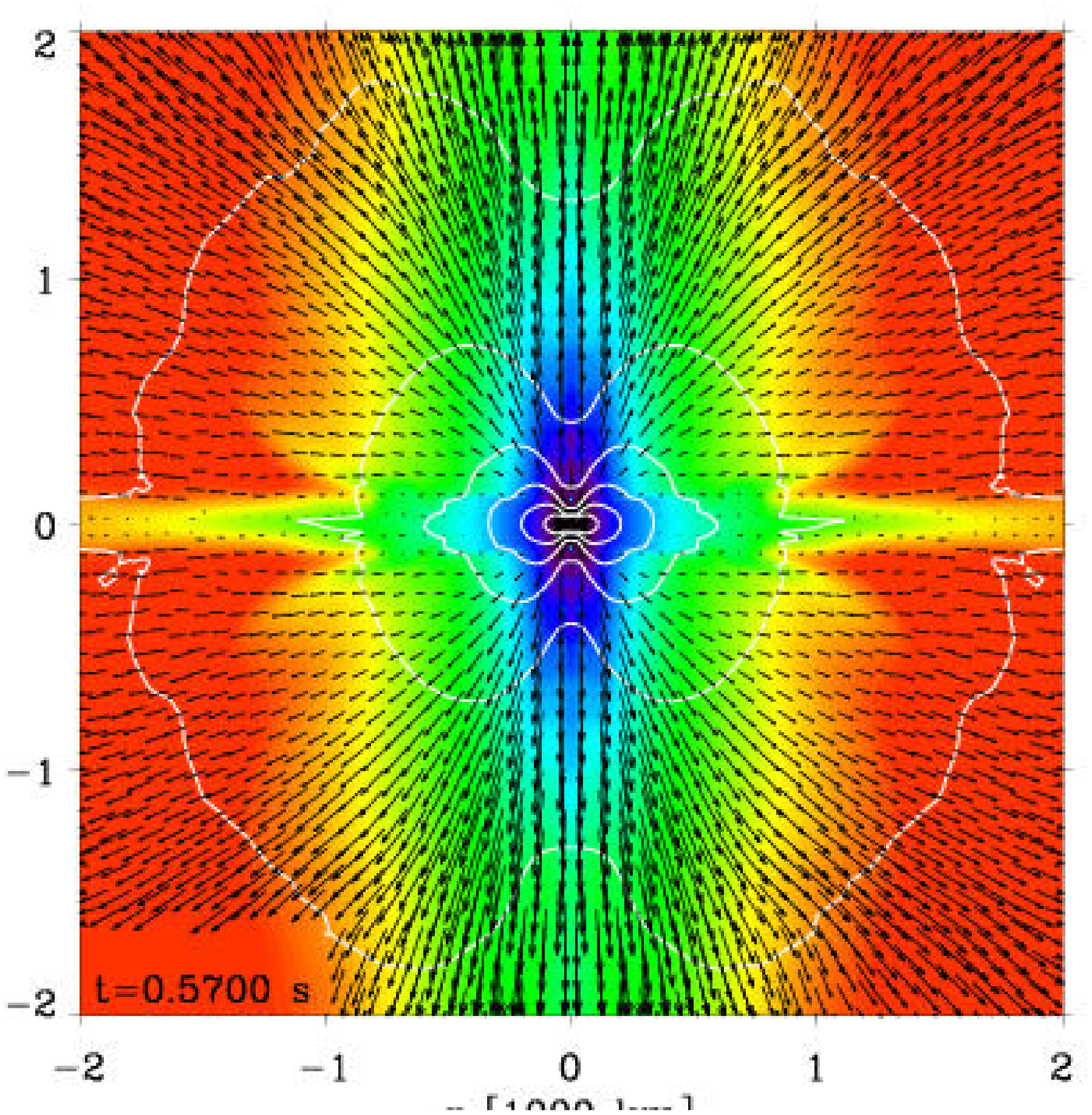}{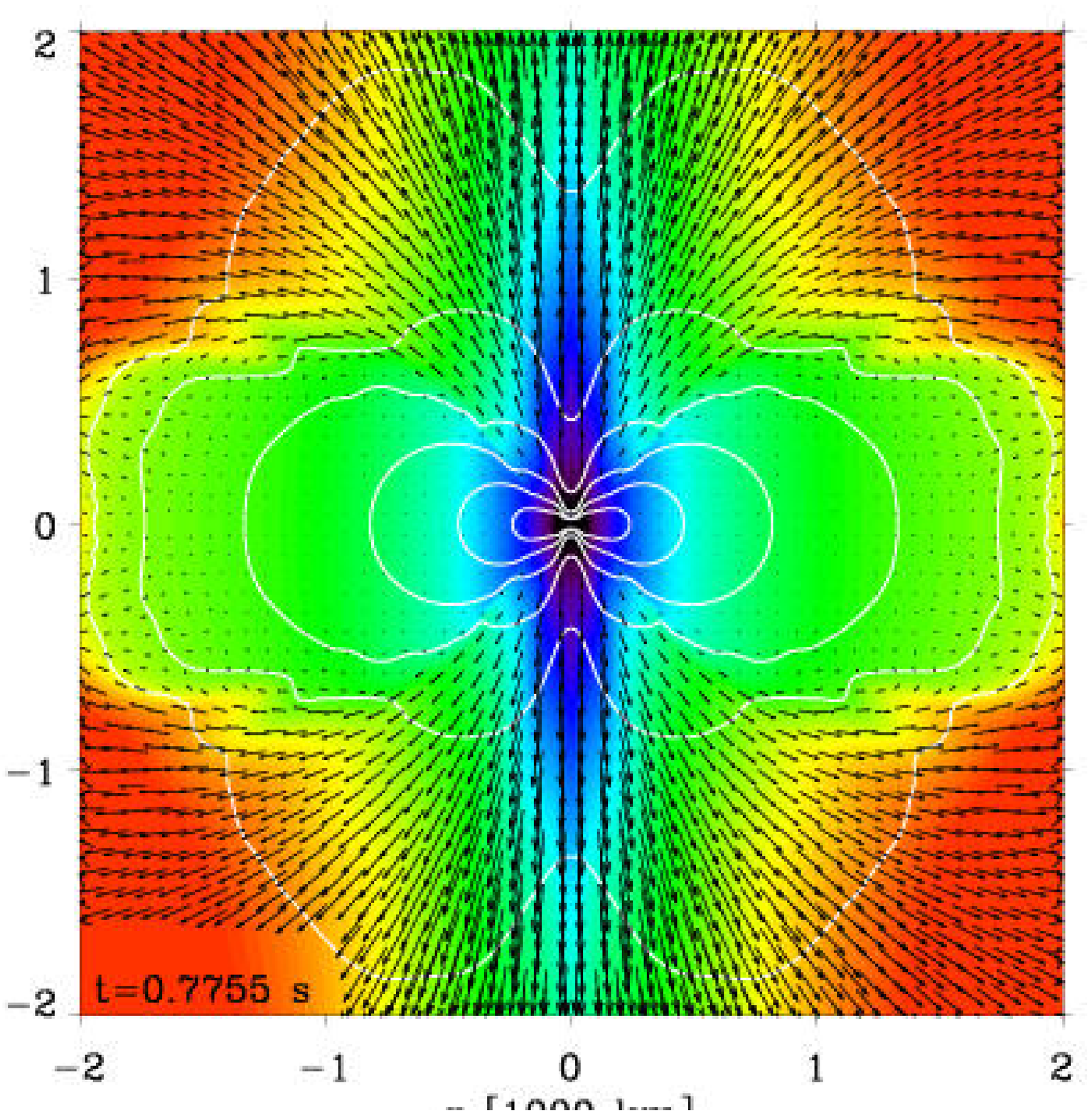}
\caption{
Color map of the logarithm of the angular velocity, at the last simulated time (570\,ms
for the 1.46-\mo model, and 775\,ms for the 1.92-\mo model), out to a 
maximum radius of 2000\,km, for the 1.46-\mo (left) and 1.92-\mo (right) progenitors. 
The initial angular velocity profiles $\Omega(r,z)$, functions of the cylindrical 
radial coordinate only (YL05), are given for the equatorial direction in Fig.~\ref{fig_w_j} 
(bottom curve). To enhance the visibility, we have reduced the colorbar range 
(see Fig.~\ref{fig_w_j} for the unsaturated values).We also overplot in white the 
corresponding contours of the density for every decade between 10$^{4}$ and 
10$^{10}$\,g\,cm$^{-3}$. The vector of maximum length corresponds to a velocity 
magnitude of 29000\kms (outflow) in both panels. The spatial scale is 4000\,km 
in width.}
\label{fig_omega2d}
\end{figure*}
%\clearpage
   Let us now turn our discussion to the angular momentum and angular velocity budget and 
profiles in our simulations. As discussed in \S\S\ref{sect_results}--\ref{sect_pns}, 
we find that the early neutron stars have comparable masses in the two simulations, 
the rest residing not so much in the outflow than in a substantial amount of 
``circum-neutron star'' disk material, rotating fast, but having little outflow or 
inflow velocity (see Fig.~\ref{fig_vr_radslice}).

   In Fig.~\ref{fig_w_j}, we plot a temporal sequence of the equatorial radial profile of the
angular velocity (top row) and specific angular momentum (bottom row), for the 1.46-\mo
model (left column) and 1.92-\mo model (right column).
In the 1.46-\mo model, the central angular velocity is $\sim$0.1\,rad\,s$^{-1}$ 
(or a period $P=63$\,s) at the start of the simulation (initial conditions in the YL05 progenitor), 
and $\sim$1000\,rad\,s$^{-1}$ ($P=6.3$\,ms) at the end the simulation, a spin-up factor of 10000.
In the 1.92-\mo model, we start with a much higher angular rotation rate of $\sim$20\,rad\,s$^{-1}$ 
($P=0.3$\,s), but the final values are comparable with those of the 1.46-\mo model, 
being $\sim$2800\,rad\,s$^{-1}$ ($P=2.2$\,ms). 
Thus, both simulations lead to the formation of a neutron star with a period of a few milliseconds, although
we expect the neutron star formed in the 1.92-\mo model to further accrete mass and angular momentum,
which may spin-up the residue to even shorter periods.
The general angular velocity and specific angular momentum profiles for both models are quite similar.
Despite wiggles observed in the 1.46-\mo model in the inner 10 kilometers (which we associate with 
slight numerical artifacts along the axis - this problem is not present in the 1.92-\mo model, whose
grid covers only 90$^{\circ}$), the neutron star is close to solid-body rotation out to 30\,km,
showing a steady and smooth decline with radius beyond.
In all four panels, we overplot as a broken and black line the corresponding local Keplerian angular velocity,
$\Omega_{\rm Keplerian}(r) = \sqrt{GM/r^3}$, where $G$ is the gravitational constant
and $M$ is the mass interior to the radius (cylindrical, or 
spherical).  (For these plots, we employ the corresponding neutron star mass.)
The angular velocity or specific angular momentum profiles beyond 30\,km graze the corresponding
line for the Keplerian value, being always lower by a few tens of percent.
In the 1.46-\mo model, the profiles evolve significantly toward this Keplerian limit, angular momentum being
gained along the equatorial direction through the radial infall of non-zero latitude material.
The constancy of the angular velocity with $z$ allows a significant gain from such infall.
In the 1.92-\mo model, the rotational properties along the equator are originally 
closer to the Keplerian values, but, accordingly, evolve little.
In both models, angular momentum is transported outwards, first in material blown away
by the shock wave initiated at core bounce, and then in the neutrino-driven wind.  
This occurs as the outflowing material wraps around the
progenitor white dwarf and eventually meets along the equator.
There is no ``physical'' viscosity in the code that would permit a proper
modeling of the accretion disk.
Mass accretion should occur in partnership with outward transport of angular momentum
over a longer timescale, yet to be determined.

In the 1.92-\mo model, at the end of the simulation, the near-Keplerian disk extends from 30\,km
out to $\sim$1800\,km, covering a range of densities (temperatures) from 10$^{13}$\,g\,cm$^{-3}$
down to 10$^{8}$\,g\,cm$^{-3}$ (3\,MeV down to 0.1\,MeV; Fig.~\ref{fig_rho_radslice}). 

\section{Energetics and the neutrino-driven wind}
\label{sect_ener}
%\clearpage
\begin{figure*}
% \plottwo{aic_1pt46_e_vcut10000kms.ps}{aic_1pt92_e_vcut10000kms.ps}
\plottwo{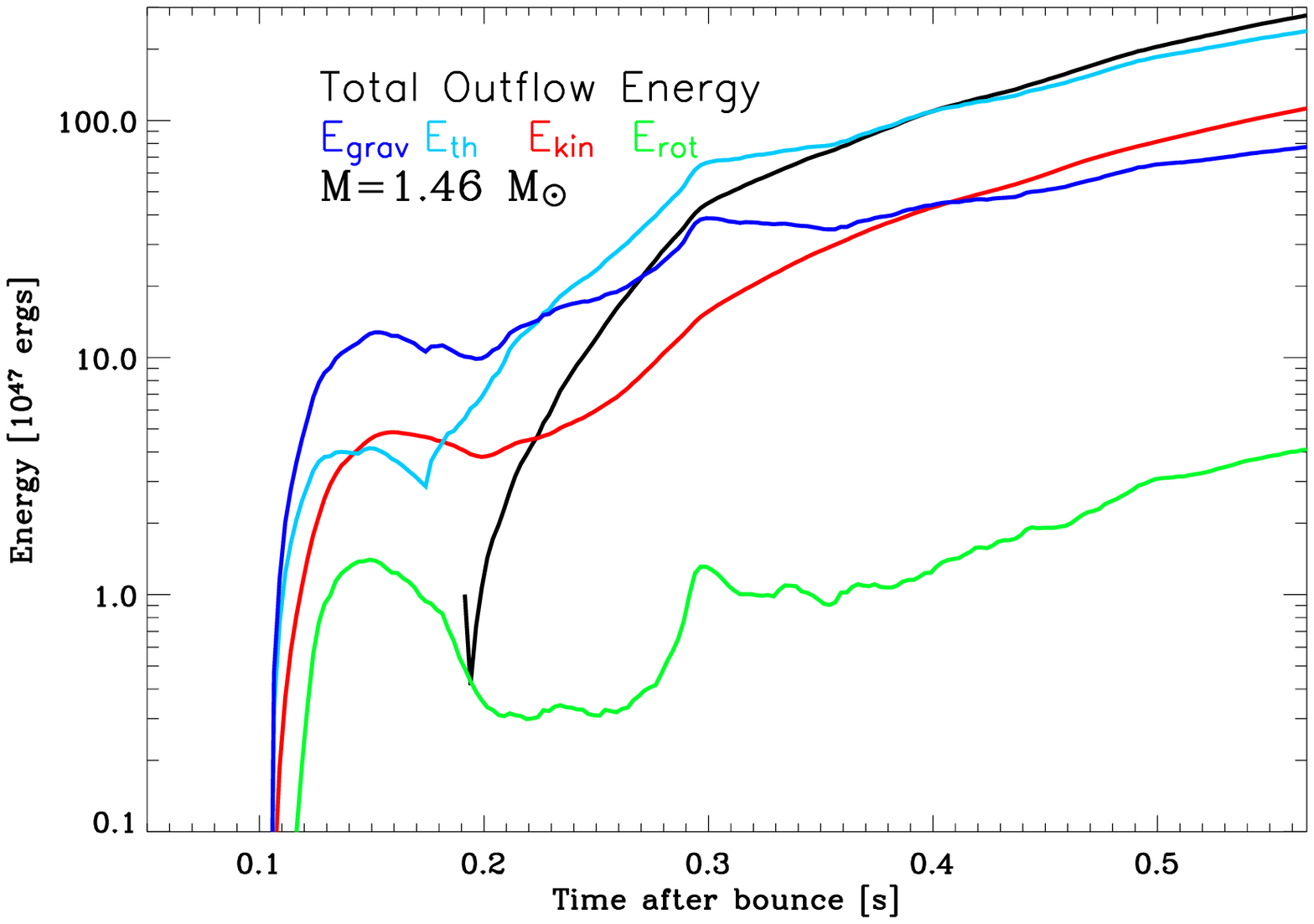}{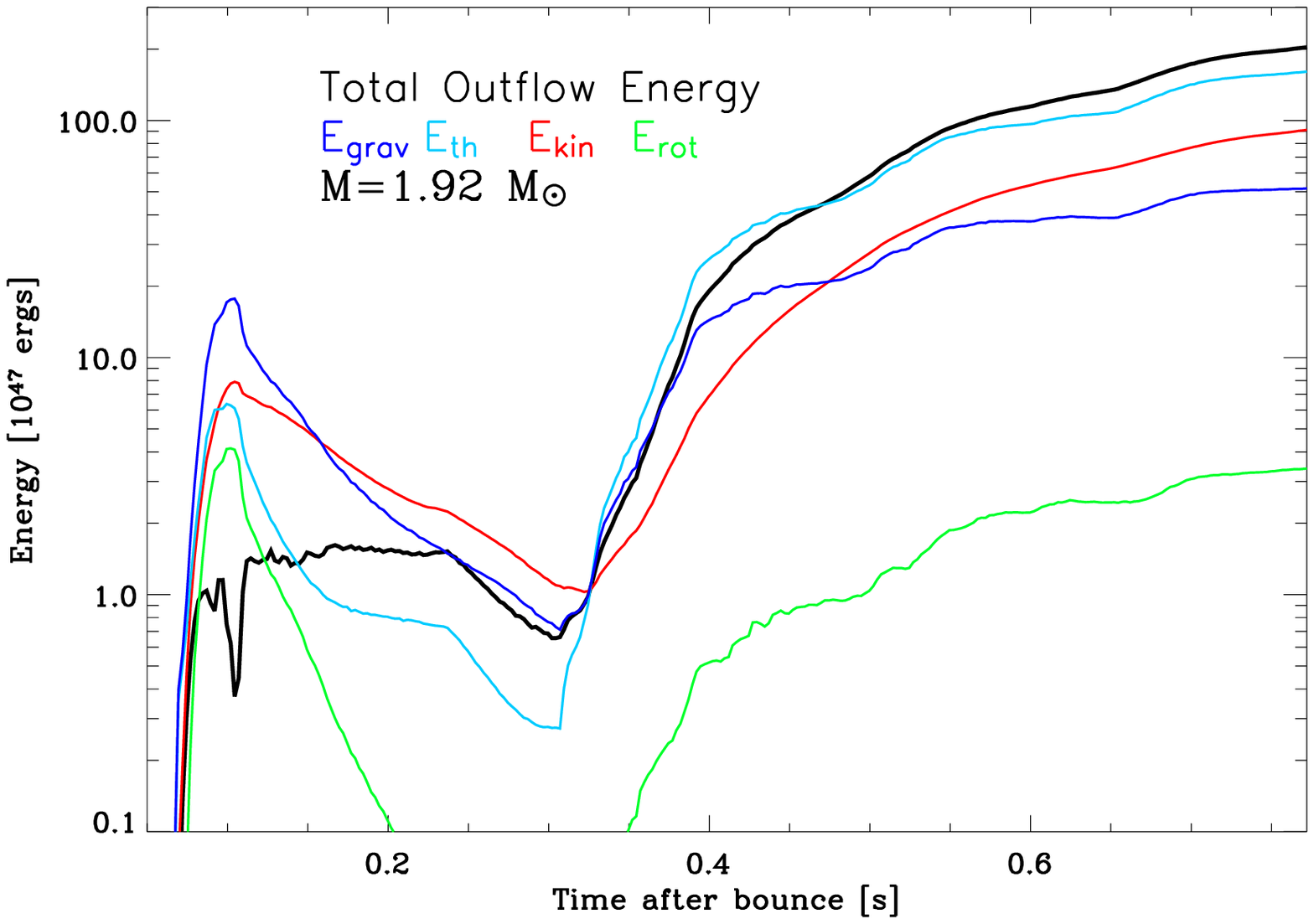}
\caption{
Time evolution of outflow gravitational (blue), thermal (cyan), 
($r,z$)--kinetic energy (red), and rotational (green) energy (on a log scale)
for the 1.46-\mo model (left) and 1.92-\mo model (right),
accounting for all the material with radial outflow velocity higher than 
10000\,\kms. The explosion energy at the end of the simulation is 2.7$\times$10$^{49}$erg 
for the 1.46-\mo model and 2$\times$10$^{49}$erg for the 1.92-\mo model, with final 
expected energies of $\sim$10$^{50}$erg and $\sim$5$\times$10$^{49}$erg, in the same order.
See text for details.
}
\label{fig_energy}
\end{figure*}

\begin{figure}
% \plotone{aic_1pt92_mass_loss.ps} 
\plotone{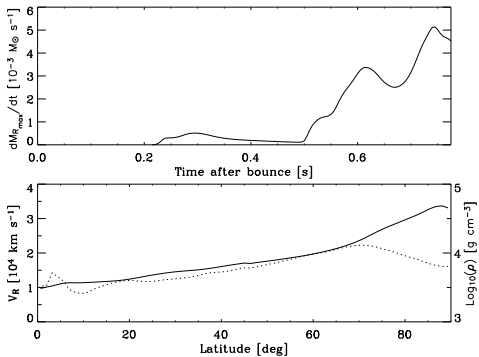} 
\caption{
{\it Top}: Time evolution of the total mass loss rate through the outer grid radius, 
$dM_{R_{\rm max}}/dt$, in the 1.92-\mo model.
{\it Bottom}: Latitudinal dependence of the radial velocity (solid line) and density (broken line)
in the 1.92-\mo model {\it at the latest time} computed, when the neutrino-driven wind
has reached a steady-state.
}
\label{fig_mdot}
\end{figure}

\begin{figure*}
% \plotone{colorbarflux.ps}
\plotone{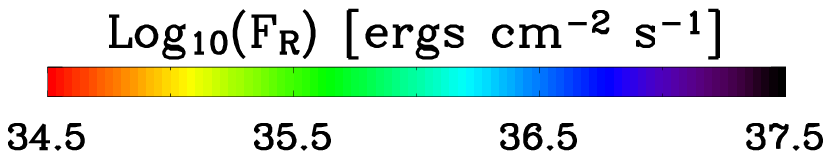}
%\vspace{-0.8cm}
% \plottwo{aic_m1pt46_flux_t=0.0160s.ps}{aic_m1pt92_flux_t=0.0165s.ps}
\plottwo{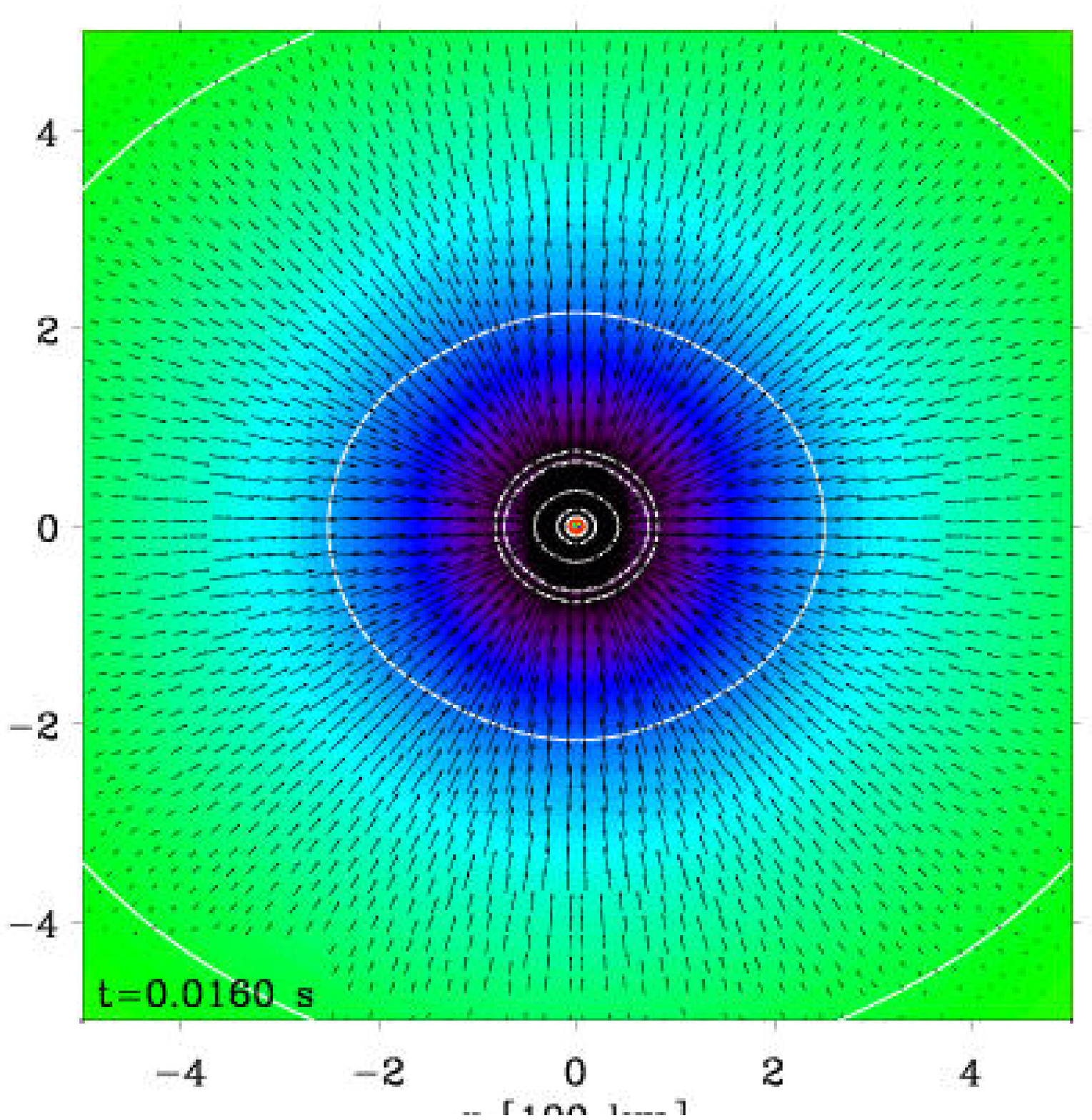}{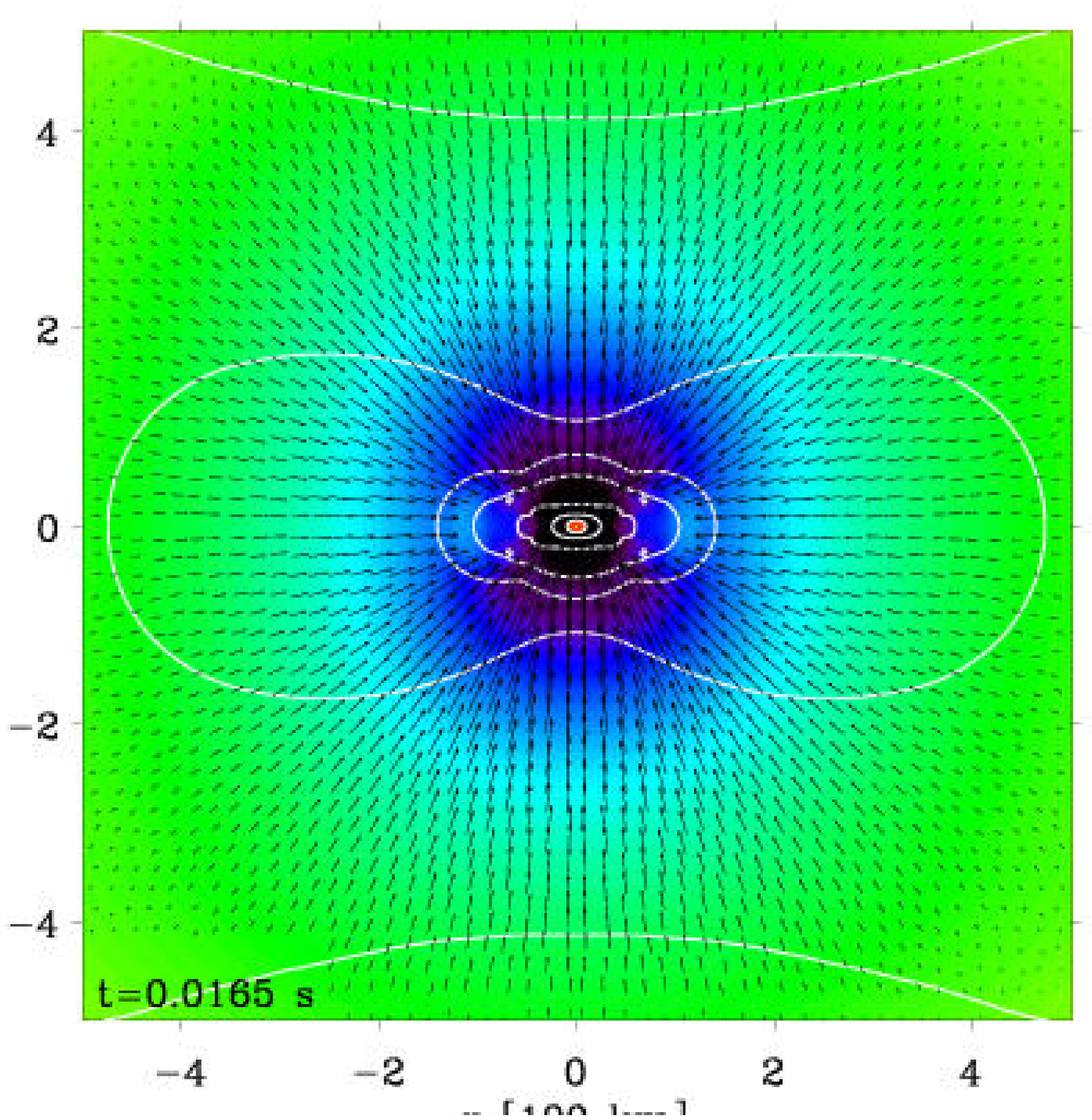}
\vspace{-0.3cm}
% \plottwo{aic_m1pt46_flux_t=0.5700s.ps}{aic_m1pt92_flux_t=0.7755s.ps}
\plottwo{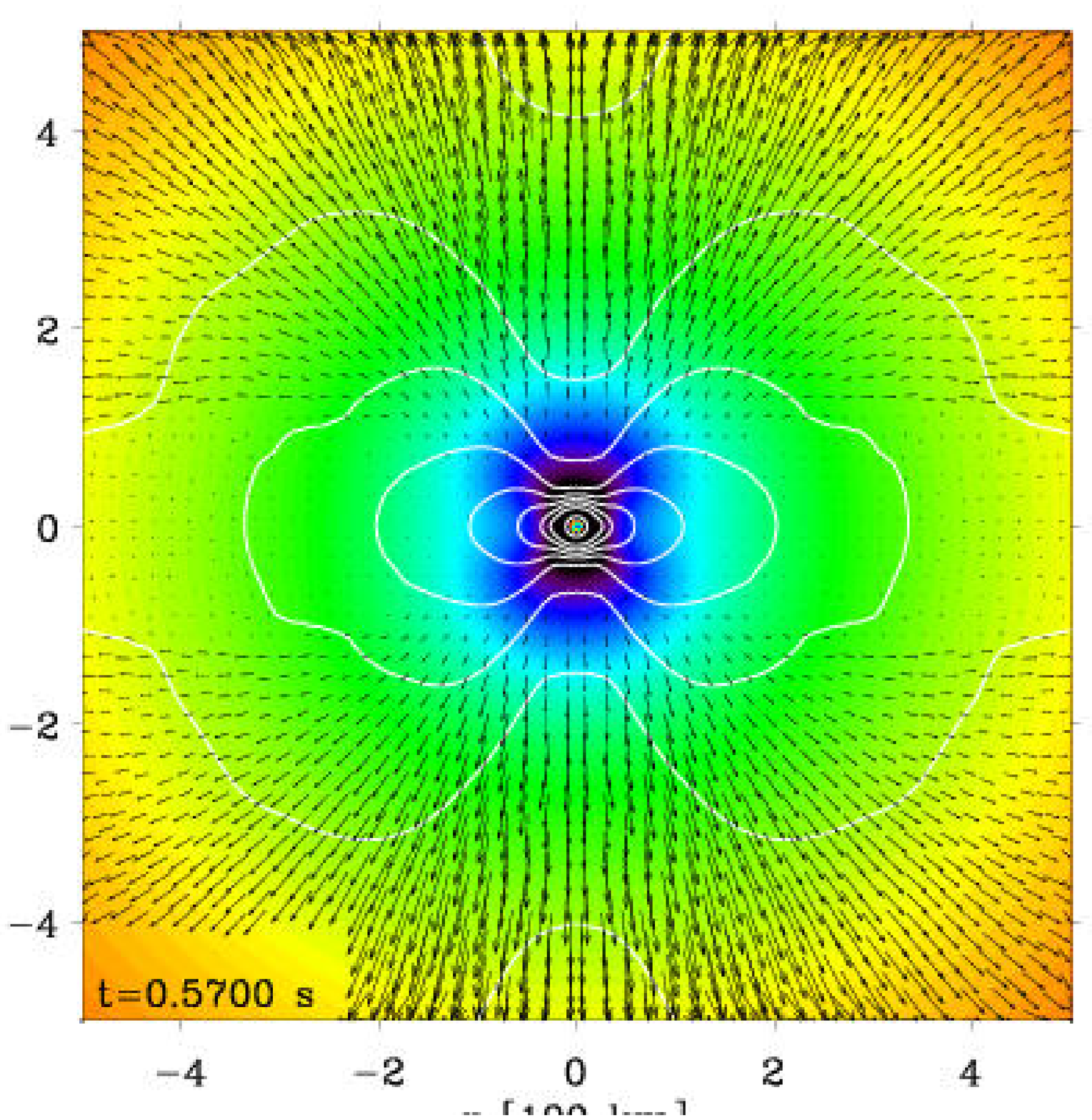}{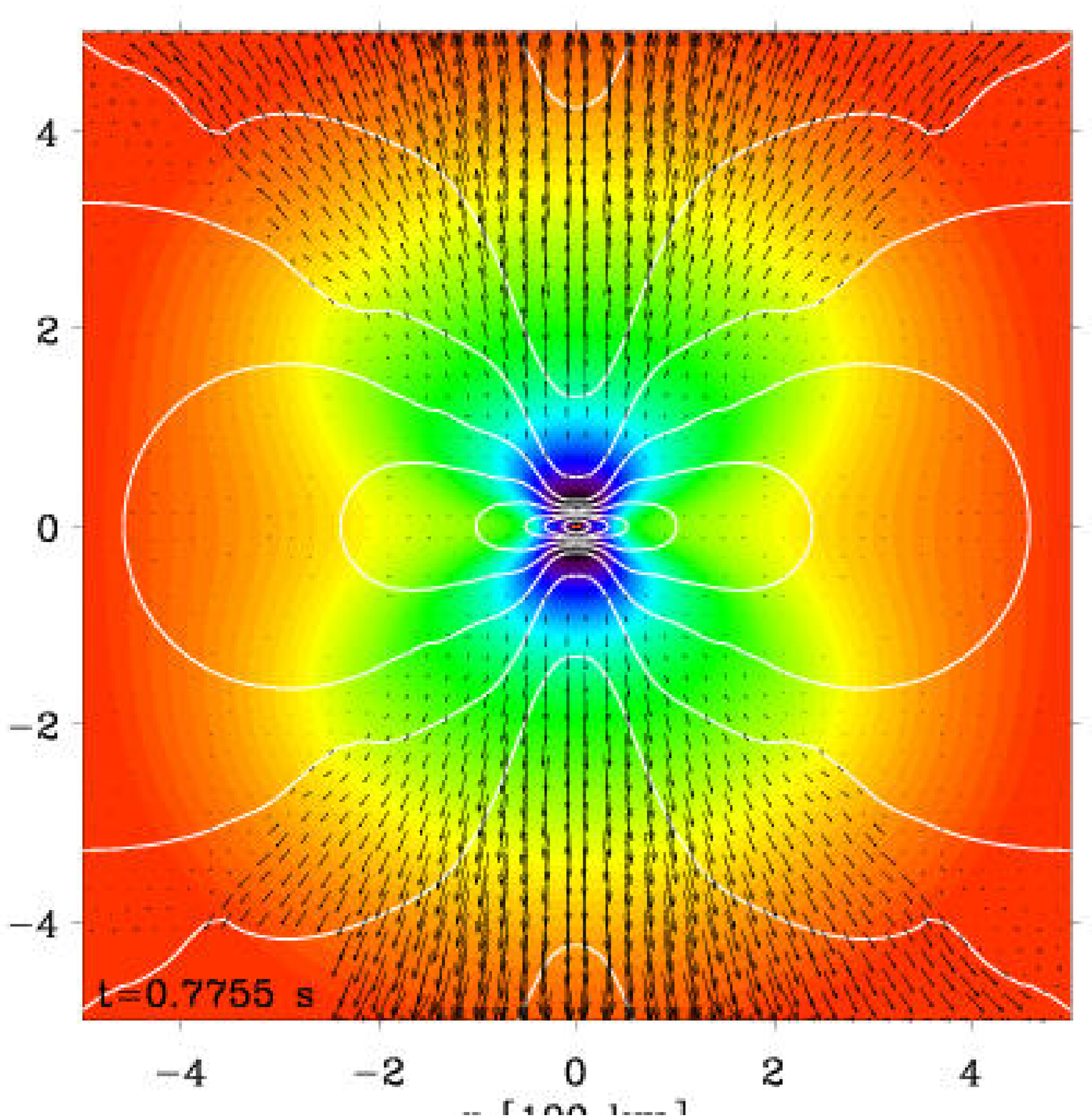}
\caption{    
Color maps of the total neutrino flux soon after bounce (top row) and at the last simulated
time (bottom row), for the 1.46-\mo model (left column) and 1.92-\mo model (right column),
with velocity vectors and density contours (for every decade starting from 10$^{14}$\,g\,cm$^{-3}$).
Note the growing flux anisotropy with time, more manifest for the 1.92-\mo model, with regions
of high flux coinciding with regions of low density, following primarily the angle-dependent
surface area of the neutrinosphere and, secondarily, the
latitudinal dependence of the temperature (Fig.~\ref{fig_temp}) near such neutrinospheres. 
Note the gravity darkening effect {\it \`a la von Zeipel}.
From top to bottom, left to right, the vector of maximum length corresponds to a velocity 
magnitude of 52500\,\kms (infall), 53000\,\kms(infall), 18500\,\kms (outflow), and
15500\,\kms (outflow).
}
\label{fig_nuflux_2d}
\end{figure*}

\begin{figure*}
% \plotone{aic_m1pt92_rho_nue_t=0.7755s.ps}
\plotone{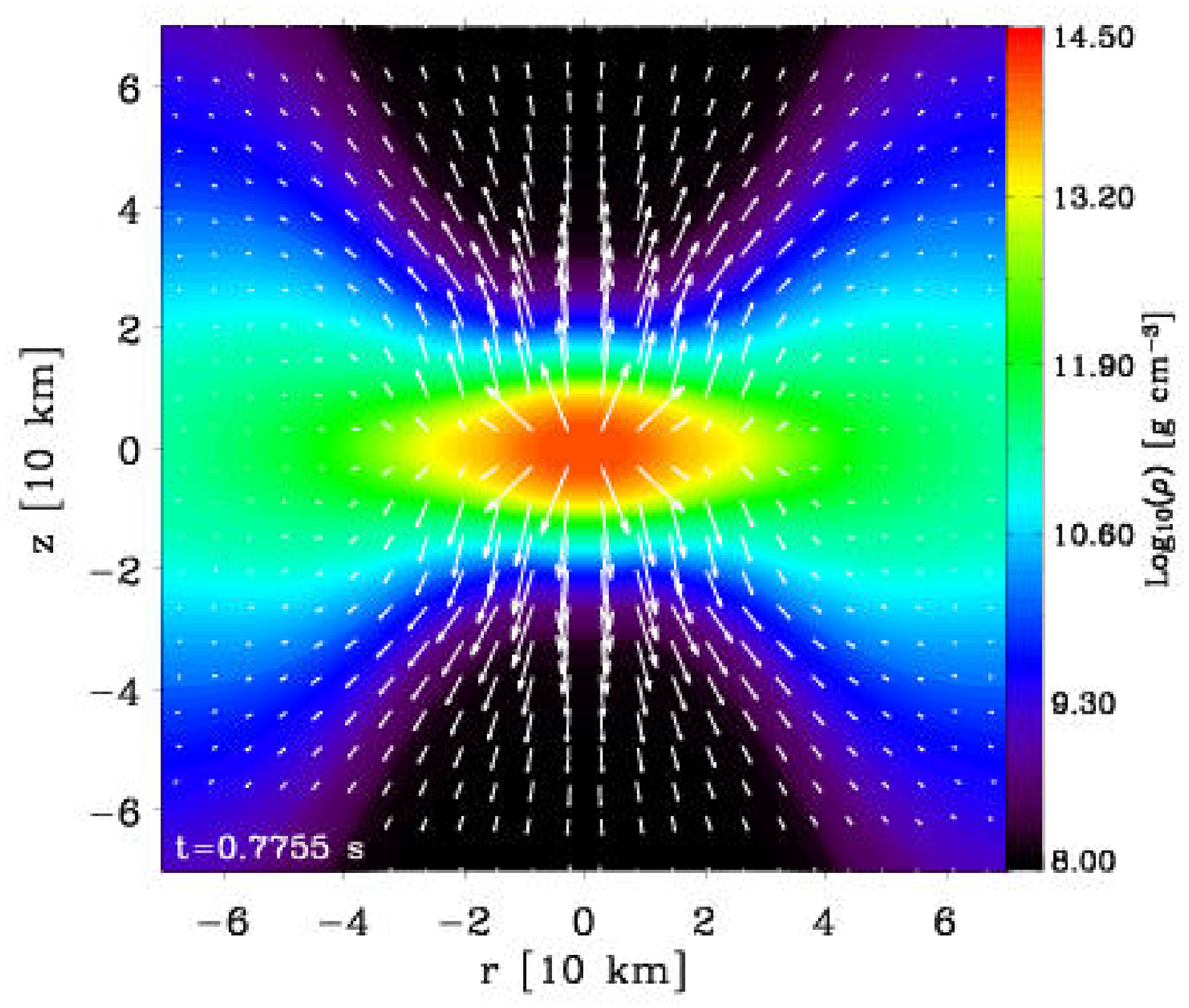}
\caption{
Color map of the density, in the inner $\sim$70\,km, overplotted with white vectors 
representing the energy-integrated 
electron-neutrino flux at the last time simulated for the 1.92-\mo model, presented for
comparison with the bottom-right panel of Fig.~\ref{fig_nuflux_2d}.
Note how the flux vectors tend to point in a direction perpendicular to the local 
isochores, modulo variations, e.g., due to the non-uniform temperature distribution, 
but in general conformity with the von Zeipel theorem which stipulates that the radiative 
flux should be aligned with the local effective gravity.
The vector of maximum length corresponds to a flux magnitude of 
3$\times$10$^{37}$\,erg\,cm$^{-2}$\,s$^{-1}$. 
For a more quantitative assessment of the electron-neutrino flux magnitude, see Fig.~\ref{fig_nuflux}
and Fig.~\ref{fig_nuflux_2d}.
}
\label{fig_nu_flux_vec}
\end{figure*}
%\clearpage
  Given that it leads to the formation of a $\sim$1.4-\mo neutron star, the AIC of a 
1.4--2.0\,\mo white dwarf is expected to result, in the case of a successful explosion, to an 
outflow of modest mass. Furthermore, due to the similarity with the core collapse
of massive star progenitors, their explosion kinetic energy should be lower than 
the $\sim$10$^{51}$ erg inferred, e.g., for SN1987A (Arnett 1987).

   As discussed in \S\ref{sect_results} and \S\ref{sect_disk}, the total sum of 
the neutron star and disk masses is very close to the original progenitor mass, leaving 
typically a few 0.001\,\mo for the ejected material. Integrating all the mass that has 
left the grid over the course of the simulation, as well as all the material outflowing 
with a positive radial velocity greater than 10000\,\kms (which is of the order of the 
escape velocity at a radius of 3000\,km) we find a value of 4$\times$10$^{-3}$\,\mo for the 1.46-\mo 
model and 3$\times$10$^{-3}$\,\mo for the 1.92-\mo model. 
In Fig.~\ref{fig_energy}, we show the evolution of the 
corresponding gravitational (blue), thermal (cyan), and kinetic (2D planar: red; 
rotational: green) energies for this outflowing material as solid lines, 
including the total energy as a black dotted line. 
At the last computed time, the total energy is indeed lower than that inferred for standard 
core collapse. Adopting a radial velocity cut of 10000\,\kms, we find an energy at the 
last simulated time of 2.7$\times$10$^{49}$erg for the 1.46-\mo model and
2$\times$10$^{49}$erg for the 1.92-\mo model.
Note that energy is still being pumped into the wind by the slowly decaying neutrino
luminosity emanating from the neutron star; the trend of the total energy curve suggests that the
total energy of the explosion will be 2-3 times higher, thus $\sim$10$^{50}$erg for the 1.46-\mo model
and $\sim$5$\times$10$^{49}$erg for the 1.92-\mo model.

This is over one order of magnitude smaller than the 
explosion energy inferred for normal core-collapse supernovae.
The AIC of white dwarfs is likely to lead generically to underenergetic explosions because
there is too little mass to absorb neutrinos, most of it being quickly accreted while
the rest is centrifugally-supported at large radii, far beyond the region
where there is a positive net gain of electron-neutrino energy.
Interestingly, similar underenergetic explosions are obtained by Kitaura et al. (2005)
and Buras et al. (2005b) for initial main sequence stars of 8.8\,\mo (Nomoto 1984, 1987) 
and 11.2\,\mo (Woosley, Heger, \& Weaver 2002). Echoing the properties of AIC progenitors, 
the low envelope mass and fast declining density (and, therefore, accretion rate) are 
key beneficial components for the success of neutrino-driven explosions, but the same properties are
also why the explosion is necessarily underenergetic.

The various curves also show a few dips and bumps. The first bump, most pronounced 
in the 1.92-\mo model, is associated with an early outflow that eventually fell back 
to smaller radii, while subsequent bumps are caused by episodic mass loading of the 
neutrino-driven wind, which sets in $\sim$200\,ms after core bounce and drives a 
8$\times$10$^{-3}$\,\mo\,s$^{-1}$ (5$\times$10$^{-3}$\,\mo\,s$^{-1}$) mass loss rate 
in the 1.46-\mo (1.92-\mo; see also Fig.~\ref{fig_mdot}, top panel) model. 
The higher-mass flux in the 1.46-\mo model
results from the higher neutrino luminosity, higher mean neutrino energies, and 
bigger opening angle of escape for the neutrino-driven wind.
Interestingly, this mass flux is strongly angle-dependent, varying by a factor of a 
few between the pole and the angle that grazes the pole-facing side of the disk.
As described in \S\ref{sect_results}, the dynamical effects of the neutrino-driven 
wind are to entrain the material lying along this interface, tearing the disk 
via Kelvin-Helmholtz shear instabilities and 
mass-loading the wind along the corresponding latitudes.
Further in, neither this wind nor the neutrinos have an appreciable dynamical impact in driving
the disk material outwards, a feature only excacerbated by the reduced neutrino flux at low latitudes. 
In Fig.~\ref{fig_mdot}, we show in the bottom panel the latitudinal variation of the asymptotic 
velocity (solid line) and the density (dotted line).
Both show an overall decrease towards lower latitudes by a factor of three. The dip in density and higher 
values of the velocity along the pole to around 70$^{\circ}$ latitude are possibly 
due to wind mass loading, in combination with centrifugal support at the neutrinosphere
for off-polar latitudes.
In Fig.~\ref{fig_nuflux_2d}, we show at early times (top row) and at the last simulated
time (bottom row) for the 1.46-\mo model (left column) and the 1.92-\mo model (right column)
color maps of the total neutrino flux in the radial direction, with isodensity contours
overplotted as white curves, and velocity vectors as black arrows.
First, due to the history of the collapse, the neutron star is relatively devoid of overlying material
in the polar direction, while for the higher-mass progenitor a massive ($\sim$0.6\,\mo), 
dense (10$^{6-10}$\,g\,cm$^{-3}$ ), near-Keplerian disk obstructs the neutron star at latitudes 
$\sles\pm$40$^{\circ}$. 
Given this configuration, conditioned essentially by the mass distribution of the progenitor 
white dwarf, the dynamical effect of a spherically-symmetric neutrino flux would be enhanced along
the ``excavated'' polar direction. Indeed, we see a strong neutrino-driven wind in the polar 
direction that does not exist in directions within $\sim\pm$40$^{\circ}$ of the equator.
However, even in the absence of this anisotropic matter distribution, Fig.~\ref{fig_nuflux_2d}
reveals the strong latitudinal variation of the neutrino flux at a given Eulerian radius, a variation
that is established independently of the configuration of the circum-neutron star disk material.
What controls the flux geometry is the combination of two effects. First, the exceptional elongation of 
the neutrinospheres along the equatorial direction leads to a decoupling radius (surface) about 10 (100)
times bigger for a polar observer than for an equatorial observer. The angle-dependent decoupling
radius of neutrinos mitigates this result (Walder et al. 2005), but, as shown in 
Fig.~\ref{fig_nuflux_2d}, the latitudinal
variation along different directions persists in the total neutrino flux. 
Similarly, Fig.~\ref{fig_nu_flux_vec} shows the anisotropy of the $\nu_e$ neutrino flux,
rendered by the corresponding flux vectors. 
Notice how the base of the flux vectors in the high-density central regions is perpendicular
to the local isodensity (or, equivalently, equipotential) contour. 
Second, as shown in Fig.~\ref{fig_temp}, the temperature and its radial-gradient along 
a given isodensity contour are both significantly lower along the equatorial direction, leading 
to reduced diffusive fluxes.
These properties are reminiscent of the effect of gravity darkening (von Zeipel 1924) in fast rotating 
(non-compact) stars and the associated scaling of the radiative flux with the local effective gravity
(see Owocki et al. 1996), although this may be the first time it is 
reported in the context of a protoneutron star (but see Walder et al. 2005).
With such a polar-enhanced wind, the angular momentum loss rate is 
reduced, with consequences for the spin evolution of the PNS.

\section{Ejecta composition}
\label{sect_ye}
%\clearpage
\begin{figure}
% \plotone{ye_dist_aic_vcut10000kms.ps}
\plotone{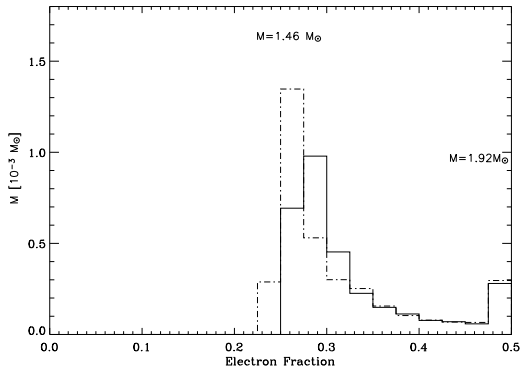}
\caption{
Distribution of the total outflow mass (cells with outward radial velocity of at least 10000\,\kms) 
as a function of electron fraction, taken at the
last time for both models (solid line: 1.92-\mo model; broken line: 1.46-\mo model).
The total mass in the outflow, making allowance for any mass lost through the outer grid radius,
is 0.003\,\mo for the 1.92-\mo model and 0.004\,\mo for the 1.46-\mo model.
}
\label{fig_ye_dist}
\end{figure}

\begin{figure*}
% \plottwo{aic_m1pt92_ye_t=0.7755s_inner_region.ps}{aic_m1pt92_gain_nue_t=0.7755s_inner_regions.ps}
\plottwo{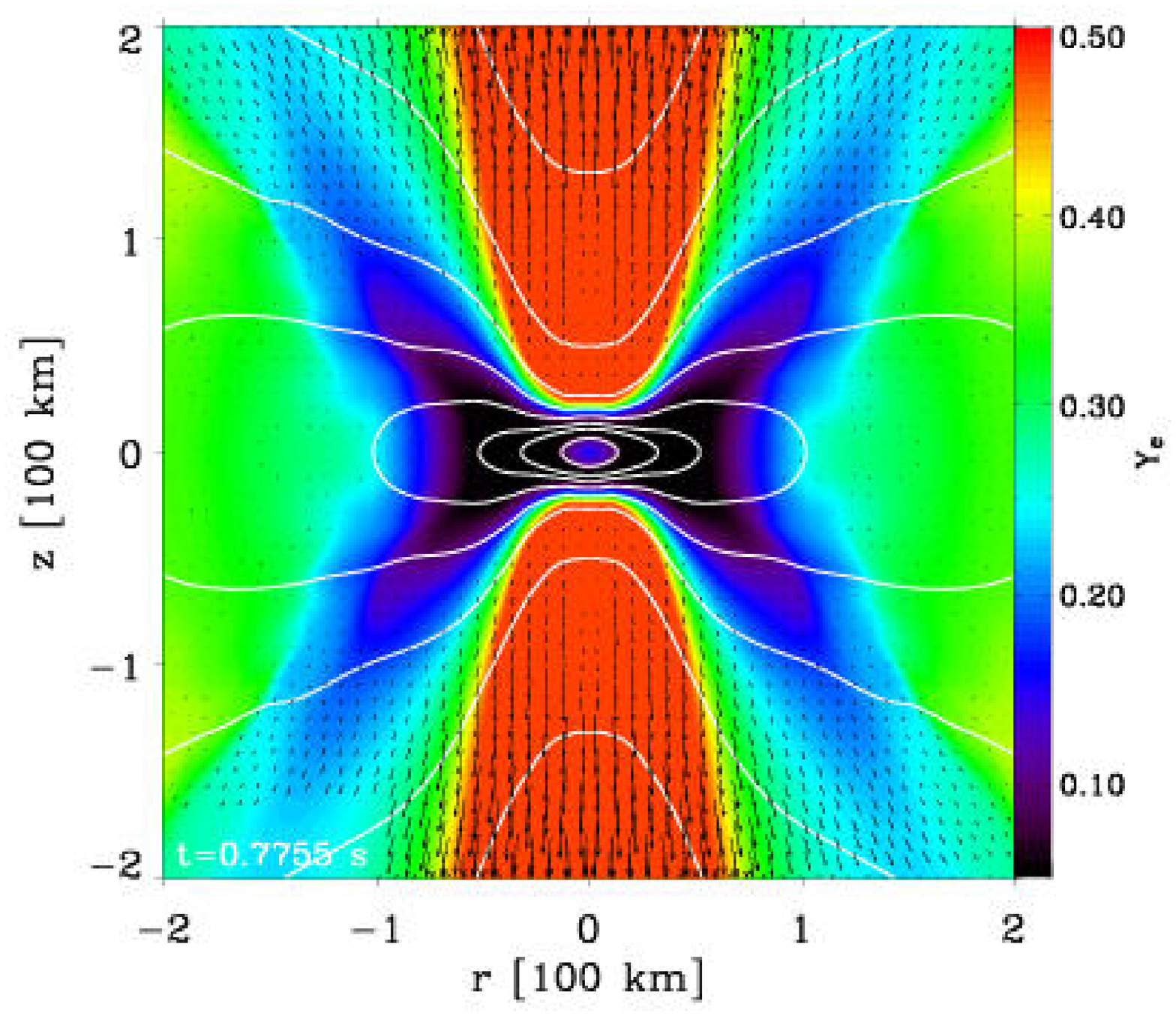}{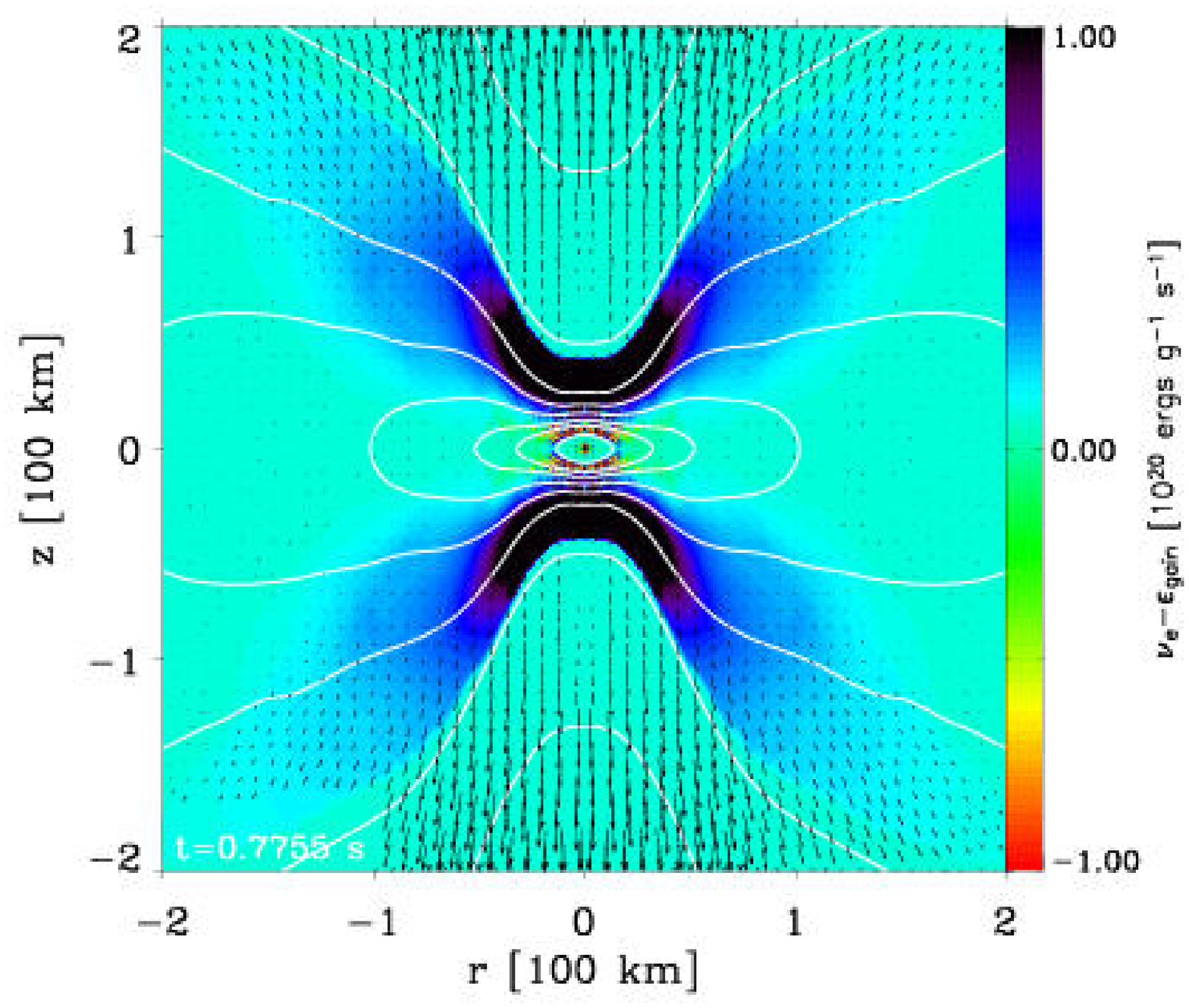}
\caption{
{\it Left}: Color map of the electron fraction in the inner regions of the 1.92-\mo model at the 
last computed time, with isodensity contours 
(white lines; every decade, starting at 10$^{14}$\,g\,cm$^{-3}$) and velocity vectors 
(black arrows) overplotted, showing in particular the locations of the 
deleptonized regions and the neutrinospheres.
Notice the $Y_{\rm e}$ change in the material as it moves away from the neutrinosphere 
(roughly the 10$^{10-11}$\,g\,cm$^{-3}$ contour). It rises steeply from 0.05 to 0.5
in just a few tens of kilometers along the pole, while this rise is more gradual
at $\sim$45$^{\circ}$ latitude, saturating at $\sim$0.25.
{\it Right}: Same as left, but for the net energy gain due to absorption (positive) and
emission (negative) of electron neutrinos (the colorbar range used saturates the gain values 
in the inner regions). Notice how much stronger (and positive) the gain is 
along the poles, causing the $Y_{\rm e}$ of the material to rise steeply to 0.5, while its more modest
value away from the pole alters the $Y_{\rm e}$ of the corresponding material only up to $\sim$0.25.
Note that the anti-electron neutrino flux is one order of magnitude lower than that of the
electron neutrinos, and thus, too weak to cause the reverse process of a decrease of the 
$Y_{\rm e}$ of the material.
The vector of maximum length corresponds to a velocity magnitude of 9130\,\kms. 
}
\label{fig_ye_gain}
\end{figure*}
%\clearpage
In Fig.~\ref{fig_ye_dist}, we show for the two baseline models the electron fraction ($Y_{\rm e}$) distribution
of the material in the ejecta (material outside the neutron star moving outwards with a radial 
velocity greater than 10000\,\kms), accounting as well for the mass loss through the
outer grid radius. Such cumulative outflow amounts to 
4$\times$10$^{-3}$\,\mo for the 1.46-\mo model and 3$\times$10$^{-3}$\,\mo for the 1.92-\mo model.
We obtain double-peak profiles, the first blast propelling symmetric material 
($Y_{\rm e}=0.5$), subsequently followed after 200\,ms by progressively neutron-rich material, i.e., 
$Y_{\rm e} =0.25-0.35$, in the neutrino-driven wind. Note that for these runs we enforced an upper limit of 0.5
to the computed $Y_{\rm e}$ values.
Fryer et al. (1999) obtained an ejecta mass in the vicinity of 0.2\,\mo, two orders of 
magnitude larger than our values. Because our ejecta masses are much smaller, we find that 
the mass loss rates and the kinetic energies associated with the neutrino-driven 
wind are relatively more important for the global energetics of the AIC of white dwarfs.

At late times, the asymptotic electron fraction $Y_{\rm e}^{\rm a}$ of the neutrino-driven wind 
varies with latitude 
(despite the smooth variation of other quantities at correspondingly larger distances). 
Material ejected within 20$^{\circ}$ of the pole has an electron fraction of $\sim$0.5, while towards
the equator, this electron fraction decreases to 0.3, rising again to near 0.5 values
in regions belonging to the disk (see bottom row panels in Fig.~\ref{fig_seqye}).
The $Y_{\rm e}^{\rm a}$ values seen in our simulations are in fact already set when wind material
leaves the vicinity of the neutrinosphere, whose properties depend on the particle 
trajectory under scrutiny.
We know from previous studies (Qian \& Woosley 1996; Wheeler et al. 1998; Thompson et al. 2001;
Pruet et al. 2005; Fr\"{o}lich et al. 2005) that the
asymptotic electron fraction of the ejecta is controled by competing factors. 
The electron and anti-electron neutrino luminosities, modulated by the hardness of their respective
energy distributions, influence the electron flavor production rates via the reactions 
$\nu_e$n$\rightarrow$pe$^{-}$ and $\bar{\nu}_e$p$\rightarrow$e$^+$n and, thereby, the 
neutron-richness of the ejecta.
The expansion timescale sets the duration over which interactions between neutrinos and
nucleons can take place.

% The electron fraction at the base of the outflow will be the starting values upon which 
% the above factors act to determine the asymptotic values seen.
The starting value of the electron fraction, i.e., at the base of the outflow, is altered by
the above factors and differs from the asymptotic value seen.
In Fig.~\ref{fig_ye_gain}, we show a color map of the electron fraction in the inner 200\,km,
highlighting the butterfly shape of the deleptonized region in cross section, in stark contrast with the 
corresponding near-spherical shape seen in core-collapse simulations of both rotating and non-rotating
progenitors (Keil et al. 1996; Walder et al. 2005; Dessart et al. 2005).
Deleptonization obtains preferentially in the vicinity of the dumbbell-shaped neutrinosphere, 
and stretches outwards for off-polar latitudes. Along the equator, deleptonization
ceases at smaller radii due to the lower effective temperatures (tied to the neutrino fluxes).
Temperature and neutrino flux are in fact intertwined, since energy deposition by neutrinos
may raise the temperature locally in the so-called gain region.
This is also vividly represented in Fig.~\ref{fig_ye_gain} (right) by the net gain associated with 
electron-neutrino energy deposition in this inner region, which also shows the same butterfly shape. 
Electron neutrinos emerge from the neutron star, and, due to the dumbbell neutrinosphere morphology,
at much smaller radii along the poles than for off-polar latitudes.
The decreasing neutrino flux (dilution) reduces this energy deposition
beyond $\sim$50\,km, and even at smaller radii along the equator due to the additonal
flux reduction there (Fig.~\ref{fig_nuflux_2d}).

The asymptotic value of the material electron fraction is determined in the vicinity
of the neutrinosphere and, therefore, is directly influenced by this configuration of the
inner $Y_{\rm e}$ distribution. Along the pole, the wind carries initially low $Y_{\rm e}$ material that
absorbs electron neutrinos, whose associated luminosity is one magnitude higher than that of the
anti-electron neutrinos (see Fig.~\ref{fig_nuflux}), raising the electron fraction to the ceiling
value of 0.5 artificially adopted in these calculations.
Away from the pole, the neutrinosphere is located further
out, and despite similar neutrinosphere $Y_{\rm e}$ values, the larger distance from the neutron star implies
a reduced electron-neutrino luminosity and a reduced absorption of neutrinos, leading to asymptotic
values of the electron fraction of only $\sim$0.25, not far from the values at the 
corresponding neutrinosphere. 
To summarize, the progressive decrease of the electron fraction (and of the entropy) away from
the pole is a result of the reduced electron-neutrino luminosities in the vicinity of the 
latitudinal-dependent neutrinosphere radius (and the associated reduced heating and electron-capture 
rates).   

\section{Gravitational wave signature}
\label{sect_gw}

We estimate the gravitational wave emission from aspherical mass
motions in our models via the Newtonian quadrupole formalism as
described in M\"onchmeier et al. (1991). In addition, we compute
the gravitational wave strain from anisotropic neutrino emission
employing the formalism introduced by Epstein (1978) and developed 
by Burrows \& Hayes (1996) and M\"uller \& Janka (1997).

Due to rapid rotation and the resulting oblateness of the core,
one would expect that rotating AIC models would have significant
gravitational wave signatures.  
% This is true.  
However, though the contribution
to the metric strain in the equatorial plane, $h_+$,
of the aspherical and dynamical matter distributions is not small, we find
that that of the aspherical neutrino field is larger in magnitude, though at
much smaller frequencies.  While we calculate that $h_+$(max) for the matter in the
1.46-\mo\ model is $\sim$5.9$\times$10$^{-22}$, with a spectrum that peaks 
at $\sim$430 Hz, the corresponding $h_+$(max) due to neutrinos
is $\sim$4.6$\times$10$^{-21}$ (derived from the fluxes at 200 km), but at
frequencies between ($\sim$0.1-10 Hz).  The total energy radiated in 
gravitational waves is $\sim$5.7$\times$10$^{-10}$ \mo{$c^2$}. 
The corresponding numbers for the faster rotating and more massive 1.92-\mo\ model 
are $\sim$3.6$\times$10$^{-21}$ (matter),
$\sim$165 Hz, $\sim$2.0$\times$10$^{-20}$ (neutrinos at 300 km), and
$\sim$7.0$\times$10$^{-8}$ \mo{$c^2$}. Note that almost all of the 
energy is being emitted by mass motions (99.8\% in the 1.46-\mo\ model and 
98.4\% in the 1.92-\mo\ model), since the power scales with the time derivative of
the wave strain $h_+$, which is small for the waves emitted from the 
aspherical neutrino field.

We compare the above numbers with those obtained by Fryer, Holz \& Hughes (2002) for an
AIC model of Fryer et al.\ (1999) which was setup with a simple, solid-body
rotation law and had a final T/$|$W$|$ of $\sim$0.06 
(for our 1.46-\mo\ model: 0.059; see Table 1). They did not consider anisotropic
neutrino emission. Our more realistic initial models
yield maximum (matter) gravitational wave strains that are 1.5 to 2 orders
of magnitude smaller than those predicted by Fryer, Holz \& Hughes (2002). The total
energy emissions match within a factor of two since our models emit at higher 
frequencies. 

Based on our results, we surmise that gravitational waves from axisymmetric 
AIC events may be detected by current LIGO-class detectors if occurring anywhere 
in the Milky Way, but not out to megaparsec distances as suggested by 
Fryer, Holz \& Hughes (2002). It is, however, likely (\S 5) that at least the
1.92-\mo\ model will undergo a dynamical rotational instability leading to non-axisymmetric
deformation (which can not be captured by our 2D approach) and, hence, to copious gravitational 
wave emission over many rotation periods, greatly enhancing detectability.

\section{Discussion and conclusions}
\label{sect_conc}

   We have presented a radiation/hydrodynamic study with the code VULCAN/2D of the 
collapse and post-bounce evolution of massive rotating high-central-density white dwarfs,
starting from physically consistent 2D rotational equilibrium configurations
(Yoon \& Langer 2005). The main results of this study are: 

\begin{itemize}
\item The AIC of white dwarfs leads to a successful explosion with modest energy 
$\sles$10$^{50}$\,erg, thus comparable to the energies obtained through the collapse of 
O/Ne/Mg core of stars with $\sim$8-11\,\mo main sequence mass (Kitaura et al. 2005; 
Buras et al. 2005b).
This is, however, underenergetic, by a factor of about ten, compared with the inferred value for
the core collapse of more massive progenitors leading to Type II Plateau supernovae. 
Although less and less likely to be the engine behind most core-collapse supernova explosions, 
the neutrino mechanism can successfully power explosions of low-mass progenitors and AICs due to the
limited mantle mass and steeply declining accretion rate. 

\item Due to high-mass and angular-momentum accretion, white dwarf progenitors 
leading to AIC can have masses of up to $\sim$2\,\mo and rotate fast, with rotational
to gravitational energy ratios of up to a few percent prior to collapse.
The asphericity of such white dwarfs allows the shock generated at core bounce to
escape along the poles in just a few tens of milliseconds, opening a hole in the white dwarf
along the poles. The blast expands and wraps around the progenitor and escapes the grid
($\sim$5000 km) within a few hundred milliseconds, at which time a neutrino-driven wind has grown in
the pole-excavated region of the white dwarf. Both the original blast and the wind
show strong latitudinal variations, partly constrained by the obstructing uncollapsed equatorial
disk regions of the progenitor, whose centrifugal support prevents it from collapsing on a dynamical
timescale. Rotation in such progenitors, thus, affects both directly and indirectly the morphology
of the explosion.

\item The neutron stars formed have masses on the order of 1.4\,\mo, with rotation periods close
to a millisecond in the rigidly rotating inner $\sim$30\,km. The final rotational to gravitational
energy ratios, for our two test cases, cover 0.06 to 0.26, the latter being large enough to
grow non-axisymmetric instabilities. At the end of our simulations, 
the neutron stars are oblate and pinched along the poles, with polar and equatorial radii in the 
ratio 1:15 for the faster rotating (1.92-\mo) model. 

\item The morphology of the neutron star leads to a latitudinal variation of the 
neutrino flux, the net energy gain, and the temperature. In the faster rotating model, 
the ``$\nu_\mu$'' neutrino flux is reduced, while the anti-electron neutrino
flux is a factor of ten lower than that of the electron-neutrino. 
This raises the electron fraction of the ejected material to values close to 0.5 along the poles,
but to only $\sim$0.25 at lower latitudes, since the corresponding neutrinosphere is more remote and,
thus, the electron-neutrino flux is smaller. This introduces
a latitudinal dependence of the electron fraction of the ejected material, but more importantly
allows neutron-rich material, with entropy on the order of 20-40\,$k_{\rm B}$/baryon, to escape.
Thus, a low-entropy r-process might take place under these conditions.

\item The high original angular momentum of the progenitor follows the mass and is, thus, found 
mostly in the neutron star at the end of the simulation. However, rotational energy is
also given to the ejecta, which uses it to gain (planar) radial kinetic energy to escape
the potential well, while the rest is found in a quasi-Keplerian disk of up to 0.5\,\mo in 
the 1.92-\mo model. This disk is an essential component of these AICs; it collimates the explosion
and the neutrino-driven wind and also suggests a second stage of long-term accretion onto the compact
remnant. 

\item The total ejected mass is only of the order of a few times 0.001\,\mo, with only a quarter in
the form of $^{56}$Ni. The original blast and the short-lived neutrino-driven wind will lead
to a considerable brightening of the object, but the small ejecta mass will quickly become 
optically-thin, gamma rays leaking out rather than depositing their energy to power a durable 
light curve. Therefore, these explosions should be underluminous and very short lived.
Their appearance may also vary considerably with viewing angle, depending on the mass of
the progenitor and the presence of a sizable disk in the equatorial regions.

\end{itemize}

This study has shown that more consistent, rotating 2D models alter considerably our understanding of
accretion-induced collapse, previously obtained under the simplifying assumptions of spherical symmetry 
and/or zero rotation. Further improvements will come by including 
a consistent temperature structure for the progenitor white dwarf and by accounting for
the effects of magnetic fields.  Due to the rapid differential rotation, magnetic fields
could be amplified considerably and result in MHD jets that might alter yet again 
our overall picture of accretion-induced collapse and the energetics of the phenomenon.
Three-dimensional effects may also alter the protoneutron star properties presented here, since 
the faster rotating (1.92-\mo) model is expected to experience non-axisymmetric instabilities.

\acknowledgments

We acknowledge discussions with and help from
Jeremiah Murphy and Casey Meakin.
Importantly, we acknowledge support for this work
from the Scientific Discovery through Advanced Computing
(SciDAC) program of the DOE, grant number DE-FC02-01ER41184
and from the NSF under grant number AST-0504947.
E.L. thanks the Israel Science Foundation for support under grant \# 805/04,
and C.D.O. thanks the Albert-Einstein-Institut for providing CPU time on their
Peyote Linux cluster. We thank Jeff Fookson and Neal Lauver of the Steward Computer Support Group
for their invaluable help with the local Beowulf cluster.
This research used resources of the National
Energy Research Scientific Computing Center, which is supported by the
Office of Science of the U.S. Department of Energy under Contract No.
DE-AC03-76SF00098.

\end{document}